\RequirePackage{pdf14}

\documentclass[letterpaper]{article}
\usepackage[usenames,dvipsnames,table]{xcolor}
\usepackage{graphicx}
\usepackage{caption}
\usepackage{subcaption}
\usepackage{array,setspace,mathrsfs,amsthm,amsfonts,amsmath,yfonts,dsfont,bbm,colonequals,mathtools}
\usepackage{./aux/jheppub}
\usepackage{afterpage}
\usepackage{xspace}
\usepackage[normalem]{ulem}

\makeatletter
\def\maketag@@@#1{\hbox{\m@th\normalfont\normalsize#1}}
\makeatother
\def\Om{{\mathcal{O}}}
\def\Cm{{\mathcal{C}}}
\def\Em{{\mathcal{E}}}
\def\Am{{\mathcal{A}}}
\def\Gm{{\mathcal{G}}}
\def\Bm{{\mathcal{B}}}
\def\Nm{{\mathcal{N}}}
\def\Fm{{\mathcal{F}}}

\def\Dm{{\mathcal{D}}}
\def\Qm{{\mathcal{Q}}}
\def\Mm{{\mathcal{M}}}
\def\Pm{{\mathcal{P}}}
\def\Km{{\mathcal{K}}}
\def\Sm{{\mathcal{S}}}
\def\Rm{{\mathcal{R}}}

\def\a{{\alpha}}
\def\b{{\beta}}
\def\ad{{\dot{\alpha}}}
\def\bd{{\dot{\beta}}}

\def\ph{\phantom}

\def\nn{\nonumber}

\newcommand\be{\begin{equation}}
\newcommand\bea{\begin{eqnarray}}
\newcommand\eea{\end{eqnarray}}
\newcommand\ee{\end{equation}}

\newcommand\eg{{\it e.g.}}
\newcommand\ie{{\it i.e.}}

\def\Om{{\mathcal{O}}}
\def\Cm{{\mathcal{C}}}
\def\Em{{\mathcal{E}}}
\def\Am{{\mathcal{A}}}
\def\Gm{{\mathcal{G}}}
\def\Bm{{\mathcal{B}}}
\def\Nm{{\mathcal{N}}}
\def\Fm{{\mathcal{F}}}

\def\Dm{{\mathcal{D}}}

\def\a{{\alpha}}
\def\b{{\beta}}
\def\ad{{\dot{\alpha}}}
\def\bd{{\dot{\beta}}}

\def\ph{\phantom}

\def\nn{\nonumber}

\title{Bootstrapping $\Nm=2$ chiral correlators}
\author[1]{Madalena Lemos,}
\author[2]{Pedro Liendo}

\affiliation[1]{DESY Hamburg, Theory Group, Notkestrasse 85, D–22607 Hamburg, Germany}
\affiliation[2]{IMIP, Humboldt-Universit{\"a}t zu Berlin, IRIS Adlershof, Zum Gro{\ss}en Windkanal 6, 12489 Berlin, Germany}

\emailAdd{madalena.lemos@desy.de}
\emailAdd{pliendo@physik.hu-berlin.de}

\preprint{DESY 15-184, HU-EP-15/49, YITP-SB-15-40}

\bigskip
\abstract{We apply the numerical bootstrap program to chiral operators in four-dimensional $\Nm=2$ SCFTs. In the first part of this work we study four-point functions in which all fields have the same conformal dimension. We give special emphasis to bootstrapping a specific theory: the simplest Argyres-Douglas fixed point with no flavor symmetry. In the second part we generalize our setup and consider correlators of fields with unequal dimension. This is an example of a mixed correlator and allows us to probe new regions in the parameter space of $\Nm=2$ SCFTs. In particular, our results put constraints on relations in the Coulomb branch chiral ring and on the curvature of the Zamolodchikov metric.}


\begin{document}
\setcounter{tocdepth}{2}
\maketitle
\setcounter{page}{1}

%!TEX root = ../mixedE.tex
%%%%%%%%%%%%%%%%%%%%%%%%%%%%%%%%%%%%%%%%%%%%%
\section{Introduction}
\label{sec:intro}
%%%%%%%%%%%%%%%%%%%%%%%%%%%%%%%%%%%%%%%%%%%%%

The ``bootstrap'' is the idea that basic consistency requirements should be enough to fix the dynamical information of a theory.
It has seen renewed interest in CFTs in $d > 2$ thanks to the work of \cite{Rattazzi:2008pe}, where numerical tools were developed that allow to put constraints on the CFT data. These constraints are numerical bounds on the conformal dimensions and OPE coefficients, obtained by starting from the four-point function of identical scalars, and imposing unitarity and crossing symmetry.

Apart from numerical bounds constraining the theory space, the modern bootstrap can sometimes be used to solve specific theories. The most successful example being the critical $3d$ Ising model, where the the low-lying spectrum was obtained in \cite{ElShowk:2012ht,El-Showk:2014dwa} by studying the four-point function of the spin operator. In addition, the original approach of \cite{Rattazzi:2010gj} has been generalized from correlators of identical operators to mixed correlators, in which not all fields are identical. This has lead to important new insights. In particular, it was shown in \cite{Kos:2014bka} that the Ising model is restricted to lie in a tiny island in the parameter space of CFTs. The size of this island has been used to estimate the critical exponents of the Ising universality class to remarkable precision \cite{Simmons-Duffin:2015qma}.

The original bootstrap setup has been adapted to study CFTs with supersymmetry, where the conformal algebra is enhanced to the superconformal algebra. The superconformal bootstrap has been implemented in different dimensions with different amounts of supersymmetry, see  \cite{Poland:2010wg,Poland:2011ey,Berkooz:2014yda,Poland:2015mta,Beem:2014zpa,Beem:2013qxa,Alday:2014qfa,Chester:2014fya,Chester:2014mea,Chester:2015qca,Bashkirov:2013vya,Bobev:2015vsa,Bobev:2015jxa,Beem:2015aoa}. 
Here we will be concerned with $4d$ $\Nm=2$ SCFTs. This program was initiated in \cite{Beem:2014zpa}, and this paper is a natural continuation of that work. 

The are several motivations for bootstrapping $\Nm=2$ SCFTs. Since the remarkable paper \cite{Gaiotto:2009we}, a vast number of $\Nm=2$ SCFTs have been discovered, many of them lacking a Lagrangian description. Having such a landscape one may ask what are the underlying principles that characterize $\Nm=2$ theories, and whether a classification program is possible. The bootstrap philosophy, being based only on the operator algebra with no reference to a Lagrangian, is a natural framework in which such questions can be explored.
A second motivation is to attempt to solve specific $\Nm=2$ SCFTs. Many of the known interacting $\Nm=2$ theories are strongly coupled and standard perturbative techniques cannot be applied. Some protected data is known, but the bootstrap seems to be the only tool available for the study of quantities not protected by supersymmetry. 

In this work, we will study the bootstrap for correlators of chiral multiplets. These multiplets are built by the action of supercharges of the same chirality on a scalar superconformal primary $\phi$. More specifically, we will bootstrap correlators of the scalar $\phi$. Our setup will consider correlators of operators with identical conformal dimensions, but also the more general case of two unequal dimensions. It is therefore the first attempt at bootstrapping mixed correlators in four dimensions.

On the single correlator front, namely, the bootstrap for identical operators, we will complement the analysis of \cite{Beem:2014zpa} by presenting additional bounds. We will also attempt to corner a particular theory: the simplest Argyres-Douglas fixed point \cite{Argyres:1995jj,Argyres:1995xn}, which is sometimes denoted by $H_0$. As we will review in section 3, there are several features that make this theory an interesting candidate for the bootstrap. 

The $H_0$ theory is interesting enough to justify a bootstrap analysis, but we would also like to use it as a proof of principle, to test how practical the bootstrap is for theories that are not naturally selected by the numerics. The best understood example so far, the $3d$ Ising case, relies on the fact that the theory appears as a ``kink'' saturating one of the numerical bounds. Only a handful of theories have been bootstrapped in this way, and it is not clear how to proceed for generic CFTs. The $H_0$ theory does not have an associated kink, and in order to corner it we will have to resort to several tricks, using known data like its central charge and the dimension of $\phi$.

In the mixed correlator system, bootstrapping fields that have unequal dimensions allows us to probe regions of the parameter space that are inaccessible with the single correlator bootstrap. In particular, we will be able to test relations in the Coulomb branch chiral ring. It is believed that this ring is freely generated, and we will be able to check this numerically for small values of the conformal dimension. Also, the OPE coefficient of one of the multiplets appearing in our equations can be related to the curvature of the Zamolodchikov metric of conformal manifolds. The bootstrap then, will translate into geometric constraints for $\Nm=2$ theories with relevant deformations.

The outline of the paper is as follows. In section 2 we will review aspects of chiral fields in $\Nm=2$ theories and write the crossing equations. Section 3 presents our results for the single correlator system with special emphasis on the $H_0$ theory. Finally, section 4 deals with the mixed correlator system, and discusses its implications for the Coulomb branch and the Zamolodchikov metric.

%!TEX root = ../mixedE.tex
%%%%%%%%%%%%%%%%%%%%%%%%%%%%%%%%%%%%%%%%%%%%%
\section{The chiral four point functions}
\label{sec:fourpt}
%%%%%%%%%%%%%%%%%%%%%%%%%%%%%%%%%%%%%%%%%%%%%

In this section we will set up the crossing symmetry equation for $\Nm=2$ chiral correlators of unequal dimension. This is an example of a mixed correlator, and will allow us to explore regions inaccessible with the single correlator bootstrap. The bootstrap program for mixed correlators was developed in \cite{Kos:2014bka} for the 3$d$ Ising model, and extended in \cite{Kos:2013tga} for the more general $O(N)$ models. The remarkable success of the mixed correlator bootstrap in three dimensions is one of the main motivations for considering mixed correlators in $\Nm=2$ SCFTs.

We start with a brief introduction to chiral operators, and then proceed to study the multiplets being exchanged in the OPE, as well as the superconformal blocks that capture the contributions of each superconformal family to the four-point function. Chiral correlators for identical operators were studied in \cite{Poland:2010wg,Poland:2011ey} and in \cite{Beem:2014zpa} for four-dimensional $\Nm=1$ and $\Nm=2$ theories respectively. The bootstrap for chiral correlators in three dimensions was studied in \cite{Bobev:2015vsa,Bobev:2015jxa}.

%!TEX root = ../mixedE.tex
%%%%%%%%%%%%%%%%%%%%%%%%%%%%%%%%%%%%%%%%%%%%%

%%%%%%%%%%%%%%%%%%%%%%%%%%%%%%%%%%%%%%%%%%%%%
\subsection{Chiral fields}
%%%%%%%%%%%%%%%%%%%%%%%%%%%%%%%%%%%%%%%%%%%%%

The $\Nm=2$ superconformal algebra consists of the conformal generators $\{\Pm_{\a \ad},\, \Km^{\a \ad},\, \Mm_{\a}^{\ph{\a} \b},\, \bar{\Mm}^{\ad}_{\ph{\ad}\bd},\, D\}$, the $R$-symmetry $SU(2)_R \times U(1)_r$ generators $\{\Rm_{\ph{i}j}^{i}, \, r \}$, and the fermionic Poincar\'e and conformal supercharges $\{\Qm^i_{\a},\bar{\Qm}_{i\, \ad}, \Sm_{i}^{\a},\bar{\Sm}^{i\, \ad}\}$. Lorentz indices are denoted by $\a=\pm$ and $\ad=\dot{\pm}$, while $SU(2)_R$ indices are $i=1,2$.
Each multiplet consists of a superconformal primary annihilated by the supercharges $\Sm$ and $\bar{\Sm}$, together with superconformal descendants obtained by repeated application of the supercharges $\Qm$ and $\bar{\Qm}$.
Apart from generic long representations in which the highest weight is unconstrained, there are several possible shortening conditions consistent with the superconformal algebra. Multiplets are then labeled by the type of shortening condition they obey, together with the quantum numbers of the superconformal primary $(\Delta,j,\bar{\jmath},R,r)$. We review these conditions and list all possible unitary multiplets in appendix~\ref{app:shortening}.

In this work we are interested in chiral multiplets. These are shortened multiplets which can be built starting from a superconformal  scalar field $\phi_{r}$ that satisfies the chirality condition
\be 
[\bar{\Qm}_{\ad\, i}, \phi_{r}(0)] = 0\, ,
\ee
where $r$ is the $U(1)_r$ charge. Chiral multiplets satisfy $\Delta=r$  while for antichiral multiplets we have $\Delta=-r$, with unitarity requiring $r \geqslant 1$ and $- r \geqslant 1$ respectively.\footnote{We follow the $r$ charge normalizations of \cite{Dolan:2002zh}.} The complete multiplet is obtained by repeated action of the chiral supercharges $\Qm^{i}_{\a}$ and hence its name. Following the conventions of \cite{Dolan:2002zh} (reviewed in appendix~\ref{app:shortening}) we denote chiral multiplets as follows:
\begin{align}
\text{Chiral multiplet:} \qquad & \Em_{r} \qquad \, \, \, \, \Delta=r\, .
\\
\text{Antichiral multiplet:} \qquad & \bar{\Em}_{-r} \qquad \Delta=-r\, .
\end{align}
And label their superconformal primaries by $\phi_r$ and $\bar{\phi}_{-r}$ respectively. As already stated, in this work we will be concerned with four-point functions of superconformal primaries,
\be
\langle \phi_{r_1}(x_1) \bar{\phi}_{-r1}(x_2)  \phi_{r_2}(x_3) \bar{\phi}_{-r_2}(x_4) \rangle\, ,
\ee
where $\phi_{r_1}$ and $\phi_{r_2}$ are not necessarily equal.

\subsubsection*{The Coulomb branch}

\begin{table}
\centering
\renewcommand{\arraystretch}{1.3}
\begin{tabular}{|c||c|c|c|c|c|c|c|}
\hline
$G$		&	$H_0$	      & $H_1$ 	       & $H_2$ 	        & $D_4$ 	      & $E_6$ 		    & $E_7$ 		  & $E_8$ 			\\ \hline\hline 
$c$ 	& $\frac{11}{30}$ & $\frac{1}{2}$  & $\frac{2}{3}$  & $\frac{7}{6}$   & $\frac{13}{6}$  & $\frac{19}{6}$  & $\frac{31}{6}$  \\ \hline
$r_0$ 	& $\frac{6}{5}$   & $\frac{4}{3}$  & $\frac{3}{2}$  & $2$             & $3$             & $4$             & $6$             \\ \hline
\end{tabular}
\caption{Properties of the canonical rank one SCFTs associated to maximal mass deformations of the Kodaira singularities \cite{Argyres:2007cn,Cheung:1997id,Aharony:2007dj}. We list the values of the $c$ anomaly coefficients, and the $U(1)_r$ charge $r_0$ of the generator of the Coulomb branch chiral ring.\label{Tab:rank1theories}}
\end{table}

Chiral multiplets are particularly relevant because of their relation to Coulomb branch physics. The Coulomb branch of the moduli space of vacua of an $\Nm=2$ supersymmetric theory is parametrized by the vacuum expectation value of chiral multiplets: $\langle \phi_{r_0} \rangle$,
and the operators themselves belong to the Coulomb branch chiral ring. The \textit{rank} of an $\Nm=2$ theory is defined as the complex dimension of its Coulomb branch, as given by the number of generators. For Lagrangian theories the rank coincides with the rank of the gauge group, but the concept applies to non-Lagrangian theories as well. 

Organizing $\Nm=2$ theories by its rank is a natural way to start a classification program for $\Nm=2$ SCFTs. A systematic study of rank one SCFTs was undertaken in \cite{Argyres:2010py,Argyres:2015ffa} using Seiberg-Witten technology. The low-energy dynamics of the Coulomb branch is encoded in a family of elliptic curves (the Seiberg-Witten curve) and in a one-form differential subject to some consistency conditions. SCFTs are associated with scale-invariant singularities of the Seiberg-Witten curve, and it is known that they coincide with  a subset of Kodaira's classification of degenerations of elliptic curves over one holomorphic variable. From these singularities it is possible  to construct a ``canonical'' set of solutions corresponding to maximal mass deformations, which we list in Tab.~\ref{Tab:rank1theories}. 
We also list in Tab.~\ref{Tab:newrank1theories} some of the ``new'' rank one theories obtained by submaximal deformations \cite{Argyres:2007tq,Argyres:2010py,Chacaltana:2012ch,Chacaltana:2014nya,Argyres:2015ffa}.
A scale-invariant solution of the Seiberg-Witten curve does not guarantee the existence of an actual SCFT, however, for the list presented in Tab.~\ref{Tab:rank1theories} there is an independent construction. This set of theories appears as the low-energy description of a single $D3$ brane probing an $F$-theory singularity where the dilaton is constant \cite{Aharony:1998xz}. These theories are often labeled by the type of singularity, given by $G$ in Tab.~\ref{Tab:rank1theories}, which corresponds to the flavor symmetry of the theory, with $H_i \to A_i$ (the $H_0$ theory has no flavor symmetry). Seeing that understanding the rank one case is already a significant challenge, our hope is that the bootstrap philosophy will work as a complementary approach by finding extra consistency conditions.\footnote{We point out that a rank $N$ version of the theories in Tab.~\ref{Tab:rank1theories} is obtained by having $N$ $D3$ branes probing the singularity \cite{Aharony:1998xz}, in this case the Coulomb branch generators have dimensions given by $r_0\,, 2r_0\,, \ldots\,, N r_0$.}

\begin{table}
\centering
\renewcommand{\arraystretch}{1.3}
\begin{tabular}{|c||c|c|c|c|c|c||c|c|c|c|}
\hline
$G$	&	$USp(4)\times U(1)$ & $USp(6)\times SU(2)$ &	$USp(10)$ 	\\ \hline\hline 
$c$ 	&  $\frac{19}{12}$& $\frac{29}{12}$ & $\frac{49}{12}$ \\ \hline
$r_0$ & 	$3$&	 $4$ & 		$6$      \\ \hline
\end{tabular}
\caption{Properties of some of the ``new'' rank one SCFTs \cite{Argyres:2007tq,Argyres:2010py,Chacaltana:2012ch,Chacaltana:2014nya,Argyres:2015ffa}. We list the flavor symmtry group $G$, the values of the $c$ anomaly coefficients, and the $U(1)_r$ charge $r_0$ of the generator of the Coulomb branch chiral ring. Unlike the rank one theories given in Tab.~\ref{Tab:rank1theories} these theories also have a mixed branch. \label{Tab:newrank1theories}}
\end{table}

Our setup for the four-point function of $\phi_r$ will be quite general, allowing for unequal chiral fields. 
Thanks to the mixed correlator system we will be able to explore SCFTs of rank two, which could only be studied indirectly with the single correlator system, by looking at the generators one at a time. The mixed correlator bootstrap opens the possibility of varying $r_1$ and $r_2$ independently, probing regions of the parameter space which were inaccessible with the single correlator bootstrap.

For example, we will be able to test numerically, for a range of $r$, if the Coulomb branch chiral ring is freely generated. In particular, we will be able to check whether relations of the form 
\be 
\phi_{r_1} \times \phi_{r_2} \sim 0\,,
\ee
are allowed. Relations of this type are not believed to exist \cite{Tachikawa:2013kta}, however, there is no rigorous proof yet. In \cite{Beem:2014zpa} absence of chiral ring relations was checked numerically for the case $\phi_{r_1} = \phi_{r_2}$ for small values of $r$, and now we will be able to extend these results for the case $\phi_{r_1} \neq \phi_{r_2}$.

The systematic study of rank two theories was started in \cite{Argyres:2005pp,Argyres:2005wx}, but it is substantially more involved than the rank one case. Nevertheless, there is a large list of known rank two (and higher) theories, see \eg, \cite{Chacaltana:2010ks,Chacaltana:2011ze,Chacaltana:2012zy,Chacaltana:2013oka,Chacaltana:2014jba,Chacaltana:2015bna, Xie:2012hs,Xie:2013jc,Xie:2015rpa,Wang:2015mra,Tachikawa:2013kta}, some of which could be accessible by the numerical techniques pursued here.

\subsubsection*{The Zamolodchikov metric}

Our results will also put constraints on the geometry of conformal manifolds of $\Nm=2$ SCFTs. Consider a moduli space $\Mm$ of exactly marginal deformations; the deformations sit in chiral multiplets of the type $\Em_2$ and $\bar{\Em}_2$. Indeed, one can check that the lowest component of this type of multiplet has the appropriate quantum numbers to be added to a Lagrangian. The concept however is more general, and we define the dimension of the conformal manifold as the number of $\Em_2$ multiplets present in a theory.
If we have $m$ such multiplets, we label their respective superconformal primaries as $\phi_a$ with $a=1, \ldots, m$, where $m=\text{dim}_{\mathbb{C}}\Mm$ is the complex dimension of the manifold. The Zamolodchikov metric is defined through the two-point function,
\be 
\langle \phi_{a}(x) \bar{\phi}_{\bar{b}}(0) \rangle = \frac{g_{a\bar{b}}}{x^4}\, .
\ee
Choosing local holomorphic coordinates it is possible to set $g_{a\bar{b}} = \delta_{a \bar{b}}$, which implies that the only vanishing four-point function is,
\be 
\langle \phi_a(x_1) \phi_b(x_2) \bar{\phi}_{\bar{c}}(x_3) \bar{\phi}_{\bar{d}}(x_4) \rangle \,.
\ee
The OPE on the $\phi_{r_1} \times \phi_{r_2}$ channel contains an $\Em_4$ multiplet, following \cite{Baggio:2014sna,Baggio:2014ioa} we write its associated three-point coupling in terms of the Riemann tensor for the Zamolodchikov metric,
\be 
\label{lambdaE4}
\lambda^2_{\Em_4\, a \bar{c} b \bar{d}} = -R_{a \bar{c} b \bar{d}} + \delta_{a \bar{c}} \delta_{a \bar{d}} 
+ \delta_{b \bar{c}} \delta_{a \bar{d}}\,.
\ee
The bootstrap set up allows to put upper and lower bounds on the coefficient $\lambda^2_{\Em_4}$, and thanks to equation \eqref{lambdaE4} they translate into constraints on the geometry of the conformal manifold.\footnote{Note that if the $S^4$ partition function of a given SCFT is known, then the prescription of \cite{Gerchkovitz:2014gta} allows the computation of the Zamolodchikov metric.}
In \cite{Beem:2014zpa}, using the single correlator setup it was possible to bound the diagonal component of the Riemann tensor $R_{a \bar{a} a \bar{a}}$ (no summation), which for one-dimensional manifolds implies bounds on the curvature scalar. Thanks to the mixed correlator bootstrap we will be able to explore non-diagonal components $R_{a \bar{a} b \bar{b}}$ as well.

%!TEX root = ../mixedE.tex
%%%%%%%%%%%%%%%%%%%%%%%%%%%%%%%%%%%%%%%%%%%%%

%%%%%%%%%%%%%%%%%%%%%%%%%%%%%%%%%%%%%%%%%%%%%
\subsection{The OPE and superconformal blocks}
%%%%%%%%%%%%%%%%%%%%%%%%%%%%%%%%%%%%%%%%%%%%%

As a first step toward obtaining the superconformal blocks we must obtain the selection rules for the various channels relevant for the double OPE of
\be
\langle \phi_{r_1}(x_1) \bar{\phi}_{-r1}(x_2)  \phi_{r_2}(x_3) \bar{\phi}_{-r_2}(x_4) \rangle\,.
\ee
We can take the OPE in three different ways: the chiral-chiral channel $\phi_{r_1}\times\phi_{r_2}$, the antichiral-chiral channel of equal dimensions $\phi_{r_1}\times\bar{\phi}_{-r_1}$, and the antichiral-chiral channel of unequal dimensions $\phi_{r_1}\times\bar{\phi}_{-r_2}$.
The multiplets appearing in the expansions are given by,
\begin{align}\label{EEbarOPE}
\begin{split}
\phi_{r_1} \times \bar{\phi}_{-r_1} & \sim \mathbf{1} + \hat{\Cm}_{0 (j,j)} + \Am^{\Delta \geqslant \ell+2}_{R=0, r=0 (j,j)}\, ,
\\
\phi_{r_1} \times \bar{\phi}_{-r_2} & \sim \left\{
  \begin{array}{l l}
    \Em_{r_1-r_2} + \Cm_{r_1-r_2(j,j)} + \Am^{\Delta \geqslant \ell+2+r_1-r_2}_{0, r_1-r_2 (j,j)} & \quad \text{if $r_1 > r_2$}\\
   \bar \Em_{r_1-r_2} + \bar \Cm_{r_1-r_2 (j,j)} + \Am^{\Delta \geqslant \ell +2 +r_2-r_1}_{0, r_1-r_2 (j,j)} & \quad \text{if $r_1 < r_2$}
  \end{array} \right. 
  \,,\\
\phi_{r_1} \times \phi_{r_2}\, \, \, &\sim \Em_{r_1+r_2}   + \Bm_{1, r_1+r_2-1 (0,0)} + \Cm_{\frac12, r_1+r_2- \frac32 (j-\frac12,j)} \\ 
& \ph{\sim}   + \Bm_{\frac12,r_1+r_2-\frac12 (0,\frac12)} +\Cm_{0, r_1+r_2- 1 (j-1,j) } + \Am_{0, r_1+r_2-2 (j,j)}  \,.
\end{split}
\end{align}
Note that unitarity requires the multiplet $\Em_{r_1-r_2}$ to have $r_1-r_2 \geqslant 1$, if not, the multiplet is absent. Similarly, $\bar \Em_{r_1-r_2}$ only appears if  $r_2-r_1 \geqslant 1$.
In the above OPE the  $\Em$ multiplets are not the only ones with a clear physical meaning. The $\hat{\Cm}_{0 (j,j)}$ multiplets include conserved currents of spin $2j+2$, with the $j=0$ multiplet containing the stress tensor, while the $\Bm_{1, r_1+r_2-1 (0,0)}$ multiplet may be identified with a mixed branch chiral ring operator (see \eg, \cite{Beem:2014zpa,Argyres:2015ffa}). Note also that if the $\Em_{r_1-r_2}$ multiplet is present in the OPE, then $\Em_{r_1}$ must be a composite operator.
These selection rules are a generalization of the single correlator case, and reduce to those if one sets $r_1=r_2$ and imposes Bose symmetry.
For the chiral-antichiral channels one can show, \eg, by a superspace calculation, or as sketched in appendix~B of \cite{Beem:2014zpa}, that multiplets can appear in the OPE if and only if their superconformal primary also appears. Then the OPE follows from listing all the multiplets whose superconformal primaries have the right quantum numbers, namely $j_1=j_2=j$ and obey the $SU(2)_R$ and $U(1)_r$ selection rules.
For the $\phi_{r_1} \times \phi_{r_2}$ channel each superconformal multiplet contributes with a single conformal family.
The selection rules are obtained simply by enumerating all superconformal multiplets that contain a conformal descendant with the appropriate quantum numbers, which is also annihilated by $\bar{\Qm}$ and $\bar{\Sm}$ (see appendix~B of \cite{Beem:2014zpa}).

%%%%%%%%%%%%%%%%%%%%%%%%%%%%%%%%%%%%%%%%%%%%%
\subsubsection*{The \texorpdfstring{$\phi_{r_1} \times \phi_{r_2}$}{phi(r1) x phi(r2)}  channel}
%%%%%%%%%%%%%%%%%%%%%%%%%%%%%%%%%%%%%%%%%%%%%

Since in the $\phi_{r_1} \times \phi_{r_2}$ channel each multiplet contributes with a single conformal family the blocks are simply the conformal blocks of that family. 
The conformal blocks that contribute in this channel are then as follows:
\begin{equation}
\begin{alignedat}{4}
&{\Am_{0, r_1+r_2-2 (j,j)}}							&:\qquad &	(-1)^\ell g_{\Delta,\ell}\,,	\qquad &\Delta \geqslant 2+r_1+r_2+\ell~, \\
&{\Bm_{1, r_1+r_2-1 (0,0)}}							&:\qquad &	g_{\Delta=r_1+r_2+2,\ell=0}\,,				\qquad &~\\
&{\Cm_{\tfrac12, r_1+r_2-\tfrac32 (j-\tfrac12,j) }}	&:\qquad &	(-1)^\ell g_{\Delta=r_1+r_2+\ell+2,\ell}\,,		\qquad & \ell \geqslant1~, \\
&{\Em_{r_1+r_2}}											&:\qquad &	g_{\Delta=r_1+r_2,\ell=0}\,,					\qquad &~ \\
&{\Bm_{\frac12,r_1+r_2-\frac12 (0,\frac12)}}											&:\qquad &	-g_{\Delta=r_1+r_2+1,\ell=1}\,,					\qquad &~ \\
&{\Cm_{0, r_1+r_2- 1 (j-1,j) }}				&:\qquad &	(-1)^\ell g_{\Delta=r_1+r_2+\ell,\ell}\,,			\qquad & \ell \geqslant 2~, \label{EE2blocks}\\
\end{alignedat}
\end{equation}
where $\ell=2j$ and the blocks $g_{\Delta,\ell}$ are given in \eqref{eq:bos_block}. These reduce to the selection rules of \cite{Beem:2014zpa} when $r_1=r_2$ after imposing Bose symmetry.
Note that the first two classes of short representations lie at the unitarity bound for long multiplets, and their blocks are simply the specializations of the long multiplet block to appropriate values of $\Delta$ and $\ell$. This follows directly from the decomposition of the long multiplet at the unitarity bound given in \eqref{eq:AC_dec} and \eqref{eq:CisB}. On the other hand, the last three classes of short representations are separated from the continuous spectrum of long multiplets by a gap. Whenever $r_1 \neq r_2$ (or if $r_1=r_2$ but the two operators are different) both even and odd spins can appear. 

There are additional short multiplets allowed when $r_1=r_2=1$:
\begin{equation}
\Dm_{1 (0,0)}\,, \qquad \hat{\Cm}_{\frac12 (j-\frac12,j)}\,,\qquad \Dm_{\frac12 (0,\frac12)}\,, \qquad \hat{\Cm}_{0 (j-1,j)}~.
\end{equation}
The conformal blocks for these representations are $g_{\Delta=4,\ell=0}$,  $g_{\Delta=\ell+4,\ell}$ with $\ell \geqslant 1$, $ g_{\Delta=3,\ell=1}$ and $g_{\Delta=\ell+2,\ell}$ with $\ell \geqslant 2$, respectively. These blocks are actually identical to some of those appearing in \eqref{EE2blocks}. This means their contributions are indistinguishable from the blocks we already have and we can therefore ignore them.

%%%%%%%%%%%%%%%%%%%%%%%%%%%%%%%%%%%%%%%%%%%%%
\subsubsection*{The \texorpdfstring{ $\phi_{r_1} \times \bar{\phi}_{-r_2}$}{phi(r1) x bar(phi)(r2)}  channel}
%%%%%%%%%%%%%%%%%%%%%%%%%%%%%%%%%%%%%%%%%%%%%

As stated above, there are several channels in which the OPE decomposition can be implemented. However, the chiral-antichiral channel $\phi_{r_1} \times \bar{\phi}_{-r_2}$ is the only one that has an associated superconformal block. That is, a block that can be written as a finite sum of bosonic blocks with the coefficients fixed by supersymmetry.
The superconformal block for the $\phi_{r_1} \times \bar{\phi}_{-r_1}$ channel was computed in \cite{Fitzpatrick:2014oza}, and while the one for the $\phi_{r_1} \times \bar{\phi}_{-r_2}$ channel is not in the literature, it can be obtained using the same techniques. We now give an outline of this calculation and refer the reader to appendix \ref{app:superblocks} for more details. 
The most efficient way to do this calculation is using the embedding space techniques of \cite{Fitzpatrick:2014oza}, where the block is obtained by solving an eigenvalue equation for the Casimir operator of the superconformal group. For chiral multiplets with unequal dimensions we expect an expansion of the form
\bea
&&\langle \phi_{r_1}(x_1) \bar{\phi}_{-r_2}(x_2) \phi_{r_2}(x_3) \bar{\phi}_{-r_1}(x_4)  \rangle = \nn\\
\qquad && \frac{1}{|x_{12}|^{\Delta_1+\Delta_2}|x_{34}|^{\Delta_1 + \Delta_2}} \left|\frac{x_{24}}{x_{14}}\right|^{\Delta_{12}} \left|\frac{x_{14}}{x_{13}}\right|^{-\Delta_{12}}\sum_{\Om} (-1)^\ell \lambda_{1 \bar{2} \bar{\Om}} \lambda_{2 \bar{1} \Om}\Gm_{\Delta_{12}}(u,v)\, ,
\label{eq:doubleope}
\eea
where $\Delta_{ij}= \Delta_i-\Delta_j$.
Acting with the Casimir
\be
L^2 \Gm(u,v) = C_{r,\Delta,\ell}\, \Gm_r(u,v)\, ,
\ee
where $r=\Delta_{12}$ and $C_{r,\Delta,\ell}$ is the eigenvalue of a multiplet with labels $(\Delta,\ell,R=0,r)$:
\be 
C_{r,\Delta,\ell} = \ell(\ell + 2)+\Delta^2 - r^2\, .
\ee
The solution to this differential equation can be obtained by writing the Casimir in the embedding space where the superconformal generators are linear operators. The key observation of \cite{Fitzpatrick:2014oza} is that the Ward identities constrain this correlator to be a function of two superconformal invariants, which are supersymmetric versions of the standard $u$ and $v$ cross-ratios. Once a differential equation is obtained, it is easy to project to four dimensions and the resulting differential operator can be identified with the one studied by Dolan and Osborn in \cite{Dolan:2003hv}. The calculation then mimics the purely bosonic case and the answer is a bosonic block with shifted arguments:
\begin{align}
\label{crossedsuperblock}
\begin{split}
\Gm_r(u,v) & = \frac{z \bar{z}}{z-\bar{z}}\left( k_{\Delta+\ell}(z) k_{\Delta-\ell-2}(\bar{z})-z \leftrightarrow \bar{z} \right)\, ,
\\
k_{\beta}(z) & = z^{\frac{\beta}{2}}{_2}F_1\left(\frac{\beta-r}{2},\frac{\beta-r+4}{2},\beta+2;z\right)\, .
\end{split}
\end{align}
As a first consistency check, it is easy to see that for $r=0$ this expression reduces to the superconformal block obtained in \cite{Fitzpatrick:2014oza}. For this function to describe a superconformal block one should also be able to decompose it as a finite sum of bosonic blocks, this is indeed possible and we present the expansion in appendix~\ref{app:superblocks}.
The above superconformal block therefore encodes the contribution of all the conformal families belonging to the $ \Am^{\Delta \geqslant \ell+2+r_1-r_2}_{0, r_1-r_2 (j,j)} $ multiplet contributing to the four-point function at hand.
Note that the superconformal blocks for the $\Cm_{r_1-r_2 (j,j)}$ representation are obtained by simply specializing \eqref{crossedsuperblock} to the case $\Delta=r_1-r_2+\ell+2$ (as follows from \eqref{eq:AC_dec}), while the block for $\Em_{r_1-r_2}$ is obtained when setting $\Delta=r_1-r_2$ and $\ell=0$.
For $r=0$ the superconformal block encodes the contribution of the $\Am^{\Delta \geqslant \ell+2}_{R=0, r=0 (j,j)}$ multiplet, and specializing to $\Delta=\ell+2$ one obtains a  $\hat{\Cm}_{0 (j,j)}$ multiplet, as follows from the decomposition given in \eqref{eq:AChat_dec}. In the case $r=0$ the identity multiplet also appears, and it contributes with a constant block set to one (which can also be found from the superblock by taking the limit $\Delta=\ell=0$).
All in all, expression \eqref{crossedsuperblock} encodes the blocks for all the multiplets being exchanged in the $\phi_{r_1} \times \phi_{r_2}$ channel.

It will be useful for later to work out the superconformal block with a different ordering of the fields,
\bea
&&\langle \bar{\phi}_{-r_2}(x_1)  \phi_{r_1}(x_2) \phi_{r_2}(x_3) \bar{\phi}_{-r_1}(x_4)  \rangle = \nn\\
\qquad && \frac{1}{|x_{12}|^{\Delta_1+\Delta_2}|x_{34}|^{\Delta_1 + \Delta_2}} \left|\frac{x_{24}}{x_{14}}\right|^{-\Delta_{12}} \left|\frac{x_{14}}{x_{13}}\right|^{-\Delta_{12}}\sum_{\Om} (-1)^\ell \lambda_{\bar{2} 1 \bar{\Om}} \lambda_{2 \bar{1} \Om}\tilde{\Gm}_{\Delta_{12}}(u,v)\,.
\label{eq:doubleope2}
\eea
In this case the superconformal blocks are
\begin{align}
\label{superbraidedblock}
\begin{split}
\tilde{\Gm}_r(u,v) & =  \frac{z \bar z}{z-\bar z } \left[ \tilde{k}_{\Delta+ \ell} (z) \tilde{k}_{\Delta -\ell-2} (\bar z) - z \leftrightarrow \bar{z}\right]\,,\\
\tilde{k}_\beta&= z^{\frac{\beta}{2}} {}_2F_1\left( \frac{\beta + r}{2}, \frac{\beta - r}{2};\beta + 2; z\right)\,.
\end{split}
\end{align}
%!TEX root = ../mixedE.tex
%%%%%%%%%%%%%%%%%%%%%%%%%%%%%%%%%%%%%%%%%%%%%

%%%%%%%%%%%%%%%%%%%%%%%%%%%%%%%%%%%%%%%%%%%%%
\subsection{Crossing equations}
%%%%%%%%%%%%%%%%%%%%%%%%%%%%%%%%%%%%%%%%%%%%%

We are finally ready to write down the system of crossing equations.
As shown in \cite{Kos:2014bka}, in order to have positivity (in the sense described in the following subsection) in the $(\langle \phi_{r_1}(x_1) \bar{\phi}_{-r1}(x_2)) \to ( \phi_{r_2}(x_3) \bar{\phi}_{-r_2}(x_4))$ channel of the double OPE, we must also consider the crossing equations for the $\langle \bar \phi_i \phi_i \bar \phi_i \phi_i \rangle$ correlator, with $i=1,2$.
These were derived in \cite{Beem:2014zpa}, and for convenience are reproduced in \eqref{eq:single_corr_cross}. (Note that all of the mixed correlator crossing equations obtained in this subsection reduce to the three crossing equations for a single correlator if we take the dimensions to be equal.)

We now focus on obtaining the crossing equations for the mixed correlator.  For that we must take all possible double OPE limits, corresponding to the different orderings of the operators. 
To easily obtain all the different orderings, we write the correlator with all dimensions arbitrary, constrained only by $-\Delta_4+\Delta_3+\Delta_1-\Delta_2=0$, and then set the dimensions equal pairwise in different ways.
As a result we only need to consider the following two orderings of the operators:
\begin{align}
&\langle \bar \phi_2 (x_1) \phi_1(x_2) \phi_3 (x_3) \bar \phi_4(x_4) \rangle  \nn \\
&\qquad=  \frac{1}{x_{12}^{\Delta_1+\Delta_2}x_{34}^{\Delta_3+\Delta_4}} \left(\frac{x_{24}}{x_{14}}\right)^{-\Delta_{12}} \left(\frac{x_{14}}{x_{13}}\right)^{\Delta_{34}} \sum_{\Om \in \bar \phi_2 \phi_1, \Om \in \phi_3 	\bar \phi_4} (-1)^\ell  \lambda_{\bar \phi_2 \phi_1 \Om} \lambda_{\phi_3 \bar \phi_4 \Om} \tilde{\Gm}^{-\Delta_{12},\Delta_{34}}(u,v) \nn \\
&\qquad = \frac{1}{x_{14}^{\Delta_2+\Delta_4}x_{23}^{\Delta_1+\Delta_3}}\left(\frac{x_{24}}{x_{12}}\right)^{\Delta_{24}}\left(\frac{x_{12}}{x_{13}}\right)^{-\Delta_{13}} \sum_{\bar \Om \in \bar \phi_2 \bar \phi_4, \Om \in \phi_3  \phi_1} (-1)^\ell \lambda_{\phi_3 \phi_1 \Om} \lambda_{\bar \phi_2 \bar \phi_4 \bar \Om} g^{\Delta_{24}, -\Delta_{13}}(v,u)\,,
\label{crossing1}
\end{align}
and
\begin{align}
&\langle \phi_1 (x_1) \bar \phi_2(x_2) \phi_3 (x_3) \bar \phi_4(x_4) \rangle  \nn \\
&\qquad=  \frac{1}{x_{12}^{\Delta_1+\Delta_2}x_{34}^{\Delta_3+\Delta_4}} \left(\frac{x_{24}}{x_{14}}\right)^{\Delta_{12}} \left(\frac{x_{14}}{x_{13}}\right)^{\Delta_{34}} \sum_{\Om \in  \phi_1 \bar \phi_2, \Om \in \phi_3 	\bar \phi_4}  \lambda_{ \phi_1 \bar \phi_2 \Om} \lambda_{\phi_3 \bar \phi_4 \Om} (-1)^\ell \Gm^{\Delta_{12},\Delta_{34}}(u,v) \nn \\
&\qquad = \frac{1}{x_{14}^{\Delta_1+\Delta_4}x_{23}^{\Delta_2+\Delta_3}}\left(\frac{x_{24}}{x_{12}}\right)^{\Delta_{14}}\left(\frac{x_{12}}{x_{13}}\right)^{-\Delta_{23}} \sum_{ \Om \in \phi_1 \bar \phi_4, \Om \in \phi_3  \bar \phi_2} \lambda_{\phi_1 \bar \phi_4 \Om} \lambda_{ \phi_3 \bar \phi_2 \bar \Om} (-1)^\ell \Gm^{\Delta_{14}, -\Delta_{23}}(v,u)\,.
\label{crossing2}
\end{align}
Here the superconformal blocks are defined as in \eqref{crossedsuperblock} and in \eqref{superbraidedblock}, with the addition that we keep track of the external dimensions, but once these are set equal pairwise, all the blocks reduce to the ones obtained in the previous subsection.
We recall that in the chiral-chiral channel each superconformal multiplet contributes with a single conformal primary, and the superconformal blocks are simply conformal blocks.

To get the first set of equations we take  $\Delta_1=\Delta_4$ and $\Delta_2=\Delta_3$, and find the following two crossing equations:
\begin{align}
u^{\frac{\Delta_1+\Delta_2}{2}} \sum \lambda_{\phi_2 \phi_1 \Om} \lambda_{\bar \phi_2 \bar \phi_1 \bar \Om} (-1)^\ell g^{-\Delta_{12},-\Delta_{12}} (v,u) &= v^{\frac{\Delta_1+\Delta_2}{2} }\sum \lambda_{\bar \phi_2 \phi_1 \Om} \lambda_{\phi_2 \bar \phi_1 \Om} (-1)^\ell \tilde{\Gm}^{-\Delta_{12}, -\Delta_{12}}(u,v) \,, \label{eq1}\\
 v^{\Delta_2} \sum \lambda_{\phi_1 \bar \phi_2 \Om} \lambda_{\phi_2 \bar \phi_1 \Om}(-1)^\ell \Gm^{\Delta_{12}, -\Delta_{12}} (u,v) 
&= u^{\frac{\Delta_1+\Delta_2}{2}} \sum \lambda_{\phi_1 \bar \phi_1 \Om} \lambda_{\phi_2 \bar \phi_2 \Om}(-1)^\ell \Gm^{0,0}(v,u)\,.
\label{eq2}
\end{align}
Note that we have left the OPE coefficients  as they came out of \eqref{crossing1} and \eqref{crossing2}, but we know how to re-order the fields using braiding: $\lambda_{12 \mathcal{O}}=(-1)^\ell \lambda_{2 1 \mathcal{O}}$. (We take $\lambda_{123}$ real for real fields, which implies $\lambda_{123}^\ast=\lambda_{\bar{1} \bar{2} \bar{3}}$ for complex fields.)
The superblocks $\Gm^{0,0}$ is simply \eqref{crossedsuperblock} with $r=0$, \ie, the same block as for the correlator of identical fields. 

If we instead take $\Delta_1=\Delta_2$ and $\Delta_3=\Delta_4$ (and rename $\Delta_3 \to \Delta_2$) we find:
\begin{align}
\label{eq3}
&u^{\Delta_2} \sum \lambda_{\phi_2 \phi_1 \Om} \lambda_{\bar \phi_1 \bar \phi_2 \bar \Om}(-1)^\ell g^{\Delta_{12},-\Delta_{12}}(v,u) = v^{\frac{\Delta_1+ \Delta_2}{2}} \sum \lambda_{\bar \phi_1 \phi_1 \Om} \lambda_{\phi_2 \bar \phi_2 \Om} (-1)^\ell\tilde{\Gm}^{0,0}(u,v)\,, \\
&v^{\frac{\Delta_1+\Delta_2}{2}} \sum \lambda_{\phi_1 \bar \phi_1 \Om} \lambda_{\phi_2 \bar \phi_2 \Om} (-1)^\ell\Gm^{0,0}(u,v) = u^{\Delta_2} \sum \lambda_{\phi_1 \bar \phi_2 \Om} \lambda_{\phi_2 \bar \phi_1 \Om} (-1)^\ell\Gm^{\Delta_{12}, -\Delta_{12}}(v,u) \,.
\label{eq4}
\end{align}
Note that Eq.~\eqref{eq2} and \eqref{eq3} are the same, up to $u \leftrightarrow v$.
All in all, the full set of crossing symmetry equations for the mixed correlator consists of \eqref{eq1},~\eqref{eq2}~and~\eqref{eq3}, together with the same equations with $u \leftrightarrow v$.

Combining these equations with the crossing equations for a single correlator \eqref{eq:single_corr_cross} we get a system of $12$ crossing equations, given explicitly in \eqref{eq:fullcrossing}, in which positivity is manifest in all channels. This is the final system of crossing equations that is ready to be analysed numerically, with the methods reviewed in the following subsection.

As a final note we point out that in the double OPEs that include the exchange of the identity operator we fix its OPE coefficient to one, thereby fixing the two point functions to have unit normalization, as is conventional.
As a result  of this normalization the OPE coefficient of the stress tensor is related to the $c$ central charge of the theory.
Making use of the decomposition of the stress tensor superconformal block in conformal blocks (given in \eqref{eq:short_scb_in_bb}), which relates the OPE coefficient of the stress tensor to that of its superconformal primary, we fix the OPE coefficient of the stress tensor multiplet in terms of the central charge.
In the end one finds that the $\hat{\Cm}_{0,(0,0)}$ block contributes to the crossing equations as:
\be
\lambda_{\phi_i \bar{\phi_i} \Om_{\Delta=2,\ell=0}} \lambda_{\phi_j \bar{\phi_j} \Om_{\Delta=2,\ell=0}} = \frac{\Delta_i \Delta_j}{6 c}\,.
\ee
Moreover, since we are interested in interacting theories, we can set to zero the OPE coefficient of the $\hat{\Cm}_{0,(j,j)}$ multiplets with $j > 0$, as these include conserved currents of spin greater than two \cite{Maldacena:2011jn,Alba:2013yda}.
In practice this is tricky to achieve if we make no assumptions about the spectrum of long operators with spin $j$, as a long multiplet arbitrarily close to the unitarity bound mimics precisely these multiplets due to the decomposition \eqref{eq:AChat_dec}.

%!TEX root = ../mixedE.tex
%%%%%%%%%%%%%%%%%%%%%%%%%%%%%%%%%%%%%%%%%%%%%

%%%%%%%%%%%%%%%%%%%%%%%%%%%%%%%%%%%%%%%%%%%%%
\subsection{Numerical Implementation}
\label{sec:numericalimpl}
%%%%%%%%%%%%%%%%%%%%%%%%%%%%%%%%%%%%%%%%%%%%%

To constrain the CFT data featuring in the crossing equation \eqref{eq:fullcrossing} we proceed numerically. In this section we simply provide a short summary of how this is accomplished, and refer the reader to \cite{Kos:2014bka} where the numerical bootstrap techniques for mixed correlators were first developed.
Schematically, the crossing equation takes the following form
\be 
 \sum_i \sum_{\Om_i} \vert \lambda_{\Om_i} \vert ^2 \overrightarrow{V}_{i,\Om_i}(z, \bar z) + \sum_\Om \left( \lambda_{1 \Om}^\ast \quad \lambda_{2 \Om}^\ast \right) \overrightarrow{M}_\Om (z, \bar z) \begin{pmatrix}
\lambda_{1 \Om }\\
\lambda_{2 \Om}
\end{pmatrix} = -  \overrightarrow{V}_{\mathrm{fixed}}(z, \bar z) \,,
\label{eq:sch_cross}
\ee
where $i$ runs over the four different channels comprised in the first the two lines of \eqref{eq:fullcrossing}, and $\Om_i$ denotes the operators exchanged in channel $i$. Here $\overrightarrow{V}_{i,\Om_i}$ are $12-$dimensional vectors, and $\overrightarrow{M}_\Om$ a $12-$dimensional vector of $2\times2$ matrices.
When considering a single correlator only the first type of terms appear.
The contributions to $\overrightarrow{V}_{\mathrm{fixed}}$, given in \eqref{eq:id+st}, come from the identity operator and from the stress tensor multiplet, if the central charge is being fixed.

Then the idea, pioneered in \cite{Rattazzi:2008pe}, is to rule out a trial CFT data by showing the the crossing equations can never be satisfied.
In the original work the crossing equation was analysed using linear programming techniques, and as shown in \cite{Kos:2014bka}, a crossing equations of the form \eqref{eq:sch_cross} can be studied using semi-definite programming.
An ansatz about the spectrum of operators can be ruled out if we can find a non-trivial linear functional such that
\bea 
\overrightarrow \Phi \cdot \overrightarrow{V}_{i,\Om_i}(z, \bar z) & \geqslant & 0 \,, \qquad \text{for all $\Om_i$ in the trial spectrum} \,, \nn \\
\overrightarrow \Phi \cdot \overrightarrow{M}_\Om (z, \bar z) & \succcurlyeq & 0 \,, \qquad \text{for all $\Om $ in the trial spectrum} \,,  \nn \\
\overrightarrow \Phi \cdot \overrightarrow{V}_{\mathrm{fixed}}(z, \bar z) &=& 1 \,,
\eea
where the functional acts on each entry of the matrix $\overrightarrow{M}$, and $\succcurlyeq$ means the resulting matrix is positive semi-definite.
If such a functional exists we have reached a contradiction: all the terms on the left hand side of \eqref{eq:sch_cross} are non-negative, while the right hand side is negative.

Similarly we can obtain bounds on OPE coefficients of the operator $\Om^\star$ contributing to the crossing equations as $\overrightarrow{V}_{\Om^\star}(z, \bar z) $ by performing the following optimization:
\begin{align}
\mathrm{Maximize} \; \overrightarrow \Phi \cdot   \overrightarrow{V}_{\mathrm{fixed}}(z, \bar z) \,, \qquad \text{subject to:} \,,&  \nn \\
\overrightarrow \Phi \cdot \overrightarrow{V}_{i,\Om_i}(z, \bar z) & \geqslant  0 \,, \qquad \text{for all $\Om_i \neq \Om^\star$ in the trial spectrum} \,, \nn \\
\overrightarrow \Phi \cdot \overrightarrow{M}_\Om (z, \bar z) & \succcurlyeq  0 \,, \qquad \text{for all $\Om $ in the trial spectrum} \,,  \nn \\
\overrightarrow \Phi \cdot  \overrightarrow{V}_{\Om^\star}(z, \bar z) &= 1 \,.
\label{eq:opebound}
\end{align}
Denoting the maximum of $\overrightarrow \Phi \cdot   \overrightarrow{V}_{\mathrm{fixed}}(z, \bar z) $ by $M$, and pulling the operator $\Om^\star$ out of the sums in \eqref{eq:sch_cross}, we find
\be 
\lambda^2_{\Om^\star} \leqslant  -M \,.
\ee
In particular this allows us to obtain bounds on the central charge, by taking $\Om^\star$ to be the stress tensor multiplet, in which case $\overrightarrow{V}_{\Om^\star}$ is given by the $c-$dependent part of \eqref{eq:id+st}, and we remove its contribution from $\overrightarrow{V}_{\mathrm{fixed}}$. Since the OPE coefficient of the stress tensor multiplet depends on $1/c$ the upper bound on the OPE coefficient translates into a lower bound on the central charge. 
Note that if $M$ is positive the above bound is in contradiction with unitarity and the trial spectrum is ruled out.

To obtain a lower bound on a OPE coefficient we have to demand $\overrightarrow \Phi \cdot  \overrightarrow{V}_{\Om^\star}(z, \bar z) = - 1 $, with the same optimization problem as before yielding 
\be 
\lambda^2_{\Om^\star} \geqslant   M \,.
\ee
However this problem is only solvable if we can consistently impose the functional to be negative on  $\overrightarrow{V}_{\Om^\star}(z, \bar z)$ and positive on everything else listed in \eqref{eq:opebound}, \ie, if the block corresponding to $\Om^\star$ is isolated from the rest of the blocks.

The search space of functionals is made finite by picking a basis of derivatives with respect to $z$ and $\bar z$, evaluated at the crossing symmetric point ($z=\bar{z}=\tfrac{1}{2}$), and truncating it as
\be
\overrightarrow{\Phi}  = \sum_{n,m}^{n+m \leqslant \Lambda} \Phi_{m,n} \partial_z^m \partial_{\bar{z}}^n  \vert_{z=\bar{z}=\tfrac{1}{2}}\,.
\ee
A further truncation that must be made is in the sum over the spectrum of operators. The selection rules include multiplets of arbitrarily large spin, which must be truncated at some maximum spin $\ell_{\mathrm{max.}}$. In practice, the choice of this cutoff depends on the truncation $\Lambda$, and we have checked that our choices do not affect significantly the results. A careful examination of the dependence on this truncation has been performed in \cite{Caracciolo:2014cxa}.  
To deal with arbitrary continuous dimensions, $\Delta$, of the superconformal blocks we follow the approach originally developed in \cite{Poland:2011ey} of approximating the conformal blocks by positive functions times polynomials in $\Delta$, reducing the problem of finding the functionals to a semi-definite programming problem. To obtain this polynomial approximation we series expand the blocks using the radial coordinates of \cite{Hogervorst:2013sma}.
All in all, the problem of finding functionals satisfying the various requirements has been reduced to a semi-definite programming problem, for which we use the \texttt{SDPB} arbitrary precision solver of \cite{Simmons-Duffin:2015qma}.

By finding a functional $\overrightarrow{\Phi}$ with cutoff $\Lambda$ we are effectively showing that there is no solution to the Taylor expansion of the crossing equations, evaluated at the crossing symmetric point, to order $\Lambda$. Therefore, at each order $\Lambda$ we find valid bounds on the CFT data, that improve as $\Lambda$ is increased.
Moreover, while inside the allowed region there can be multiple solutions to the truncated crossing equations, when the numerical bounds are saturated there is a \emph{unique} solution. This was first observed in \cite{Poland:2010wg,ElShowk:2012hu}, and has been used extensively in \cite{El-Showk:2014dwa} when solving the three-dimensional Ising model, and when studying the six-dimensional $(2,0)$ $A_1$ theory in \cite{Beem:2015aoa}. 
This unique solution can be reconstructed by analysing the functional $\overrightarrow{\Phi}$, at each order $\Lambda$, providing an approximate solution to the full crossing symmetry equations. 
%!TEX root = ../mixedE.tex
%%%%%%%%%%%%%%%%%%%%%%%%%%%%%%%%%%%%%%%%%%%%%
\section{Results for the single correlator bootstrap}
\label{sec:res_single}
%%%%%%%%%%%%%%%%%%%%%%%%%%%%%%%%%%%%%%%%%%%%%

We start by taking a closer look at the bootstrap for a single four-point function.  This correlator was studied in \cite{Beem:2014zpa} where the focus was mostly on the $\phi_{r_1} \times \bar \phi_{-r_1}$ OPE. From \eqref{EEbarOPE}, after imposing the absence of higher-spin currents, the only unknown information in this channel corresponds to the OPE coefficient of the stress-tensor multiplet -- related to the central charge -- and the scaling dimensions and OPE coefficients of the unprotected long multiplets $\Am_{0,0,(j,j)}^\Delta$. In \cite{Beem:2014zpa}, the CFT data was constrained by obtaining an upper bound on the dimension of the first unprotected long scalar operator, and a lower bound on the central charge. While for large external dimension the central charge bound are relatively weak, for small external dimension they become stronger, with the central charge of the $H_0$ theory of Tab. \ref{Tab:rank1theories} appearing relatively close to the numerically excluded region. This left open the possibility that, as $\Lambda \to \infty$, the bound might converge to $c \geqslant \frac{11}{30}$ for $r_1=\tfrac{6}{5}$. 

For the $\phi_{r_1} \times \phi_{r_1}$ OPE the analysis of \cite{Beem:2014zpa} only considered the coefficient of the $\Em_{2r_1}$ multiplet, in this work we will further explore this channel. For identical scalars the last line of \eqref{EEbarOPE} becomes,
\be 
\phi_{r_1} \times \phi_{r_1} \sim   \Em_{2r_1}  + \Bm_{1, 2r_1-1 (0,0)}  
+ \Cm_{0 \, 2r_1- 1 (j-1,j) } + \Cm_{\frac12, 2r_1- \frac32 (j-\frac12,j)} + \Am_{0, 2r_1-2 (j,j)}   \,,
\ee
where now by Bose symmetry only operators with even spin $\ell=2j$ can appear.
The unknown information probed by this OPE amounts to the following quantities:
\begin{itemize}
\item OPE coefficients of the short multiplets $\Em_{2r_1}$ and $\Cm_{0,2r_1-1 (j-1,j)}$. The dimensions of these multiplets are isolated from the continuum of long multiplets by a gap.
\item OPE coefficients of the short multiplets $\Bm_{1,2r_1-1 (0,0)}$ and $\Cm_{\frac12, 2r_1- \frac32 (j-\frac12,j)}$. These multiplets sit at the beginning of the continuum of long multiplets.
\item Dimensions and OPE coefficients of the long $\Am_{0,2r_1-2 (j,j)}$ multiplets.
\end{itemize}
Unlike in the $\phi_{r_1} \times \bar \phi_{-r_1}$ channel, we now have a larger number of short multiplets. It is useful to understand either by physical motivations or the bootstrap itself, which of these short multiplets are present.
While the physical interpretation of most of these multiplets is not immediately obvious, the  $\Em_{2r_1}$ and $\Bm_{1,2r_1-1 (0,0)}$ multiplets may be identified respectively with Coulomb and mixed branch chiral ring operators (see for example \cite{Beem:2014zpa, Argyres:2015ffa,Tachikawa:2013kta}) making them an interesting target for the bootstrap.\footnote{This identification is conjectural, and it might happen that these multiplets are present without corresponding to flat directions.} A noteworthy difference between these two multiplets is that, as stated above, $\Em_{2r_1}$ is isolated by a gap from the spectrum of unprotected long multiplets, while $\Bm_{1,2r_1-1 (0,0)}$ sits precisely at the long multiplet unitarity bound. As such, both upper and lower bounds on the OPE coefficient of $\Em_{2r_1}$ can be obtained, while only upper bounds can be obtained for $\Bm_{1,2r_1-1(0,0)}$.
Note that for theories without a mixed branch, as is the case for the canonical rank one theories of Tab. \ref{Tab:rank1theories}, $\Bm_{1,2r_1-1 (0,0)}$ must be absent, and therefore one should not expect to obtain a (non-trivial) lower bound on the OPE coefficient of this multiplet. On the other hand, one can hope to single out theories without a mixed branch by imposing the absence of this multiplet. This is explored in section \ref{sec:H0} when we focus on the rank one $H_0$ theory.

%!TEX root = ../mixedE.tex
%%%%%%%%%%%%%%%%%%%%%%%%%%%%%%%%%%%%%%%%%%%%%
\subsection{Bounds on \texorpdfstring{$\Cm$}{C} OPE coefficients}
\label{Sec:COPE}
%%%%%%%%%%%%%%%%%%%%%%%%%%%%%%%%%%%%%%%%%%%%%

\begin{figure}[h!]
\centering
\includegraphics[scale=0.35]{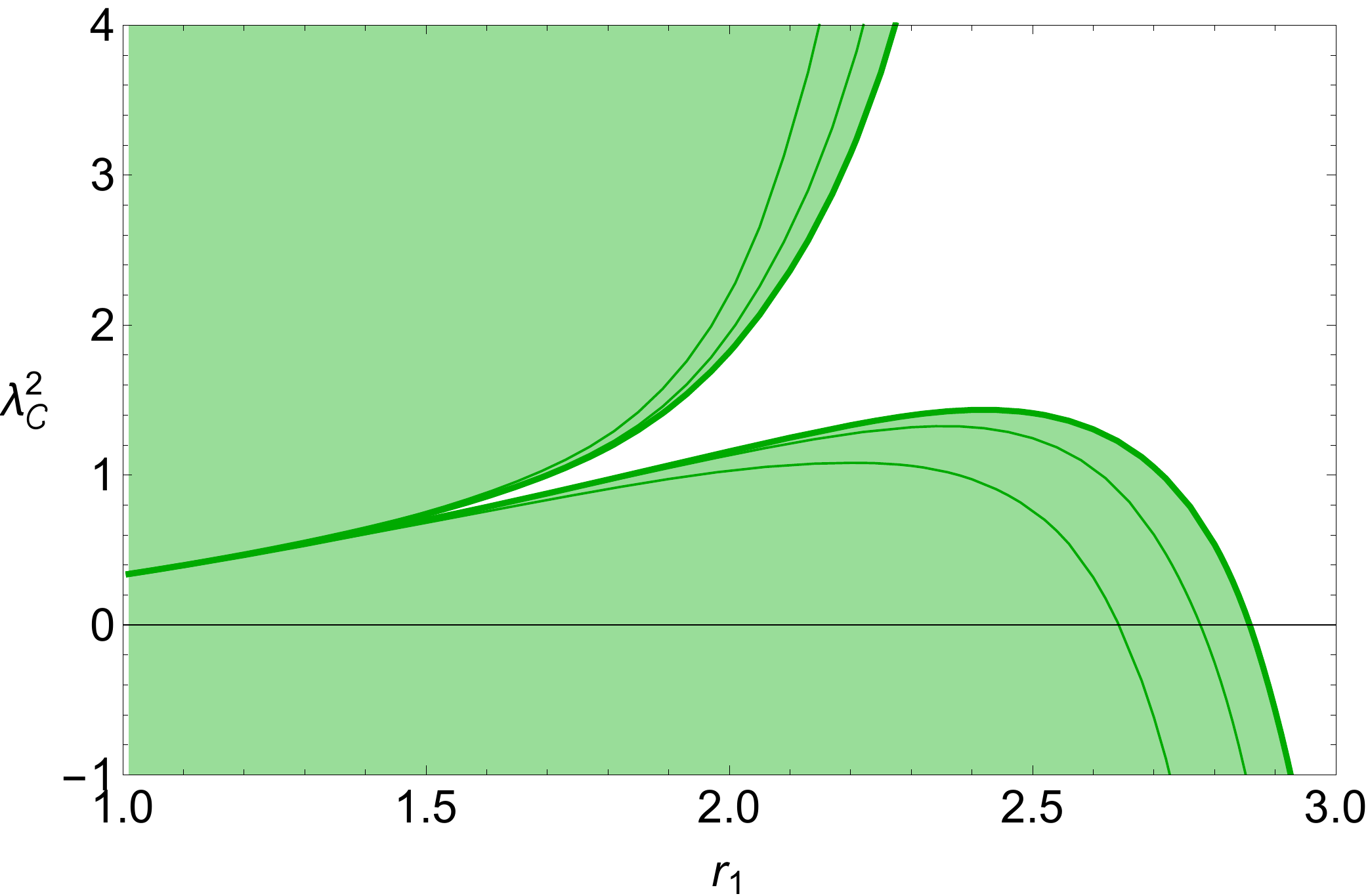}
\quad
\includegraphics[scale=0.35]{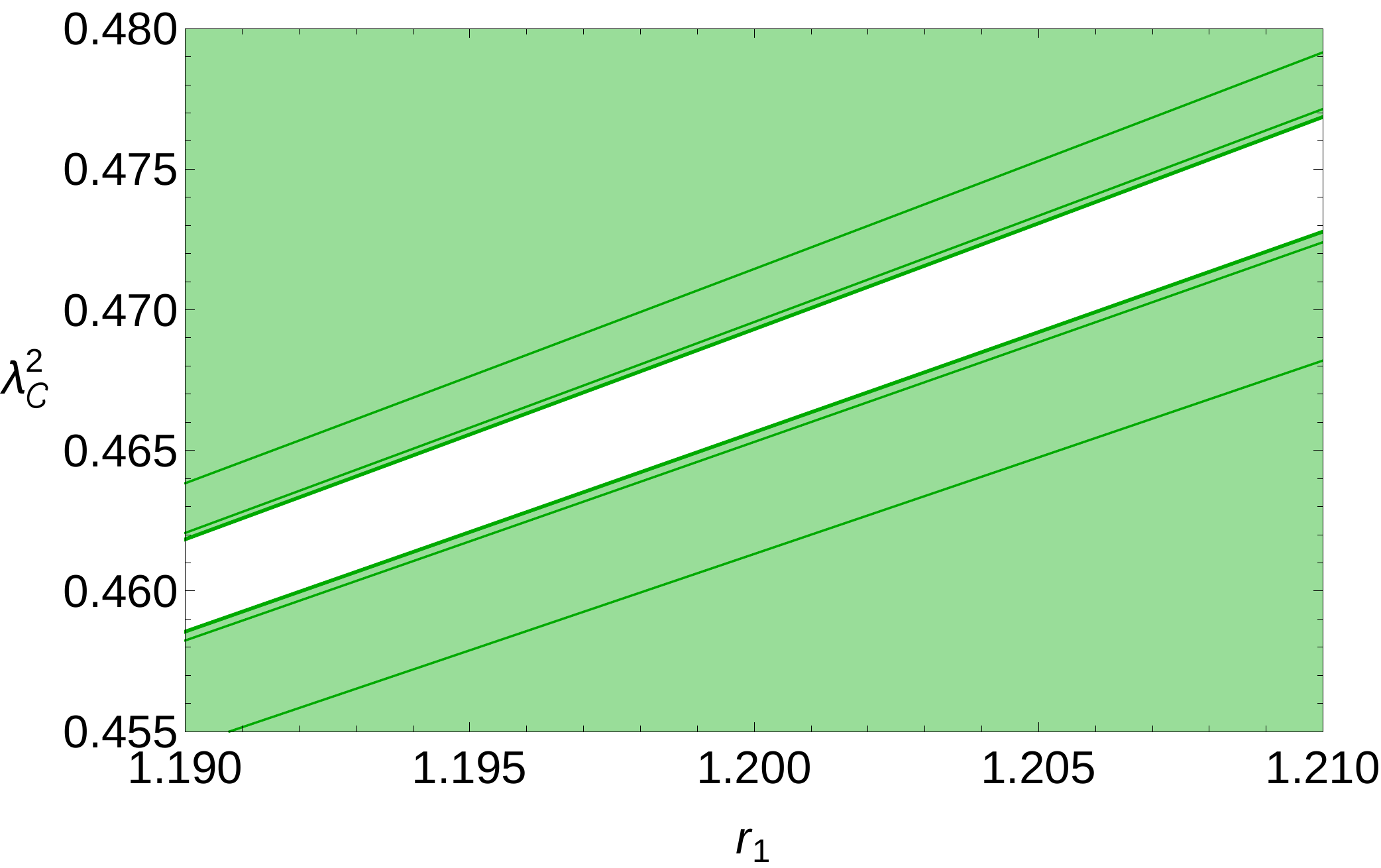}
\caption{Upper and lower bound on $\Cm_{0,2r_1- 1 (0,1)}$ for $\Lambda=12,16,20$. The panel on the right zooms in the region around $r_1=\tfrac{6}{5}$ in order to single out the $H_0$ family.}
\label{fig:Cmultipletbounds}
\end{figure}

From \cite{Beem:2014zpa} we know that the lower and upper bounds on $\lambda_{\Em_{2r_1}}^2$ are very close for small values of $r_1$, restricting the OPE coefficient to lie in a small range. Similarly, the family of multiplets $\Cm_{0,2r_1-1 (j-1,j)}$ is separated from the continuum of long operators by a gap, and their OPE coefficients are also constrained to a range. In this section we will concentrate on $\Cm_{0,2r_1-1 (0,1)}$, the first multiplet of this family. The bounds can be obtained for arbitrary central charge $c$, but one can also fix it to some particular value in order to bootstrap a specific theory. We present our results in Fig.~\ref{fig:Cmultipletbounds} for the arbitrary $c$ case, and they will be compared in section \ref{sec:H0} with the ones for $c=\tfrac{11}{30}$, which corresponds to the $H_0$ theory.

The figure shows upper and lower bounds for a range of $r$. Simple extrapolation seems to imply that both upper and lower bound will converge to the same number for $r=1$, namely, $\lambda^2_{\Cm}=\tfrac{1}{3}$ which corresponds to free field theory.\footnote{Our numerics became unstable for $r=1$, so we plotted starting from $r=1.01$ instead.}

Is also interesting to look at the value $r=\tfrac{6}{5}$ which corresponds to the $H_0$ family of theories. We have zoomed in this region on the right figure. Both upper and lower bounds appear to be close to their optimal values, and it seems that they will not converge to the same line; there will be a range of allowed values for this coefficient. To understand this, let us recall that an infinite number of theories fall within that range: all the rank $N$ $H_0$ theories. They are characterized by a lowest dimensional generator of dimension $r=\tfrac{6}{5}$ and, in the single correlator setup, can be distinguished by their central charge $c$. In Fig.~\ref{fig:Cmultipletbounds} we have been agnostic about the value of $c$, and it is therefore not surprising that the bound at $r=\tfrac{6}{5}$ allows for a range. To recap, we have proved numerically that the coefficient $\lambda^2_{\Cm}$ is constrained to lie in the range
\be 
0.465 < \lambda^2_{\Cm} < 0.470 \qquad \text{for all rank $N$ $H_0$ theories.}
\label{rankN_Cbound}
\ee
By fixing $r=\tfrac{6}{5}$ in the crossing equations we have bootstrapped the $\lambda^2_{\Cm}$ coefficient for all the rank $N$ theories to $\sim0.5\%$ precision. This coefficient is not protected by supersymmetry, and because there is no Lagrangian description for the $H_0$ family, the bootstrap is the only tool available for the calculation of this quantity.

The end point of the lower bound in Fig.~\ref{fig:Cmultipletbounds} is also of interest. One of the lessons the numerical bootstrap has taught us is that kinks are associated with the vanishing of OPE coefficients. The $3d$ Ising model kink can be identified in this way as shown in \cite{El-Showk:2014dwa}. The same phenomenon was also observed in the supersymmetric Ising model \cite{Bobev:2015vsa,Bobev:2015jxa}, and in $4d$ $\Nm=1$ theories in \cite{Poland:2010wg,Poland:2015mta}. From Fig.~\ref{fig:Cmultipletbounds} however, there does not seem to be a fixed point where the lower bound converges. The spacing of the lines is diminishing, but this seems to be associated with the derivative expansion: higher derivative terms correspond to smaller corrections. No special value of $r_1$ is being singled out, and therefore we do not expect a kink associated with the vanishing of the $\Cm_{0,2r_1- 1 (0,1)}$ multiplet.

%!TEX root = ../mixedE.tex

%%%%%%%%%%%%%%%%%%%%%%%%%%%%%%%%%%%%%%%%%%%%%
\subsection{Dimension bounds for the chiral channel}
%%%%%%%%%%%%%%%%%%%%%%%%%%%%%%%%%%%%%%%%%%%%%

We now turn to the unprotected multiplets $\Am^{\Delta_{\mathrm{primary}}}_{0,2r_1-2 (j,j)}$. We will focus only on spin zero operators, and on their dimensions, leaving the study of the respective OPE coefficients for future work. Unitarity requires the dimension of the superconformal primary of this multiplet obeys $\Delta_{\mathrm{primary}} \geqslant 2\, r_1 + 2j$, with the superconformal descendant contributing to the correlator having dimension $\Delta_{\mathrm{descendant}} = \Delta_{\mathrm{primary}} + 2$. Thus, numerical bounds on the descendant can be easily translated to bounds on the superconformal primary. In what follows we present bounds on the dimension of the descendant. As found in the previous subsection, the short multiplet $\Cm_{0,2r_1-1 (0,1)}$ must necessarily be present (at least for external dimensions smaller than $\sim 2.8$) and we therefore include it in the expansion. Similarly, the higher spin $\Cm_{0,2r_1-1 (j-1,j)}$ multiplets are expected to have lower bounds analogous to the one in Fig. \ref{fig:Cmultipletbounds}, and should be included in the analysis, at least for small values of $r_1$.

Now, the $\Bm_{1,2r_1-1 (0,0)}$ multiplet lies precisely at the unitarity bound of $\Am^{\Delta_{\mathrm{primary}}}_{0,2r_0-2 (0,0)}$, and its inclusion or exclusion can be physically motivated, depending on whether we are interested in theories with or without a mixed branch. In what follows we explore both possibilities. For the remaining short multiplets at the unitarity bound, the $\Cm_{\frac12, 2r_1- \frac32 (j-\frac12,j)}$ family for $j \geq 1$, there is no such physical interpretation. However, since no gap is being imposed in the spinning channels, and a long multiplet at the bound precisely mimics the presence of a $\Cm_{\frac12, 2r_1- \frac32 (j-\frac12,j)}$ multiplet, their presence or absence makes no difference for the numerical implementation. In what follows the central charge is also left unfixed.

\begin{figure}[htpb!]
             \begin{center}           
              \includegraphics[scale=0.35]{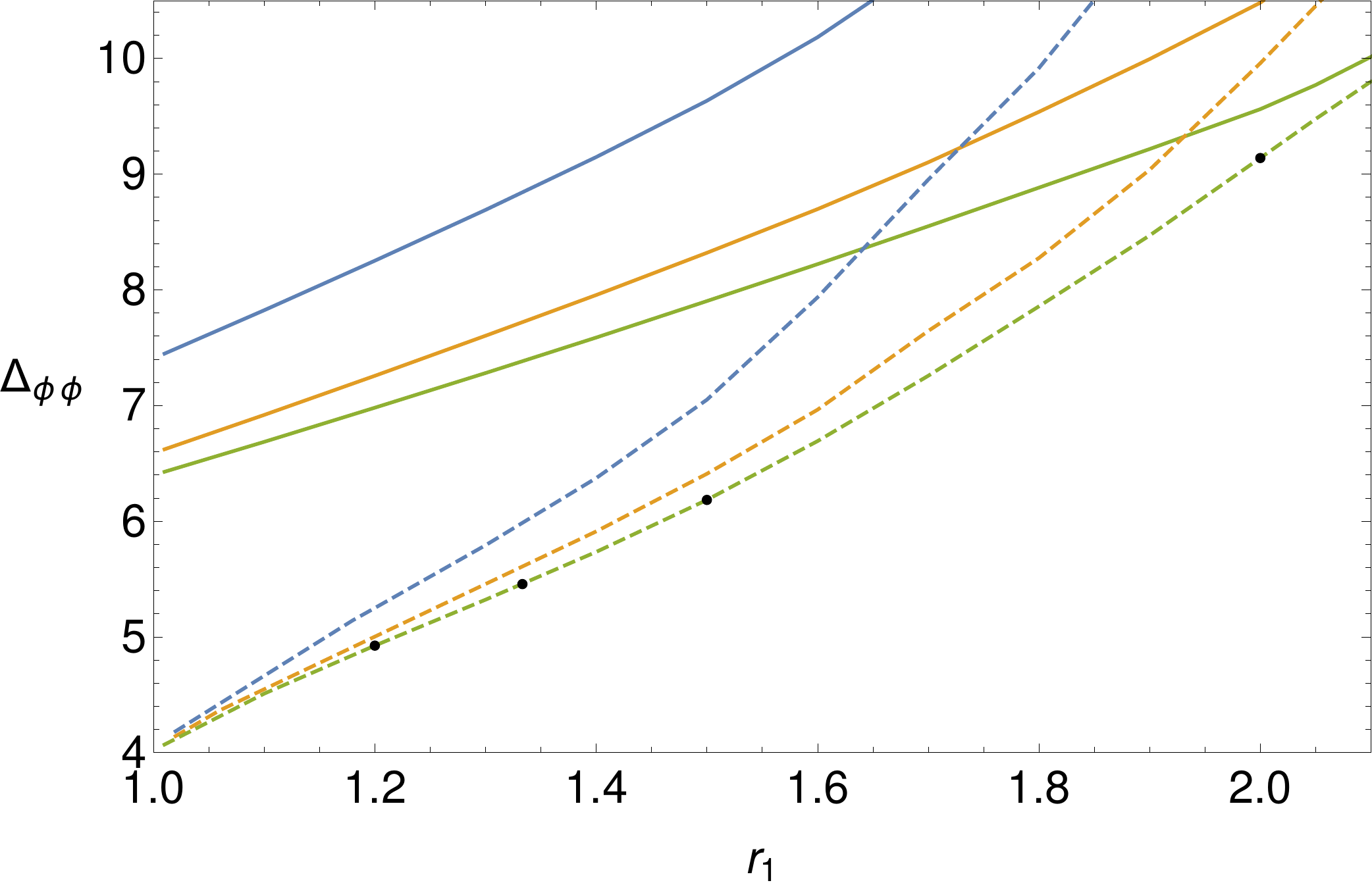}\quad \includegraphics[scale=0.35]{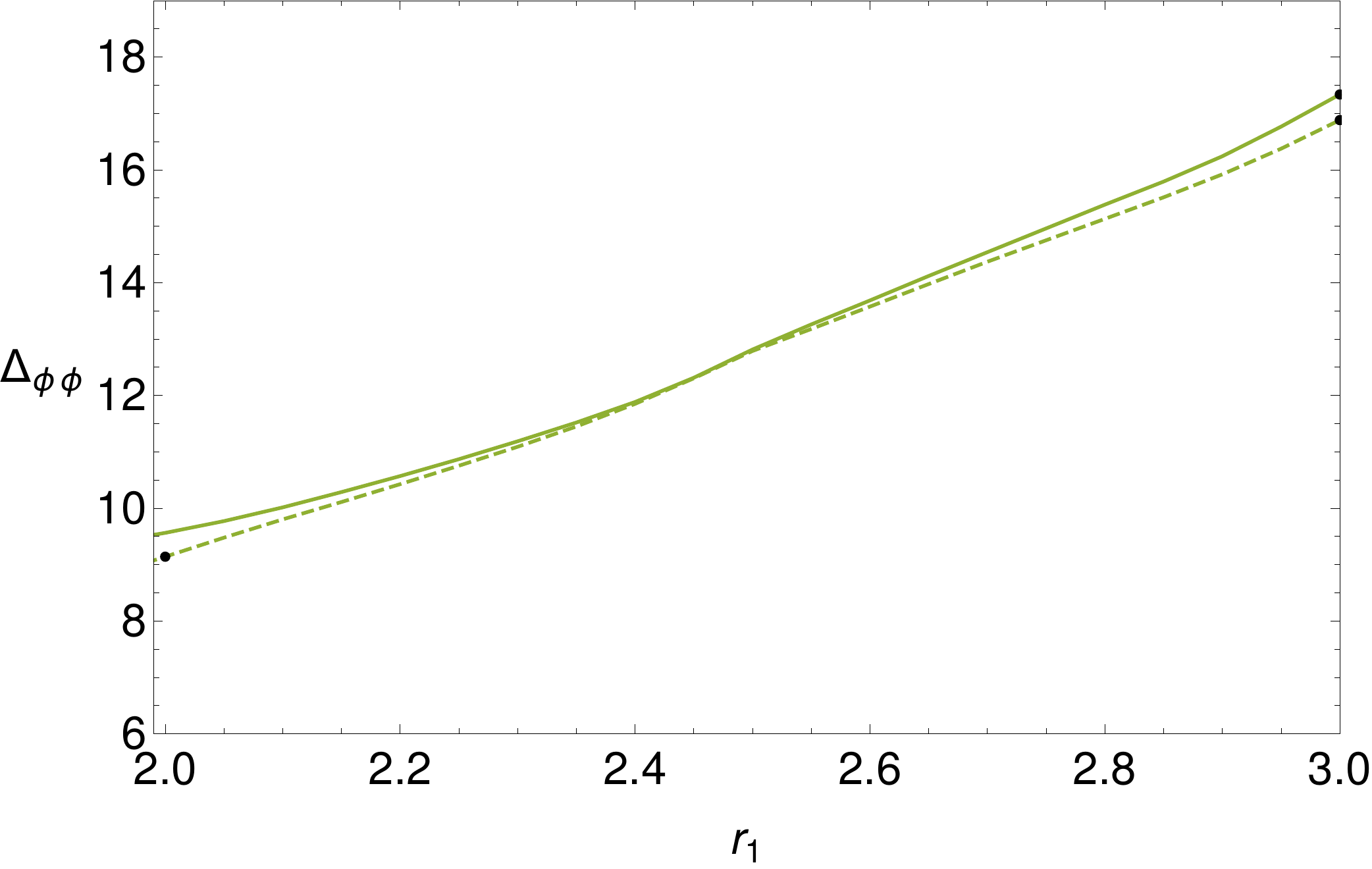}
              \caption{Bound on the dimension of the first unprotected scalar in the chiral channel for $\Lambda=12,16,20$. The solid (dashed) line allows (disallows) the short $\Bm_{1,2r_1-1 (0,0)}$ multiplet. The dimensions shown correspond to the superconformal descendant that makes an appearance in the OPE. The black dots mark the external dimension corresponding to known theories listed in Tabs.~\ref{Tab:rank1theories} and~\ref{Tab:newrank1theories}, and they are drawn on the bounds allowing or disallowing for the short multiplet according to whether or not these theories have a mixed branch.}
              \label{Fig:chiralbound}
            \end{center}
\end{figure}

The upper bound on the dimension $\Delta_{\mathrm{descendant}}$ is shown in Fig.~\ref{Fig:chiralbound} both allowing (solid) and disallowing (dashed) for the $\Bm_{1,2r_1-1 (0,0)}$ multiplet. Different ranges of the external dimension are shown on the left and right. The right plot shows only the $\Lambda=20$ curves.
We have also marked with black dots the external dimensions of known theories (see Tabs.~\ref{Tab:rank1theories} and~\ref{Tab:newrank1theories}), taking into account whether they have a mixed branch. These dots provide an upper bound for the operators in said theories.

While the bounds obtained allowing for the presence of the $\Bm_{1,2r_1-1 (0,0)}$ are very slow in convergence, if one aims to bootstrap theories for which this multiplet is not present convergence is much better, especially for small external dimensions. This is particularly relevant for the bootstrap of the $H_0$ theory discussed in the next subsection.
It is interesting to note that for external dimensions either smaller than $\sim 2.4$ or larger than $\sim 2.6$ (for $\Lambda=20$) the bound with the short multiplet removed is stronger. This is expected because disallowing the $\Bm_{1,2r_1-1 (0,0)}$ multiplet means there are less positivity conditions on the functional. There is a range around $\sim 2.5$ in which the solid bound approaches the dashed one. In the grid we are plotting they seem to coincide, although this might change with better tolerance. It would be interesting to understand what drives the solid bound and why it looses strength compared to the dashed one in that range.
%!TEX root = ../mixedE.tex
%%%%%%%%%%%%%%%%%%%%%%%%%%%%%%%%%%%%%%%%%%%%%
\subsection{The rank one \texorpdfstring{$H_0$}{H0} theory}
\label{sec:H0}
%%%%%%%%%%%%%%%%%%%%%%%%%%%%%%%%%%%%%%%%%%%%%

Until this point we have avoided making assumptions that are not shared by all $\Nm=2$ SCFTs, since our goal so far has been to obtain generic constraints on the CFT data. In this subsection we will change gears and focus on a specific theory instead. A natural candidate for this analysis is the rank one $H_0$ theory, which is the simplest of the canonical rank one $\Nm=2$ theories listed in Tab.~\ref{Tab:rank1theories}. It corresponds to the $N=1$ case in the $(A_1,A_{2N})$ family of Argyres-Douglas theories, and it also appears as the low-energy description of a single D3-brane probing an $F$-theory singularity of type $H_0$ \cite{Aharony:1998xz}.

It has been conjectured that the $H_0$ theory has a closed subsector of operators described by the two-dimensional Yang-Lee minimal model \cite{Leonardoprivate}.
It is a noteworthy fact that the $2d$ theory associated to $H_0$ can be identified as a minimal model.
This conjecture is part of the more general construction of \cite{Beem:2013sza}, where it was observed that any $\Nm=2$ SCFT has a subset of operators described by a $2d$ chiral algebra. The central charge of this $2d$ chiral algebra matches the one of the Yang-Lee minimal model, and the Schur index of the $4d$ theory seems to match the $2d$ vacuum character of the minimal model \cite{Cordova:2015nma}.
Moreover, using the $2d$ chiral algebra construction, it was shown in \cite{Liendo:2015ofa} that $H_0$ has the lowest possible value of the central charge $c$ among interacting $\Nm=2$ SCFTs.

It seems that the $H_0$ theory sits in a special place in the landscape of $\Nm=2$ SCFTs. It minimizes the central charge and does not have flavor symmetries, which simplifies its operator content. It is also suggestive that it has a subsector described by the Yang-Lee model, which is one of the paradigmatic examples of a solvable model. A further piece of information is that it does not have a mixed branch, and therefore the $\Bm_{1,2r_1-1 (0,0)}$ multiplet is absent from the $\phi_{r_1} \times \phi_{r_1}$ OPE. Finally, crucial CFT data like the external dimension $r=\tfrac{6}{5}$ and the central charge $c=\tfrac{11}{30}$ are known. Altogether, this makes the bootstrap program more likely to succeed for this theory, and we will try to leverage this information in order to corner $H_0$ and make specific statements about its operator spectrum. 

This subsection is divided in two parts. In the first part we obtain upper bounds for operator dimensions, and speculate where inside the bound $H_0$ sits. To make definite statements, we will work under the hypothesis that the numerical minimum of the central charge will converge to $\tfrac{11}{30}$ for $\Lambda \to \infty$. 
In the second part we constrain the OPE coefficients of the $\Em_{2r_1}$ and $\Cm_{0,2r_1- 1 (0,1)}$ multiplets for the rank one $H_0$ theory, now setting $c=\tfrac{11}{30}$ by hand. The allowed ranges for the last of these coefficients is significantly reduced when compared to the $\Cm_{0,2r_1- 1 (0,1)}$ range of section \ref{rankN_Cbound}. The improvement is a consequence that the central charge is now fixed, instead of being arbitrary.

Let us start by asking what characterizes the solution with the minimum central charge (for $r_1=\tfrac{6}{5}$), and if its features are consistent with the $H_0$ theory.
\begin{figure}[htbp!]
             \begin{center}
             \includegraphics[scale=0.35]{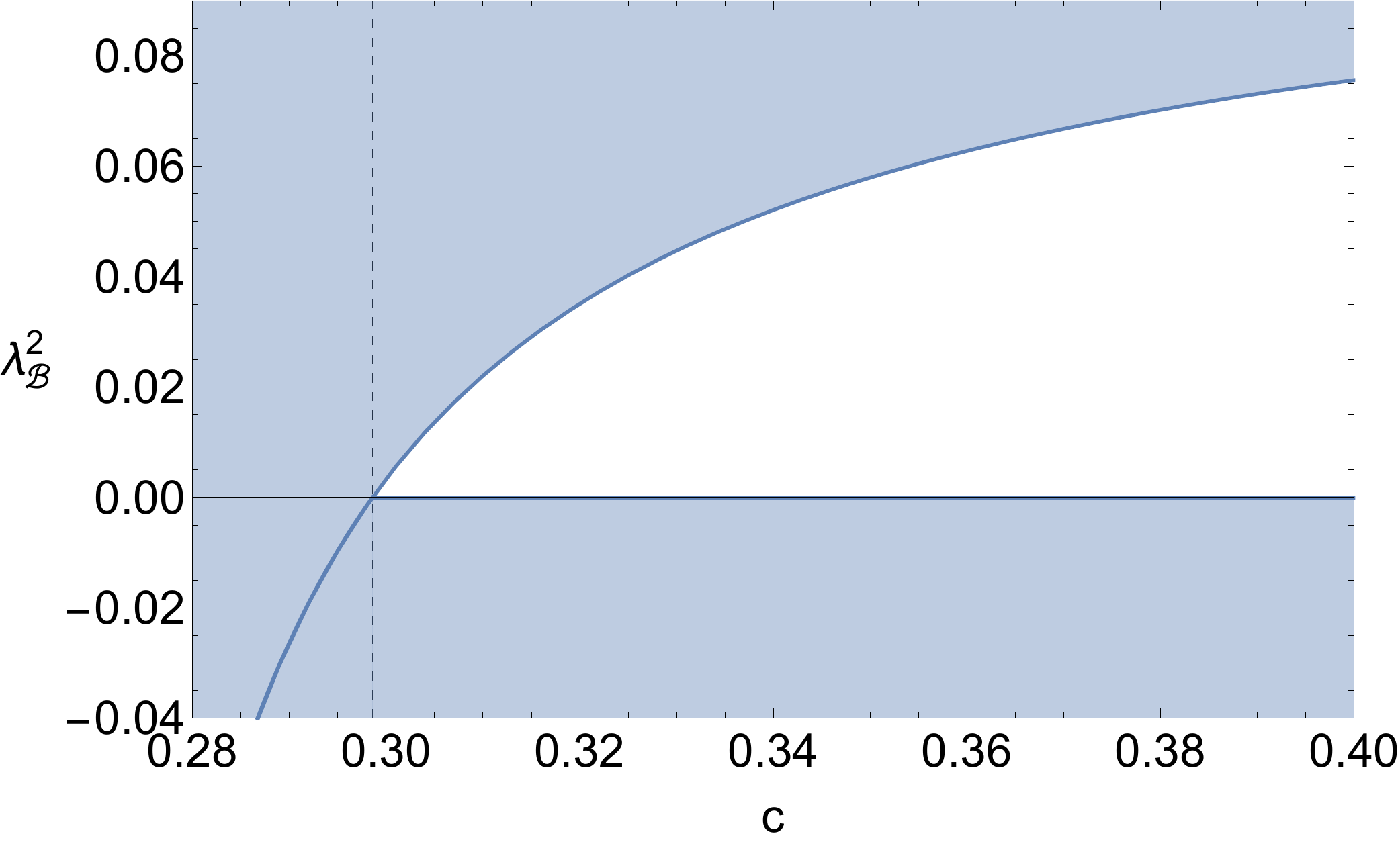}\quad        
              \includegraphics[scale=0.35]{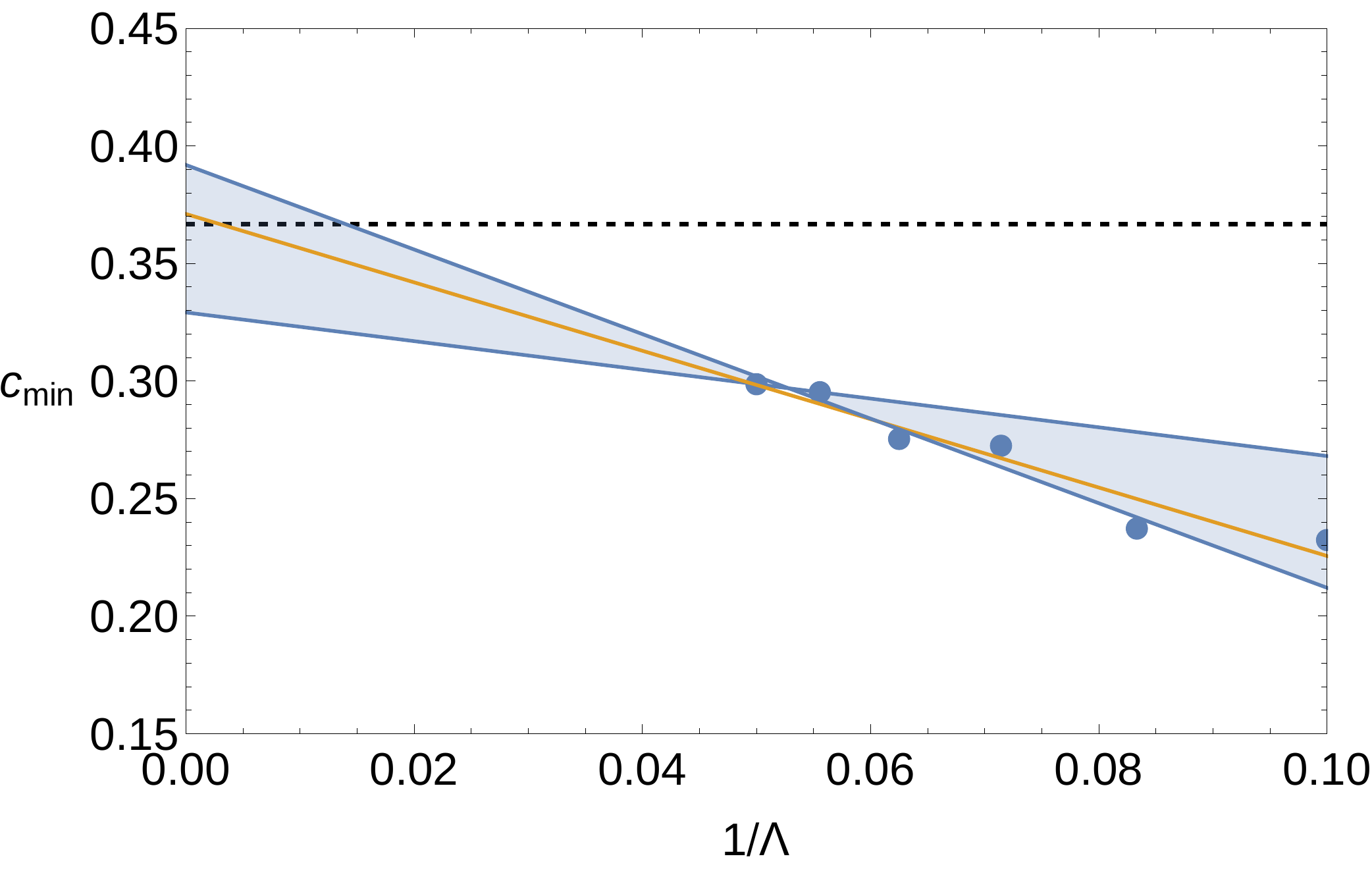}
              \caption{Left: Upper bound on the OPE coefficient of the  $\Bm_{R=1, r=2r_1-1 (0,0)}$ multiplet, for external dimension $r_1=\tfrac{6}{5}$, as a function of the central charge at $\Lambda=12$. The vertical dashed line corresponds to the minimum central charge allowed numerically with $\Lambda=12$. Right: Minimum allowed central charge for varying $\Lambda$, the dashed horizontal line marks the central charge of the rank one $H_0$ theory. The middle orange line shows a linear fit to all the data points, while the top and bottom blue lines show fits to different subsets of the points.}
              \label{Fig:BOPE_vs_c}
            \end{center}
\end{figure}
Because $H_0$ does not have a mixed branch, we actually know where to look. In Fig.~\ref{Fig:BOPE_vs_c} we present an upper bound on the OPE coefficient of the $\Bm_{1,2r_1-1 (0,0)}$ multiplet as a function of the central charge, keeping the external dimension fixed at $r_1=\tfrac{6}{5}$. Unitarity, which requires $\lambda_{\Bm}^2 \geqslant 0$, combined with the numerical upper bound restricts the coefficient to lie in the unshaded region. The upper bound, plotted for $\Lambda=12$, crosses zero precisely at the numerical central charge lower bound, $c_{min}$, for the same $\Lambda$. The bound becomes negative to the left of this point, implying that there is no unitary solution to crossing symmetry for $c < c_{min}$. The simplest interpretation is that the vanishing of this OPE coefficient is responsible for the central charge bound. This is reminiscent of what happens with the six-dimensional $A_1$ theory studied in \cite{Beem:2015aoa}. The hypothesis then is that when $\Lambda \to \infty$, $c_{min} \to \tfrac{11}{30}$, and at that point there will be a unique solution to crossing, as discussed in section \ref{sec:numericalimpl}. This solution will have $r_1=\tfrac{6}{5}$, $c=\tfrac{11}{30}$, and $\lambda_{\Bm}^2=0$, and should therefore correspond to the $H_0$ theory.

\subsubsection{Scalar bound for  \texorpdfstring{$H_0$}{H0}}

\begin{figure}[htbp!]
             \begin{center}           
              \includegraphics[scale=0.35]{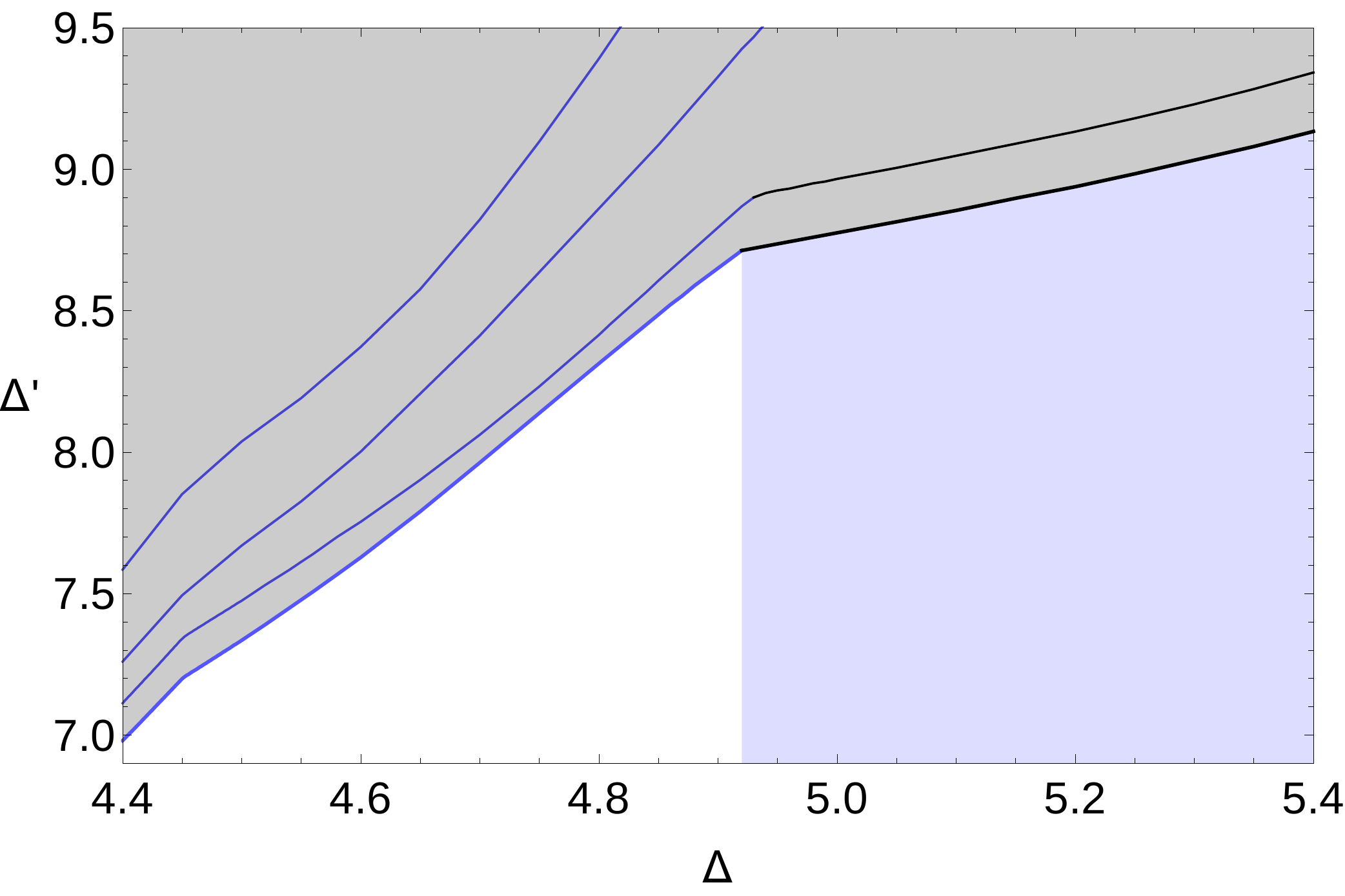}\quad
              \includegraphics[scale=0.35]{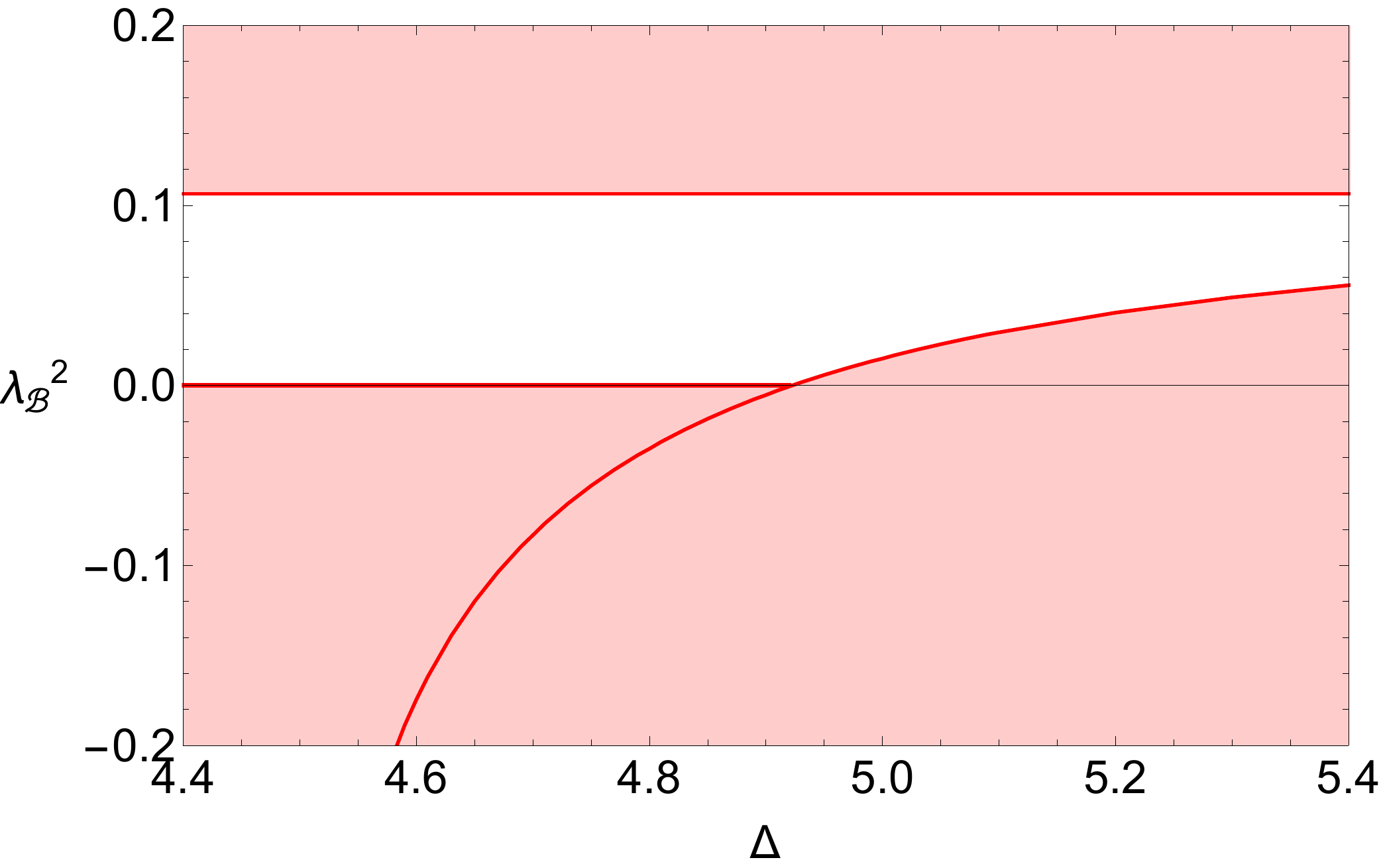}
              \caption{Left: Bound on the dimension of the second long scalar $\Delta_0^\prime$ in the $\phi \phi$ channel as a function of the dimension of the first long scalar $\Delta_0$, for external dimension $r_1=\tfrac{6}{5}$. The central charge is left arbitrary and $\Lambda=14,16,18,20$. The black curve corresponds to the region where the short multiplet is required to have a positive OPE coefficient. The shaded black region is always excluded, and the shaded blue region only if one demands the absence of the $\Bm_{1,2r_1-1 (0,0)}$ short multiplet. Right: Upper and lower bounds on the OPE coefficient of the $\Bm_{1,2r_1-1 (0,0)}$ multiplet, as a function of the dimension of the first long scalar $\Delta_0$.}
              \label{Fig:kink}
            \end{center}
\end{figure}
The dashed curve bound in Fig.~\ref{Fig:chiralbound} for $r_1=\tfrac{6}{5}$ implies the dimension of the first long scalar operator in the $\phi_{r_1} \times \phi_{r_1}$ channel must obey $4.4 \leq \Delta_{0} \lesssim 4.92$, where the lower end follows from the unitarity bound for this multiplet. As in the previous subsection we present the dimension of the superconformal descendant that appears in the OPE. We now want to understand which solution to crossing symmetry saturates the bound, and if it corresponds to the $H_0$ theory. 
For that we assume the \textit{first} long scalar $\Delta_0$ has a given dimension, lying in the allowed range, and ask what is the bound on the \textit{second} long scalar $\Delta_0^\prime$. This is shown in Fig.~\ref{Fig:kink} in black and blue.

The figure has a characteristic kink which signals the transition between two regimes. To the right of the kink (black curve), the $\Bm_{1,2r_1-1 (0,0)}$ multiplet is always present, because for $\Delta_0 \gtrsim 4.92$, its OPE coefficient is required to be positive, as it must lie in the unshaded region of the plot on the right side of Fig.~\ref{Fig:kink}.\footnote{With a gap of $\Delta_0$ being imposed we can obtain both lower and upper bounds on the $\Bm_{1,2r_1-1 (0,0)}$ OPE coefficient.} To the left of the kink (blue curve), the lower bound for $\Bm_{1,2r_1-1 (0,0)}$ disappears and the multiplet can be safely removed. This in an example of what we discussed in section \ref{Sec:COPE}: a kink in a scalar bound is a consequence of the vanishing of an OPE coefficient.

\begin{figure}[htbp!]
             \begin{center}           
              \includegraphics[scale=0.35]{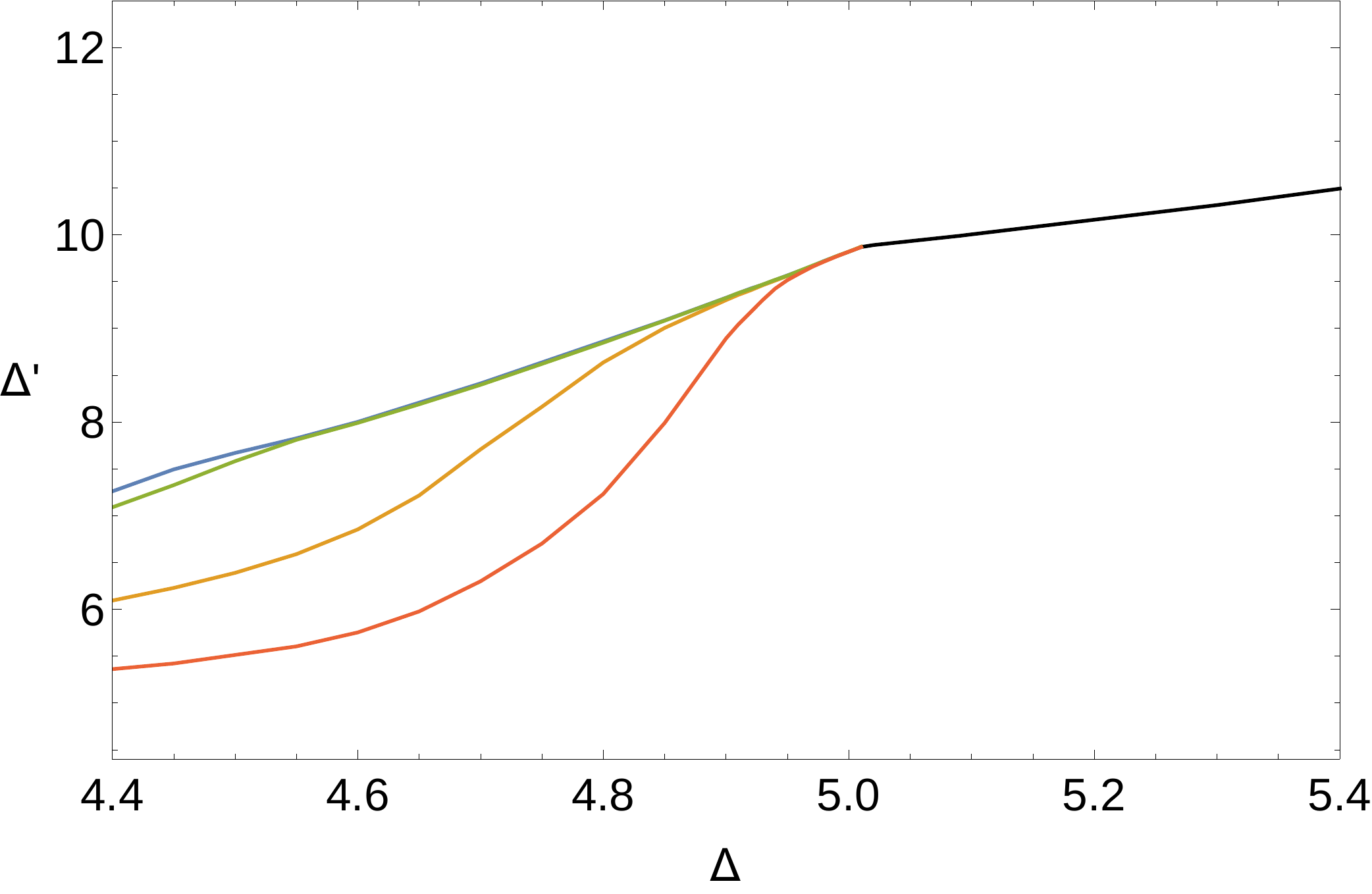}
              \caption{Bound on the dimension of the second long scalar $\Delta_0^\prime$ in the $\phi_{r_1} \times \phi_{r_1}$ channel as a function of the dimension of the first long scalar $\Delta_0$, for external dimension $r_1=\tfrac{6}{5}$ with $\Lambda=16$. The different colors correspond to different central charges.}
              \label{Fig:kink_fixedc}
            \end{center}
\end{figure}

The results so far do not clarify where inside the bound the $H_0$ theory sits. Let us repeat the analysis, but now fixing the central charge to several values. In Fig.~\ref{Fig:kink_fixedc} we show the same arbitrary central charge bound (at $\Lambda=16$) overlapped with curves in red, yellow and green obtained by fixing $c$ to $\tfrac{3}{10}$, $\tfrac{11}{30}$, and $\tfrac{17}{12}$. The first of these values is close the numerical central charge bound at $\Lambda=16$ ($\sim 0.275$), while the last two correspond to the rank one and rank two $H_0$ theories respectively. Because the long multiplet $\Am^{\Delta}_{0,0 (0,0)}$ mimics the contribution of the stress-tensor multiplet when $\Delta \sim 2$, the lines in Fig.~\ref{Fig:kink_fixedc} should be interpreted as representing a range of central charges: $c \leqslant \tfrac{3}{10}$, $c \leqslant \tfrac{11}{30}$, and $c \leqslant \tfrac{17}{12}$. 

One conclusion that can be immediately drawn from these results is that the bound for arbitrary central charge to the left of the kink (in blue) is being controlled by the large central charges. At the minimum allowed numerical central charge, a big portion to the left of the kink is ruled out. Reducing the central charge has the effect of carving the allowed region away from the kink, while keeping the region near the kink untouched.
Recalling that for the minimum numerical central charge there is a unique solution to the truncated crossing equations, and that, from Fig.~\ref{Fig:BOPE_vs_c}, the $\Bm_{1,2r_1-1 (0,0)}$ multiplet should be absent in said solution, this suggests the position of the kink corresponds to the minimum central charge theory.
If the minimum central charge reaches $\tfrac{11}{30}$ as $\Lambda \to \infty$, then this last statement applies to the rank one $H_0$ theory, implying that the theory is the \emph{unique} solution to the crossing equations for $r_1=\tfrac{6}{5}$ and $c=\tfrac{11}{30}$, with the position of the kink giving an estimate on the dimension of the two lowest operators in the chiral channel.
Altogether, the possibility that the $H_0$ theory saturates the numerical bound of $\Delta \lesssim 4.92$ seems plausible, and at the least it warrants further investigation.

\subsubsection{OPE bounds for  \texorpdfstring{$H_0$}{H0}}

We now turn to bounding the OPE coefficients of the isolated short multiplets, where lower and upper bounds can be obtained.
These OPE coefficients are bounded to very narrow ranges, confirming once again the usefulness of the numerical bootstrap program. In this section we set the central charge to $c = \tfrac{11}{30}$ and the results do not rely on any assumption regarding the $\Lambda \to \infty$ limit. The bounds obtained here are rigorous, and can only improve for higher $\Lambda$.

\begin{figure}[h!]
\begin{center} 
\includegraphics[scale=0.35]{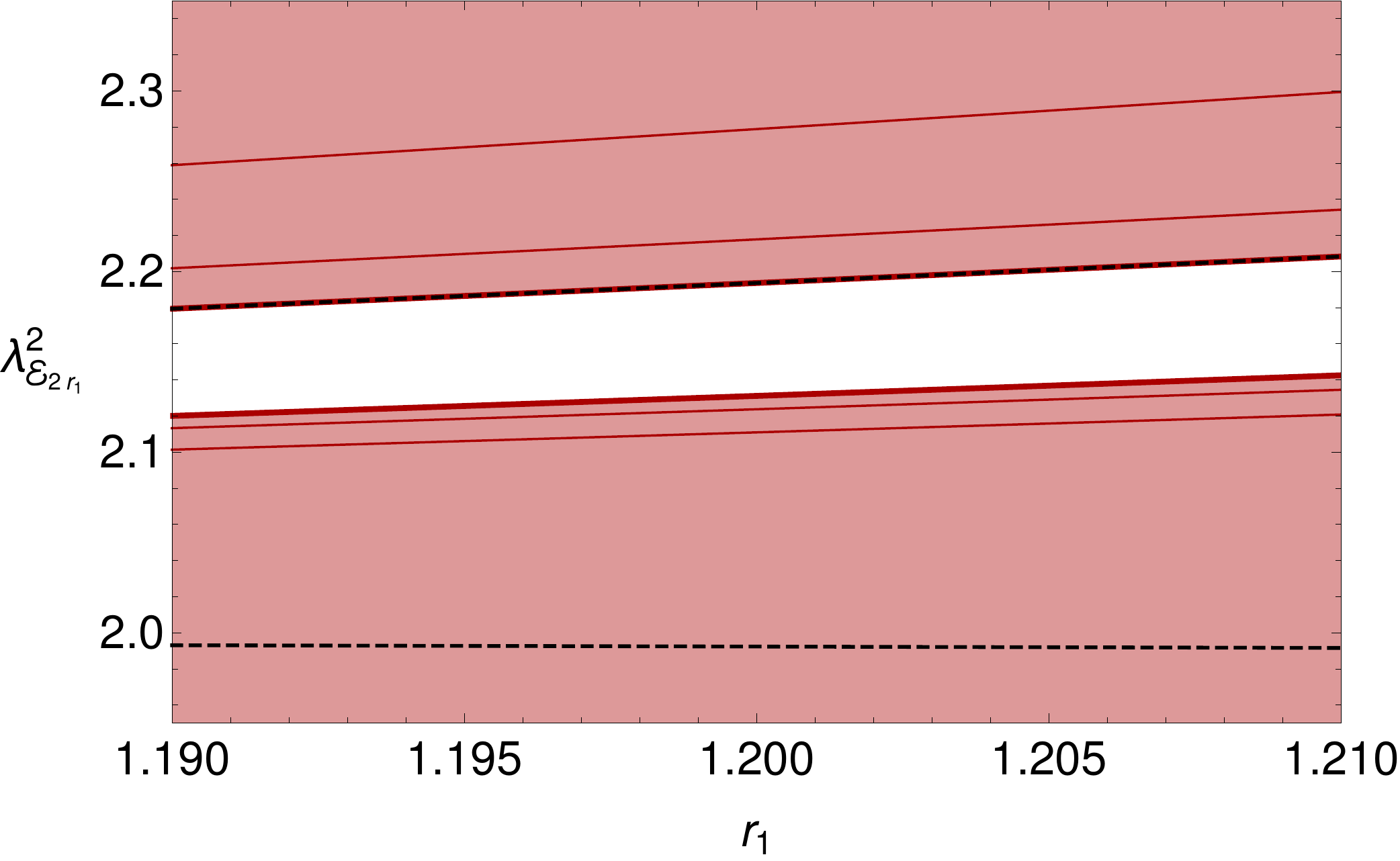}\quad
\includegraphics[scale=0.35]{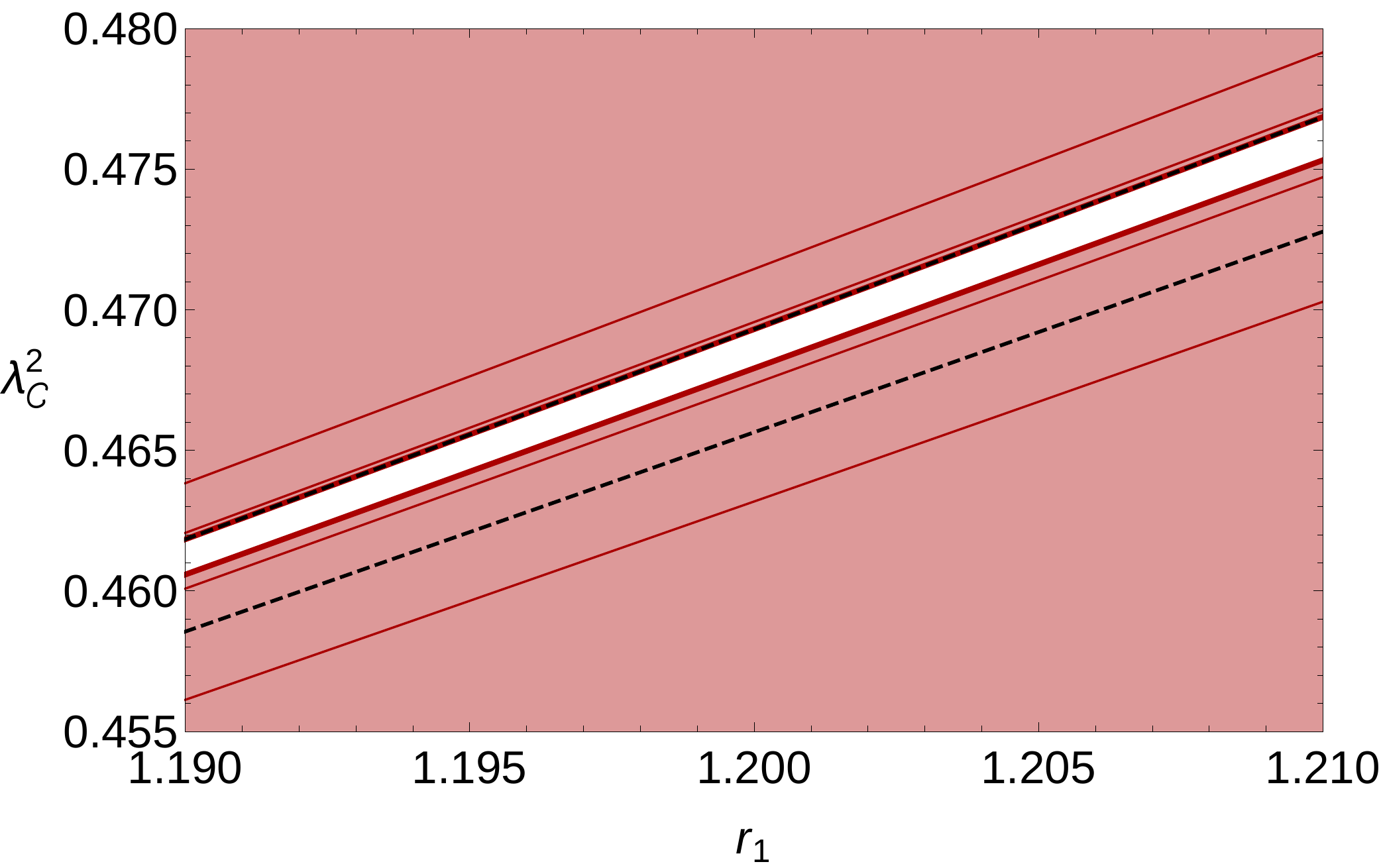}
\caption{Upper and lower bound on the $\Em_{2r_1}$ (left) and $\Cm_{R=0 \, r = 2r_1- 1 (0,1)}$ (right) OPE coefficients for $c\leqslant \tfrac{11}{30}$, with various derivatives $\Lambda=12,16,20$. The dashed black lines indicate the position of the bounds at $\Lambda=20$ with arbitrary central charge.}
\label{Fig:OPEbound_fixedc}
\end{center}
\end{figure}

The $\Em_{2r_1}$ multiplet was bounded in Fig.~24 of \cite{Beem:2014zpa} for various central charges and external dimensions. We now provide in Fig.~\ref{Fig:OPEbound_fixedc} a slice of that plot with the central charge constrained to $c \leq \tfrac{11}{30}$ (recall the discussion of the previous section regarding the inequality), and with the external dimension close to $r_1=\tfrac{6}{5}$.
This implies the  following bound:
\be 
2.13 < \lambda^2_{\Em_{2 \times 6/5}} < 2.20 \qquad \text{for the rank one $H_0$ theory.}
\ee
The situation gets better if one looks at the OPE coefficient of the $\Cm_{0 \, 2r_1- 1 (0,1)}$ multiplet,
which in Fig.~\ref{fig:Cmultipletbounds} had already been constrained to lie in a very narrow interval. If in addition we require $c \leqslant \tfrac{11}{30}$, this range gets reduced further:
\be 
0.467 < \lambda^2 < 0.470 \qquad \text{for the rank one $H_0$ theory.}
\ee
All other $\Cm_{0,2r_1-1 (j-1,j)}$ multiplets can be bootstrapped in a similar manner, and it seems plausible that they will lead to strong bounds as well. Finally, we note that if the speculations of the previous subsection prove to be true, the ranges provided here will shrink to a point for $\Lambda \to \infty$, as there will be a unique solution to crossing symmetry at $c=\tfrac{11}{30}$, $r_1=\tfrac{6}{5}$.

%!TEX root = ../mixedE.tex
%%%%%%%%%%%%%%%%%%%%%%%%%%%%%%%%%%%%%%%%%%%%%
\section{Results for the mixed correlator bootstrap}
\label{sec:res_mixed}
%%%%%%%%%%%%%%%%%%%%%%%%%%%%%%%%%%%%%%%%%%%%%

We now turn to a mixed system of correlators, considering all possible four-point functions involving two $\Nm=2$ chiral operators of dimensions $r_1$ and $r_2$, which we denote by $\phi_{r_1}$ and $\phi_{r_2}$. Since the setup is symmetric we will assume, with no loss of generality, that $\phi_{r_2}$ is the lowest dimension operator, and consider only the case $r_2 \leqslant r_1$.

As described in section \ref{sec:fourpt}, we would like these two operators to be \emph{generators} of the Coulomb branch chiral ring, meaning they cannot be written as composites, allowing us to study theories with Coulomb branches of dimension two or higher.  However, with our methods it is hard to distinguish Coulomb branch generators from more generic operators. If their dimensions are strictly smaller than two, then unitarity forces them to be generators. If not, demanding the absence of the $\Em_{r_1-r_2}$ multiplet imposes $\phi_{r_1}$ is not of the form $\phi_{r_2} \phi_{r_1-r_2}$, but it does not rule out the possibility that it is a different type of composite. In what follows we consider both the cases: with and without the $\Em_{r_2-r_1}$ multiplet. In particular, if we wish to study rank one theories, the setup must be such that $r_2=2 r_1$ and $\phi_{r_1-r_2} = \phi_{r_2}$.
It is clear that for $r_1=2 r_2$ the bounds must be at least as strong as the single correlator bounds, as the crossing equations for that system are a subset of the ones considered here.

Note that when $r_1=r_2$, the middle selection rule in \eqref{EEbarOPE} reduces to the top one, with the exception that the identity operator is absent (the $\Em$ multiplet in the middle selection rule is not present if $|r_1-r_2|<1$). The absence of the identity imposes the two operators to be distinct, even if $r_1=r_2$.\footnote{For distinct operators one should also impose that the stress-tensor multiplet is absent. However, similarly to what happens with the higher spin currents, if no gap is imposed in this channel a scalar long multiplet with $\Delta \sim 2$ mimics the stress-tensor multiplet.
}

The amount of unfixed information in this system is much larger than in the single correlator case, and so in this work we focus on a few relevant quantities: the central charge $c$, the OPE coefficients of the multiplets $\Em_{r_1 \pm r_2}$, $\Bm_{\frac12,r_1+r_2-\frac12 (0,\frac12)}$, and scalar bounds on the first non-protected scalar in the $\phi_{r_1} \times \bar{\phi}_{-r_2}$ channel.

%!TEX root = ../mixedE.tex
%%%%%%%%%%%%%%%%%%%%%%%%%%%%%%%%%%%%%%%%%%%%%
\subsection{Central charge bounds}
%%%%%%%%%%%%%%%%%%%%%%%%%%%%%%%%%%%%%%%%%%%%%

The three parameters we can tune to zoom in on different theories are the two external dimensions $(r_1,r_2)$ and the central charge $c$ of the theory.
Our first goal is to find which values of these parameters are allowed by crossing symmetry. 
Recall that any local \emph{interacting} $\Nm=2$ SCFT must obey \cite{Liendo:2015ofa},
\be
c \geq \frac{11}{30}\, .
\label{eq:analytic_c_bound}
\ee
Moreover, combining the Shapere-Tachikawa sum rule \cite{Shapere:2008zf} with the Hofman-Maldacena bound \cite{Hofman:2008ar} gives the following central charge bound for a theory with a \emph{freely generated} Coulomb branch chiral ring, with $N$ generators of dimension ${r_1, \ldots r_N}$:
\be
c \geqslant \frac{1}{6} \sum_{i=1}^N(2 r_i-1)\,.
\label{eq:ST_bound}
\ee

\begin{figure}[htbp!]
             \begin{center}           
              \includegraphics[scale=0.35]{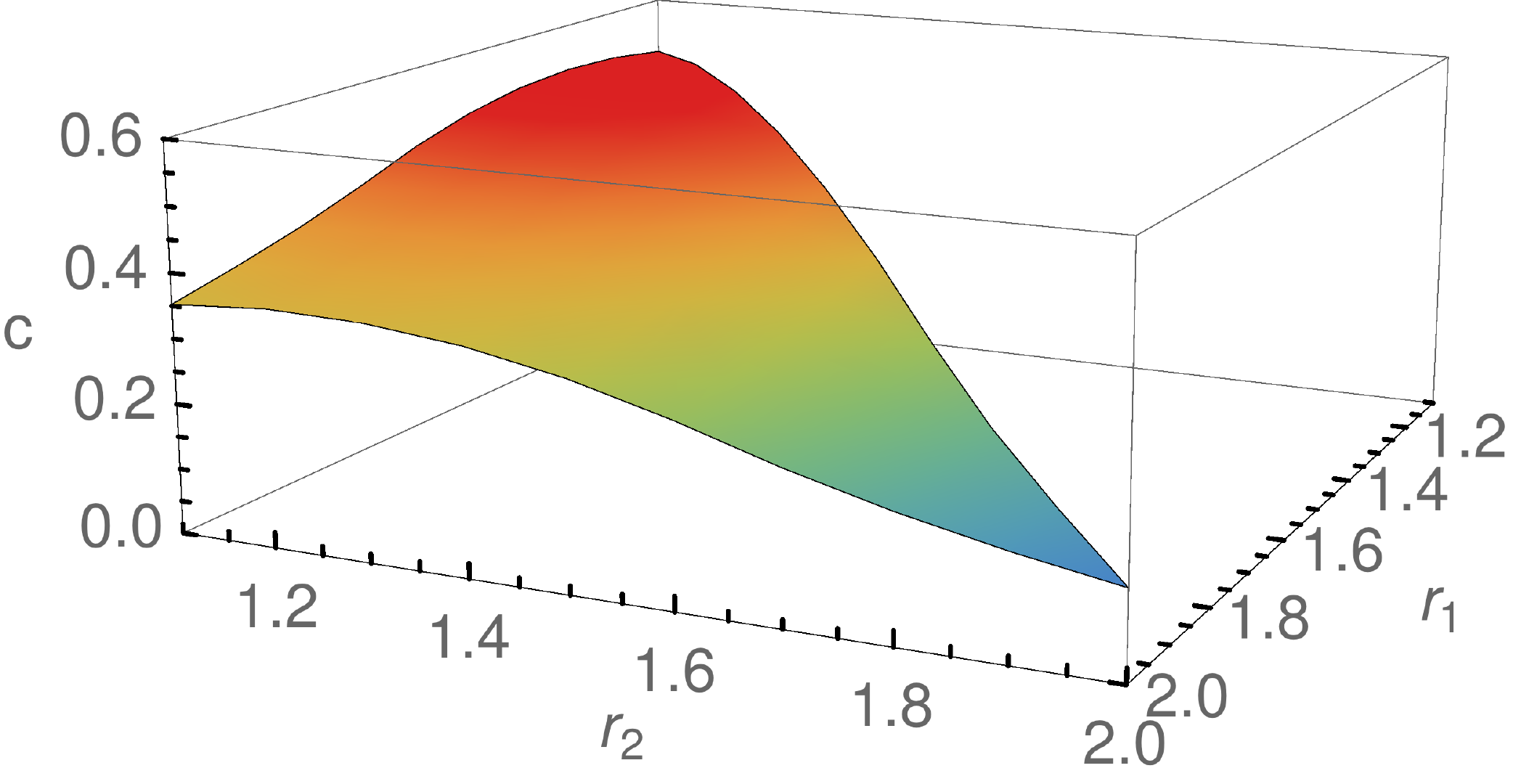}\quad
              \includegraphics[scale=0.35]{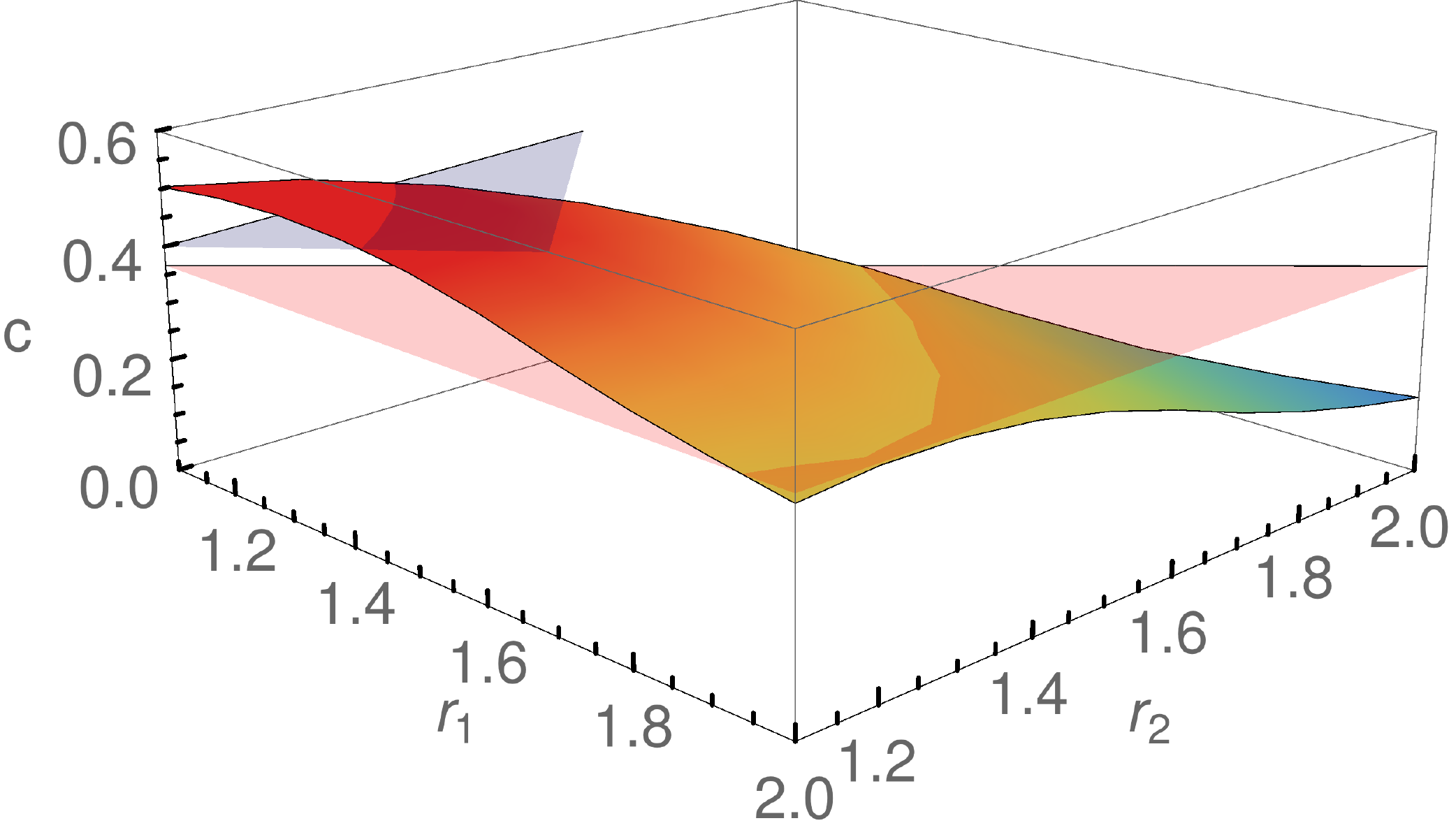}
              \caption{Central charge bound at $\Lambda=18$. On the right side the red surface corresponds to the analytic bound \eqref{eq:analytic_c_bound}, and the blue surface to \eqref{eq:ST_bound}.}
              \label{Fig:cbound}
            \end{center}
\end{figure}

We show in Fig.~\ref{Fig:cbound} the minimal allowed central charge for $\Lambda=18$, for small external dimensions $(r_1,r_2)$. 
In this region the bound is comparable to the aforementioned analytic bounds. Recall that we can interpret the slice $r_2=2r_1$ as a bound for the single correlator system for rank one theories. As already quoted, the lowest possible value for the central charge is $c=\tfrac{11}{30}$ and the mixed correlator plot of Fig.~\ref{Fig:cbound} does not improve on the single correlator bound.
If $r_2$ is independent from $r_1$, we can use our results to bound higher rank theories.
The lowest possible value for rank two theories, to the best of our knowledge, is the $(A_1,A_4)$ Argyres-Douglas SCFT, whose generators have dimensions $(r_1,r_2) = (\tfrac{8}{7},\tfrac{10}{7})$ and its central charge is $c=\tfrac{17}{21}$ \cite{Xie:2012hs,Xie:2013jc}. Although not shown, convergence is comparable to that of the single correlator bound of \cite{Beem:2014zpa}, and it does not seem likely that our bound will reach the $(A_1,A_4)$ theory.

The panel on the right shows the numerical central charge bound overlapped with the analytic bounds \eqref{eq:analytic_c_bound} in red, and \eqref{eq:ST_bound} in blue. The numerical bound shown here has not converged yet, and with improved numerics, it will get stronger than the analytic bounds in a wider range. 

\begin{figure}[htbp!]
             \begin{center}           
              \includegraphics[scale=0.35]{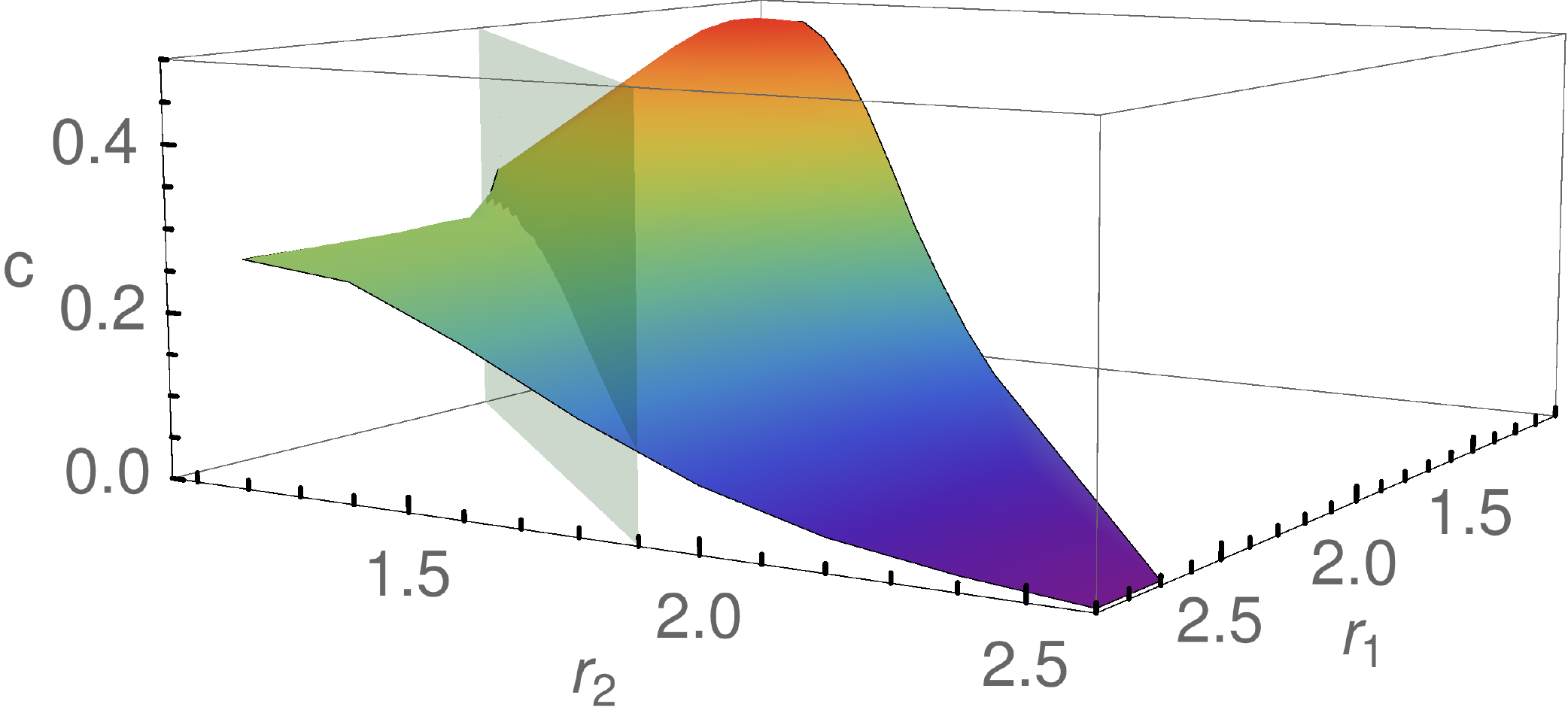}\quad \includegraphics[scale=0.35]{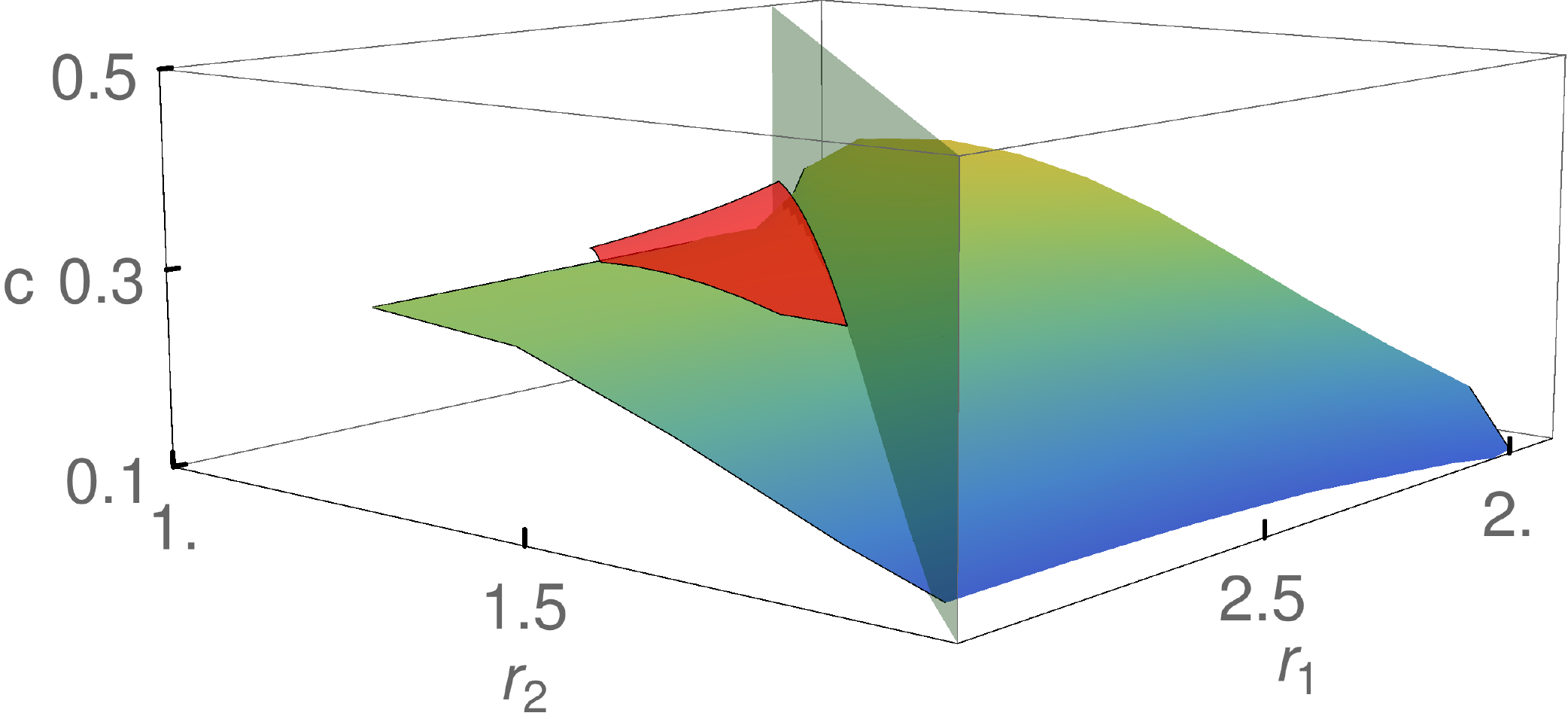}
              \caption{Central charge bound at $\Lambda=14$ for a wider range of external dimensions. The $\Em_{r_1-r_2}$ is allowed by the selection rules when $|r_1 - r_2| \geqslant 1$ (the green vertical wall) leading to the discontinuity in the bound. The red patch on the right plot shows the bound if the short $\Em_{r_1-r_2}$ multiplet is not included.}
              \label{Fig:cbound_largerange}
            \end{center}
\end{figure}

Turning to larger values of the external dimension, we show the bound for a smaller number of derivatives on the left side of Fig.~\ref{Fig:cbound_largerange}.
The bound gets substantially weaker and it asymptotes to the numerical central charge bound obtained from a single correlator in \cite{Beem:2014zpa}.
In particular, for $r_2$ small, and $r_1$ large the bound is approximately the single correlator bound for external dimension $r_2$, explaining its relatively flat behavior.\footnote{We should point out that our bound seems to rule out the free field theory value $c=\tfrac{1}{6}$. However, central charge bounds usually have a sharp drop close to $r=1$, presumably the same phenomenon happens here, but our three-dimensional grid is too rough to see it.}

Another feature of the bound shown on the right side of Fig.~\ref{Fig:cbound_largerange} is its the sudden jump at the vertical green wall, which is a purely kinematical effect.
As discussed below \eqref{EEbarOPE}, when $|r_1 - r_2| \geqslant 1$ a $\Em_{r_1-r_2}$ multiplet is allowed in the selection rules, with this sudden change being responsible for the jump in the bound.\footnote{One could hope that fixing the OPE coefficient of  $\Em_{r_1-r_2}$ equal to that of $\Em_{r_2}$ would give a stronger bound, allowing us to focus on rank one theories, with a generator of dimension $r_1$. Unfortunately from our investigations this did not appear to affect the $c$ bound.}
If this multiplet is disallowed in the selection rules, \eg, if we want $\phi_{r_2}$ not be the composite $\phi_{r_2-r_1} \phi_{r_1}$, the bound becomes smooth at this point, as shown by the red patch on the right side of Fig.~\ref{Fig:cbound_largerange}. As the values of the external dimension increase, the bounds with and without this multiplet asymptote to each other.

%!TEX root = ../mixedE.tex

%%%%%%%%%%%%%%%%%%%%%%%%%%%%%%%%%%%%%%%%%%%%%
\subsection{Bounds on the \texorpdfstring{$\Em_{r_1+r_2}$}{E(r1+r2)} OPE coefficient}
%%%%%%%%%%%%%%%%%%%%%%%%%%%%%%%%%%%%%%%%%%%%%

In the $\phi_{r_1} \times \phi_{r_2}$ OPE it is natural to bound the multiplet $\Em_{r_1+r_2}$, as the bound has clear physical consequences. Its associated conformal block is always separated by a gap in dimensions from the conformal blocks associated with all the other multiplets (see \eqref{EE2blocks}). Thus, we can obtain both lower and upper bounds on its OPE coefficient. In what follows we are interested in obtaining generic constraints, shared by many theories, and thus choose to let the central charge arbitrary.
\begin{figure}[htbp!]
             \begin{center}           
              \includegraphics[scale=0.35]{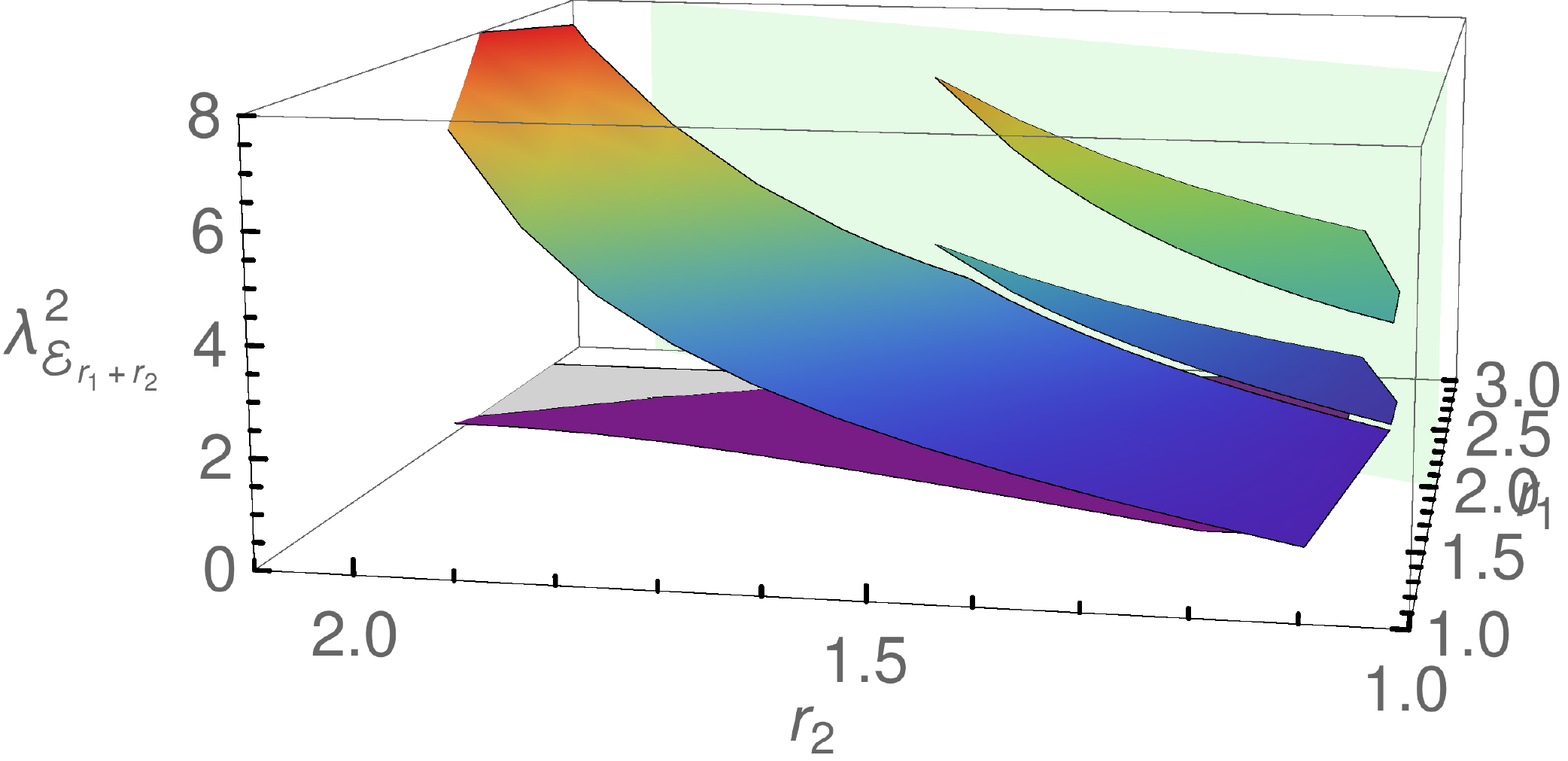} \quad \includegraphics[scale=0.35]{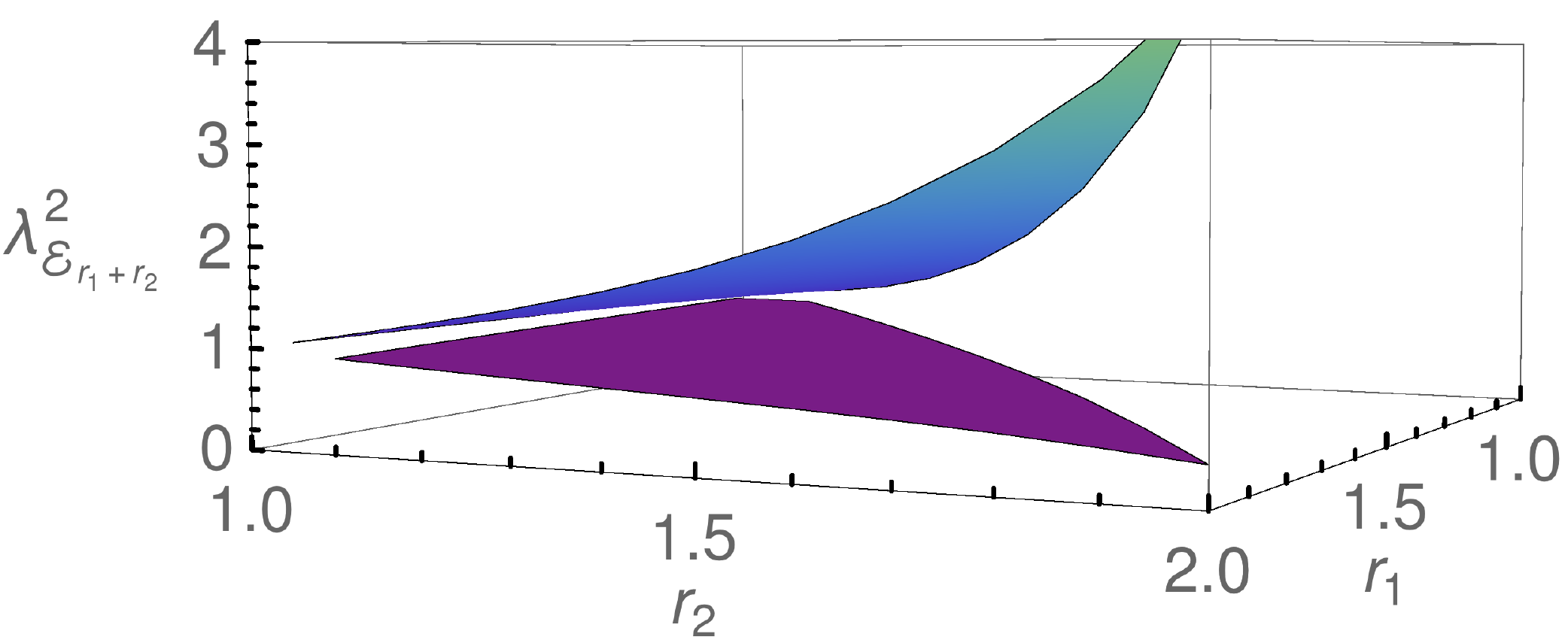}
              \caption{Upper and lower bound on the OPE coefficient of $\Em_{r_1+r_2}$ with $\Lambda=18$. On the left plot the vertical green wall marks the point at which the selection rules change due to the appearance of the $\Em_{r_1-r_2}$ multiplet. Beyond this wall there are two upper bounds shown, with the weakest one obtained allowing for the presence of said multiplet, and the strongest one forbidding its presence. The right plot shows a zoom for small external dimension.}
              \label{Fig:E1p2}
            \end{center}
\end{figure}

The OPE coefficient of $\Em_{r_1+r_2}$ must lie between the two surfaces shown in Fig.~\ref{Fig:E1p2}.
On the right side of the plot we show a zoom for small external dimensions, where the allowed range becomes more narrow.

As in the central charge bound, the discontinuity along the green wall ($r_1-r_2=1$) is caused by the drastic change in the selection rules \eqref{EEbarOPE}. If the multiplet $\Em_{r_1-r_2}$ is disallowed, then the upper bound is continuous, signalling that the discontinuity is purely a kinematical effect. The upper bound gets substantially weaker if $\phi_{r_2}$ is allowed to be a composite (by allowing the appearance of $\Em_{r_1-r_2}$). The lower bound seems insensitive to the presence or absence of this multiplet.

\paragraph{Coulomb branch relations:}

\begin{figure}[htbp!]
             \begin{center}           
              \includegraphics[scale=0.35]{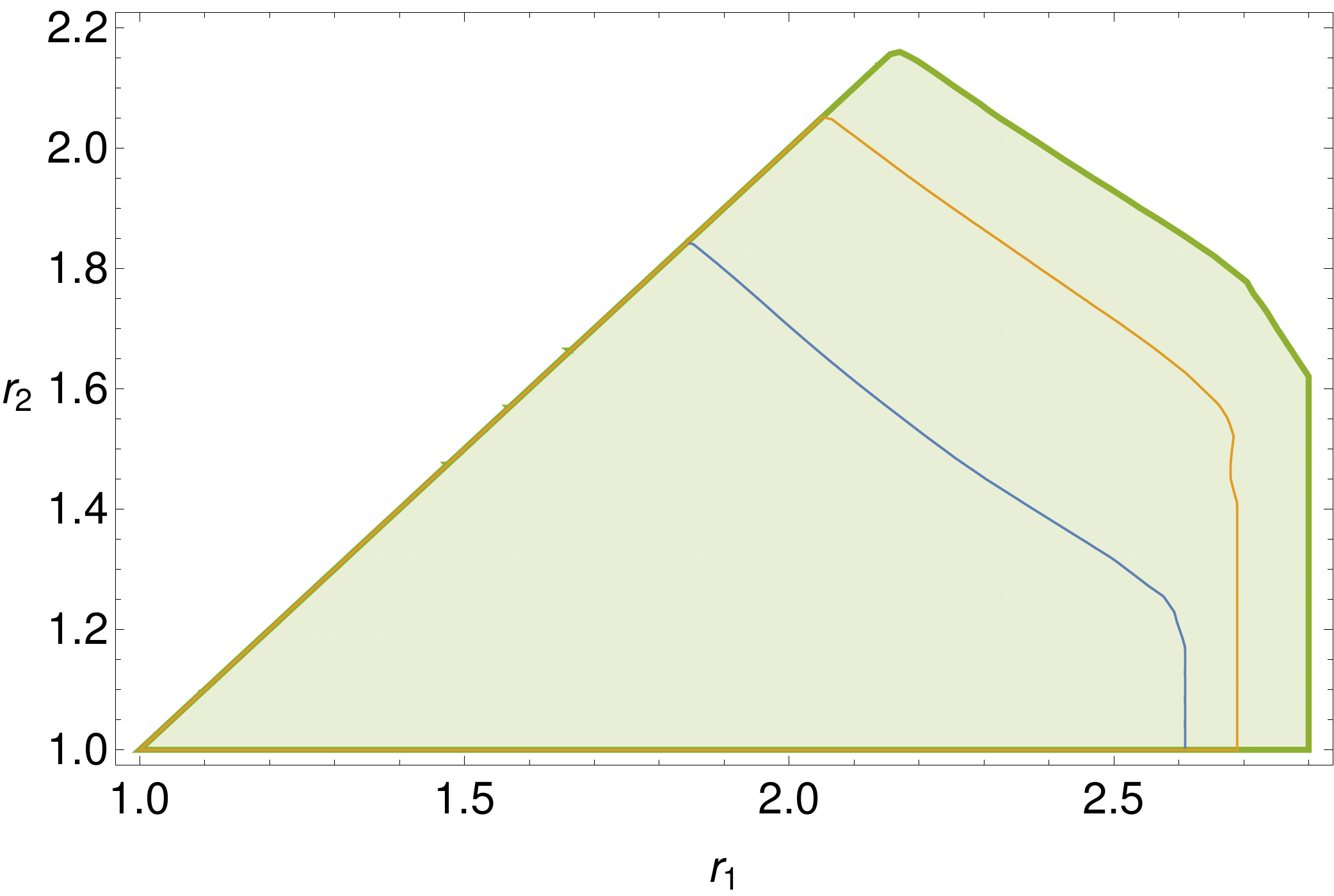}
              \caption{Projection to the $(r_1,r_2)$ plane of the bound on Fig. \ref{Fig:E1p2}, for $\Lambda=10,14,18$. Coulomb branch relations of the form $\phi_{r_1} \phi_{r_2} \sim 0$ are ruled out in the shaded region.}
              \label{Fig:zeroregion}
            \end{center}
\end{figure}

In the region where the lower bound is strictly positive we rigorously ruled out chiral ring relations of the type $\phi_{r_1} \phi_{r_2} \sim 0$. A projection to the $(r_1,r_2)$ plane is shown in Fig.~\ref{Fig:zeroregion}. This region is still increasing, as the bounds have not converged, but with the current numerical results ($\Lambda=18$) we can rule out these relations for the shaded region. We do not show surfaces for smaller $\Lambda$ in Fig.~\ref{Fig:E1p2} in order to avoid cluttering, however, we do show the locus where the upper bound is lost for several $\Lambda$ in Fig.~\ref{Fig:zeroregion}. One question that deserves further exploration is whether the numerical bound will always cross the horizontal axis as in Fig.~\ref{Fig:zeroregion}. For the the low values of $\Lambda$ considered here this is the case, however, if one hopes to rule out chiral ring relations for the whole $(r_1,r_2)$ plane this is not good enough. It might be useful then to explore the large $r_2$ region, and study crossing in the limit in which $r_1$ and $r_2$ are widely separated. Conformal blocks for disparate dimensions were studied in \cite{Behan:2014dxa}, where it was shown that the equations simplify significantly. 
Another exciting possibility is to rule out chiral ring relations analytically, at least in some approximation scheme. Apart from the highly disparate limit, crossing symmetry has been studied in several regimes (see for example \cite{Fitzpatrick:2012yx,Komargodski:2012ek,Alday:2013cwa,Fitzpatrick:2014vua,Vos:2014pqa,Fitzpatrick:2015qma,Kaviraj:2015cxa,Alday:2015eya,Kaviraj:2015xsa,Alday:2015ota} ) and maybe some, or a combination of them, are relevant for the problem at hand.
\paragraph{Zamolodchikov metric:}

\begin{figure}[htbp!]
             \begin{center}           
              \includegraphics[scale=0.35]{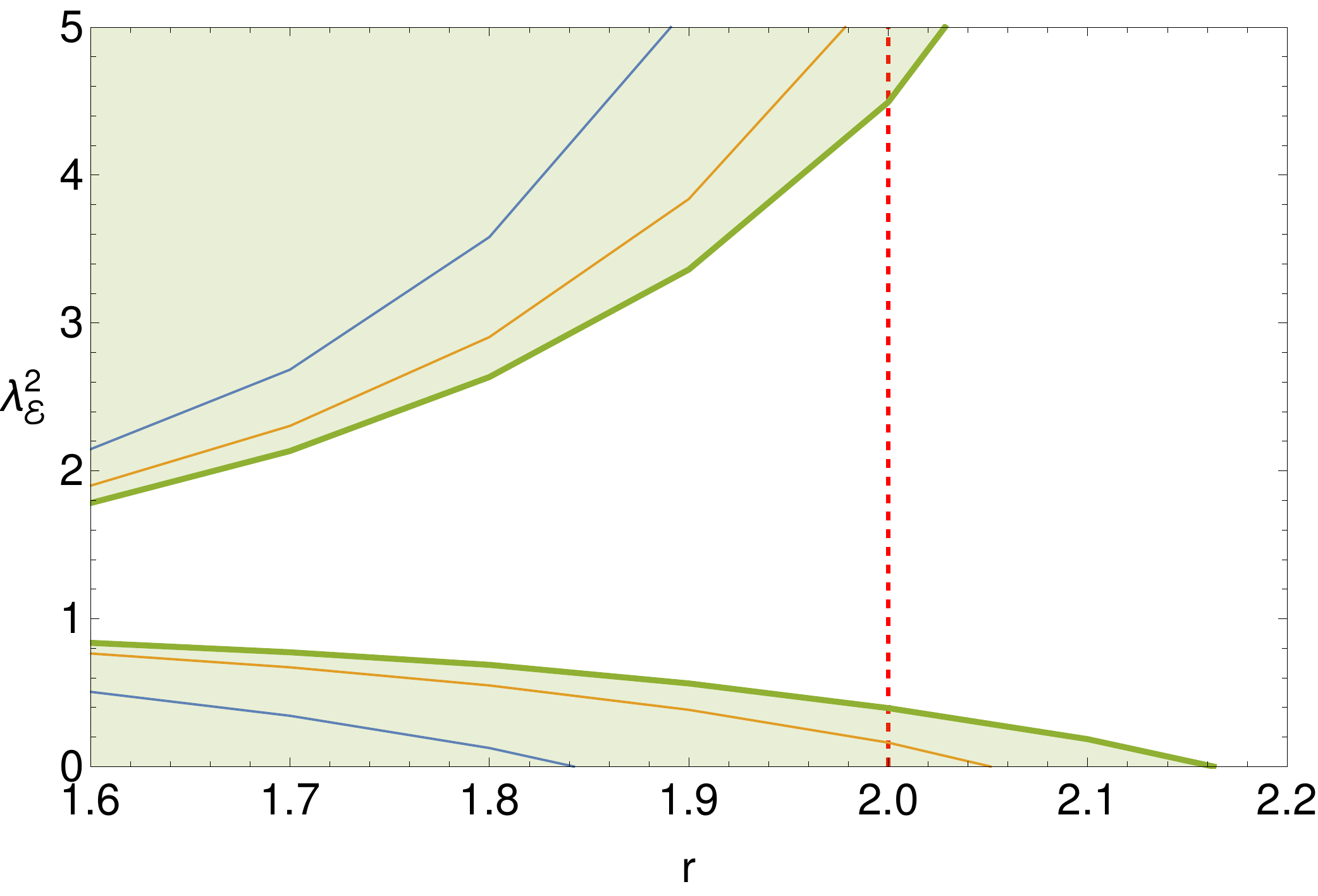}
              \caption{The $r_1=r_2=r$ slice of the bound on Fig. \ref{Fig:E1p2}, for $\Lambda=10,14,18$. The vertical dashed line corresponds to the marginal deformation multiplet $\Em_2$.}
              \label{Fig:Zmetric}
            \end{center}
\end{figure}

One final piece of information that can be extracted from Fig.~\ref{Fig:E1p2} is a bound on the curvature of the conformal manifold $\Mm$. As reviewed in section \ref{sec:fourpt}, the coefficient of the $\Em_{4}$ multiplet can be identified with a component of the Riemann tensor of the Zamolodchikov metric. For rank one theories this translated into constraints on the scalar curvature $R$ as shown in \cite{Beem:2014zpa}; here we have access to the extra component $R_{a \bar{a} b \bar{b}}$. The $r_1 = r_2$ slice of the plot in Fig.~\ref{Fig:E1p2} is shown in Fig.~\ref{Fig:Zmetric}.
This slice describes operators that have the same conformal dimension, but whose OPE in \eqref{EEbarOPE} does not have the identity operator, which implies the two operators are not the same. They can therefore be interpreted as two different relevant deformations.
For $r_1=r_2=2$ the actual bound is
\be 
-3.5 \lesssim R_{a \bar{a} b \bar{b}} \lesssim 0.6\, .
\ee
Our setup is not restricted to conformal manifolds of dimension two, and therefore this bound is valid for any two pair of marginal deformations. In \cite{Beem:2014zpa}, using the single correlator bootstrap an analogous bound was obtained for the $R_{a\bar{a}a\bar{a}}$ component of the curvature. We quote that bound for $\Lambda=18$:\footnote{In \cite{Beem:2014zpa} an analytic bound was given, assuming that the numerics will converge to known solutions to crossing. In order to have a fair comparison between single and mixed correlator results, here we have opted for the numerical result for the same $\Lambda$.}
\be 
-6.3 \lesssim R_{a \bar{a} a \bar{a}} \lesssim 0.3\, .
\ee
As before, this result holds for any deformation and in these coordinates the two components we studied tend to negative values. It would be interesting to use the prescription of \cite{Gerchkovitz:2014gta} in order to check how our bound fares with known theories, although our results are far from their optimal values (see also \cite{Gomis:2015yaa} for a recent discussion on the geometry and topology of $\Mm$).
%!TEX root = ../mixedE.tex

%%%%%%%%%%%%%%%%%%%%%%%%%%%%%%%%%%%%%%%%%%%%%
\subsection{More bounds on OPE coefficients}
%%%%%%%%%%%%%%%%%%%%%%%%%%%%%%%%%%%%%%%%%%%%%

\begin{figure}[htb!]
             \begin{center}            
			  \includegraphics[scale=0.35]{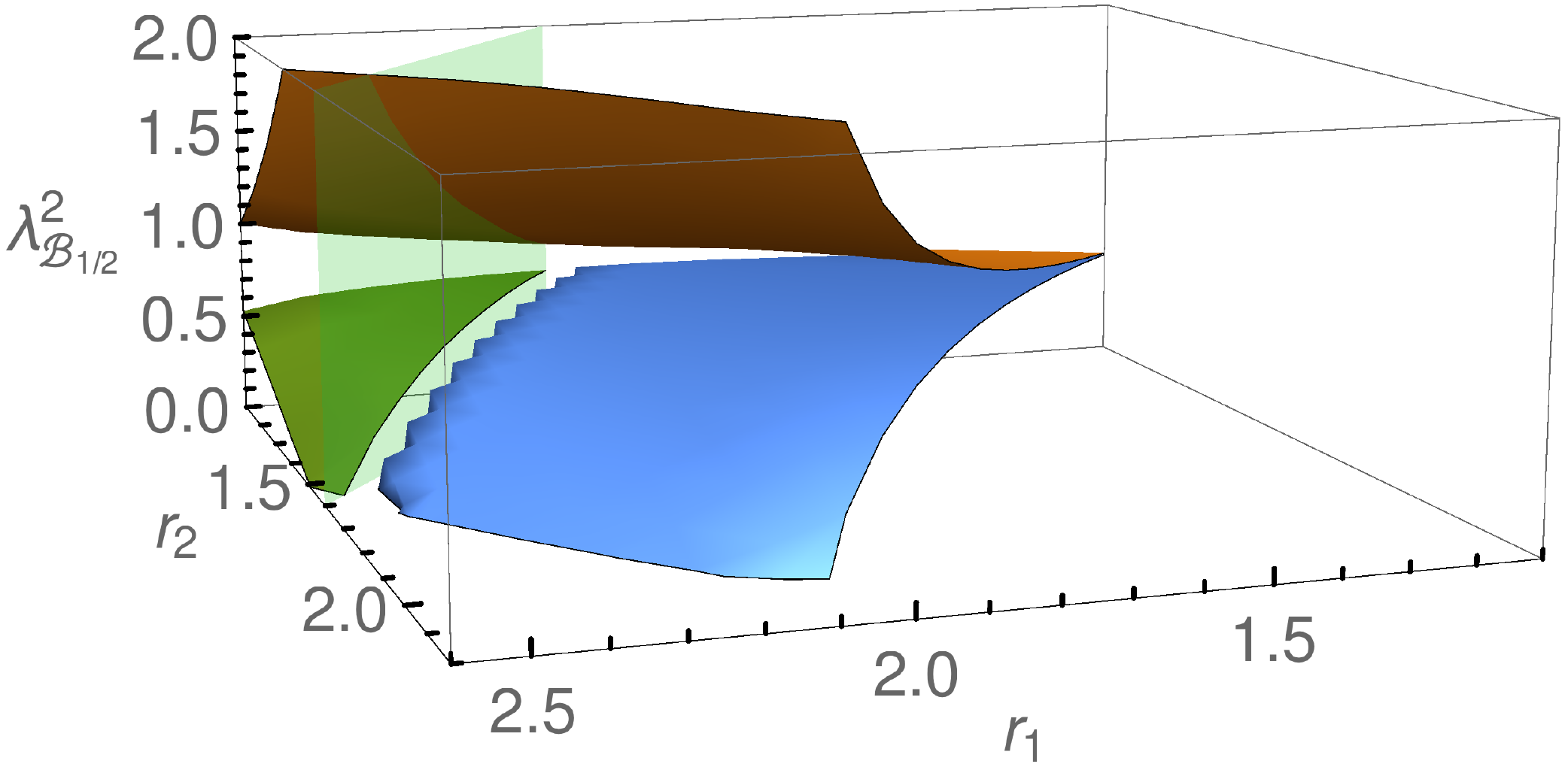}\quad
              \includegraphics[scale=0.35]{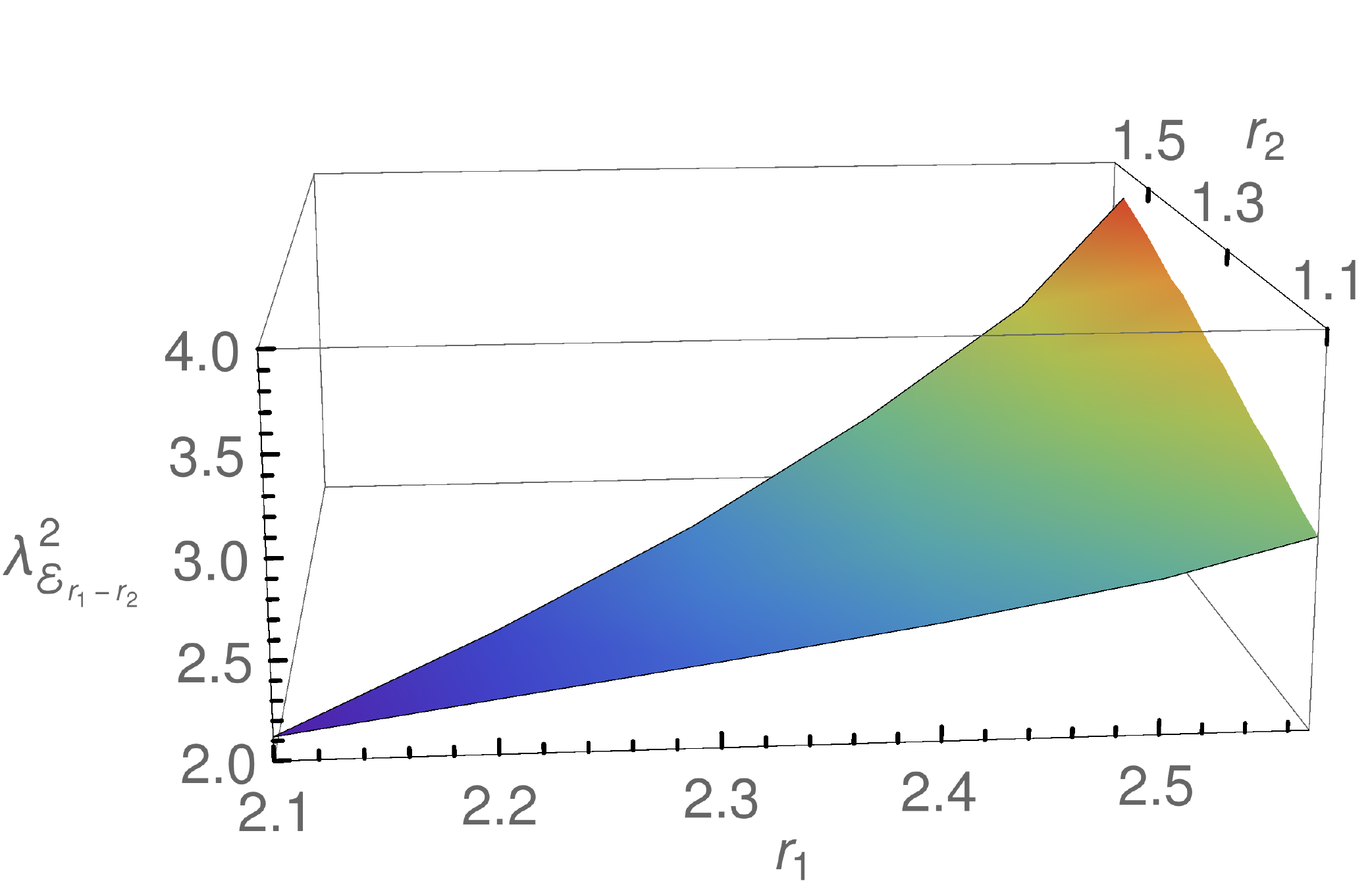}
              \caption{Left: Upper bound on the OPE coefficient of $\Em_{r_1-r_2}$ with $\Lambda=10$. The green wall marks the appearance of the short $\Em_{r_1-r_2}$ multiplet, and the green surface shows the lower bound with this multiplet removed. Right: Upper and lower bound on the OPE coefficient of $\Bm_{\frac12,r_1+r_2-\frac12 (0,\frac12)}$ with $\Lambda=10$. }
              \label{Fig:Bhalf}
            \end{center}
\end{figure}

In this section we present further bounds for OPE coefficients. Let us start with the $\Bm_{\frac12,r_1+r_2-\frac12 (0,\frac12)}$ multiplet. Because it contributes to the $\phi_{r_1} \times \phi_{r_2}$ channel with an odd spin conformal block, this multiplet was not allowed  in the OPE of two identical operators studied in \cite{Beem:2014zpa}. In the mixed correlator system, it appears separated by a gap from the continuum of unprotected multiplets, like the $\Em_{r_1+r_2}$ multiplet above, and we can again obtain upper and lower bounds for its OPE coefficient. For small external dimension there is a narrow region between the two surfaces shown on the left side of Fig.~\ref{Fig:Bhalf}, in which the coefficient is constrained to lie. As usual, although not shown, convergence is very good for small $r$.
Unlike in the previous OPE coefficients it is the lower bound that features a dependence on the presence of the $\Em_{r_1-r_2}$, which as before is allowed by the kinematics for $r_1-r_2 \geqslant 1$ (marked by the green wall in the figure). The lower bound becomes negative slightly before the multiplet is allowed. In fact, we observed that this multiplet was absent in the generalized free field theory cases we considered with $\phi_1 = \phi_2^2$, therefore a positive lower bound cannot be possible as it would rule out a known solution. On the other hand, this multiplet was present in the generalized free field theory solutions we examined for $\phi_1$ and $\phi_2$ independent.
This is consistent with the finding that the lower bound is strictly positive after removing  $\Em_{r_1-r_2}$.

Now we turn to the $\Em_{r_1-r_2}$ multiplet. It is easy to anticipate that the lower bound must be either negative or zero, as any theory for which $\Em_{r_1}$ and $\Em_{r_2}$ are generators will not have this multiplet in the OPE under consideration. Indeed, we find that the bound is always negative, and we show only the upper bound on this  multiplet on the right side of Fig.~\ref{Fig:Bhalf}. The OPE coefficient is only strictly positive if $\Em_{r_1}$ is a composite, appearing in a theory which must include the $\Em_{r_1-r_2}$ and $\Em_{r_2}$ multiplets as generators.

We also plotted the upper and lower bounds for the $\Em_{2r_1}$ and $\Em_{2r_2}$ multiplets but they turned out to be simple translations, along $r_2$ and $r_1$ respectively, of the bound obtained from the single correlator system. We noticed this phenomenon several times: the mixed correlator bootstrap cannot always improve on the single correlator bootstrap. The mixed system is of course bigger, but the functional that we obtain is the same as the single correlator functional, with zeroes in the extra channels. This was also observed for mixed correlators in the Ising model \cite{Kos:2014bka}.

%!TEX root = ../mixedE.tex

%%%%%%%%%%%%%%%%%%%%%%%%%%%%%%%%%%%%%%%%%%%%%
\subsection{Dimension bounds for the non-chiral channel}
%%%%%%%%%%%%%%%%%%%%%%%%%%%%%%%%%%%%%%%%%%%%%

\begin{figure}[htbp!]
             \begin{center}           
              \includegraphics[scale=0.35]{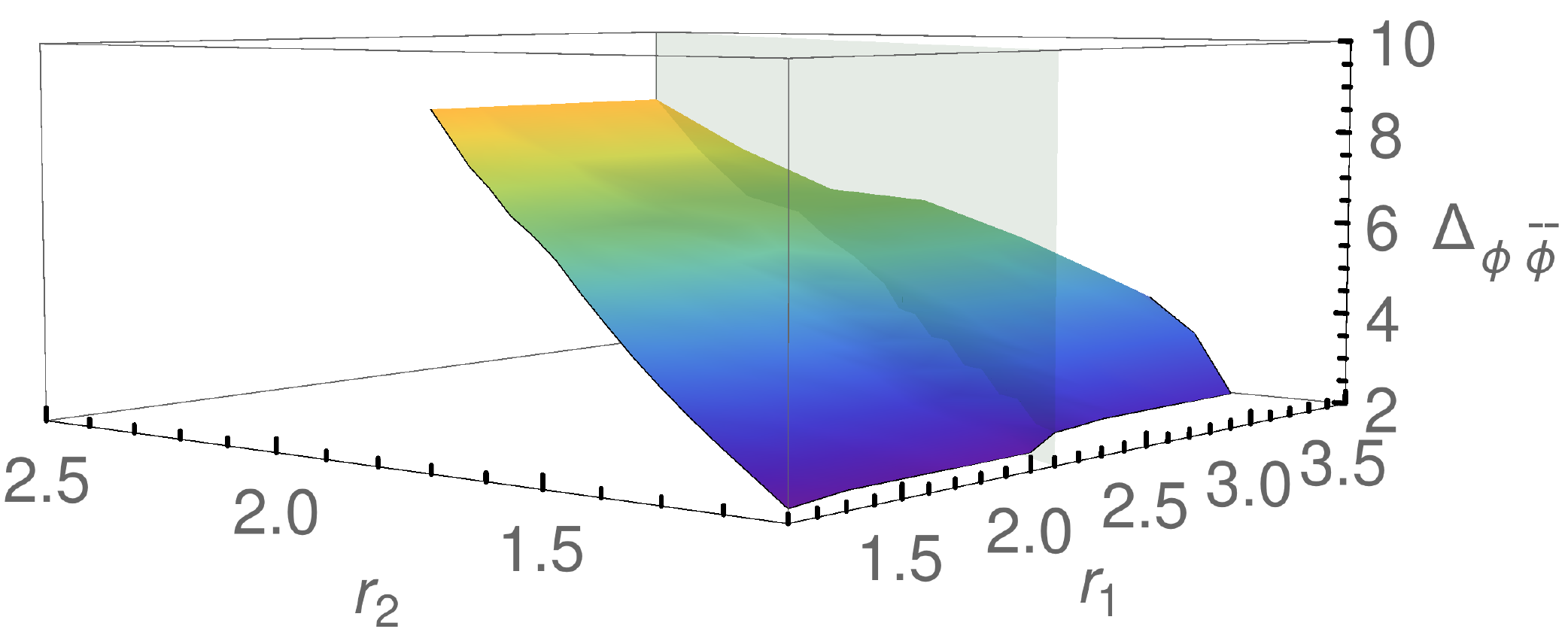} \quad
              \includegraphics[scale=0.35]{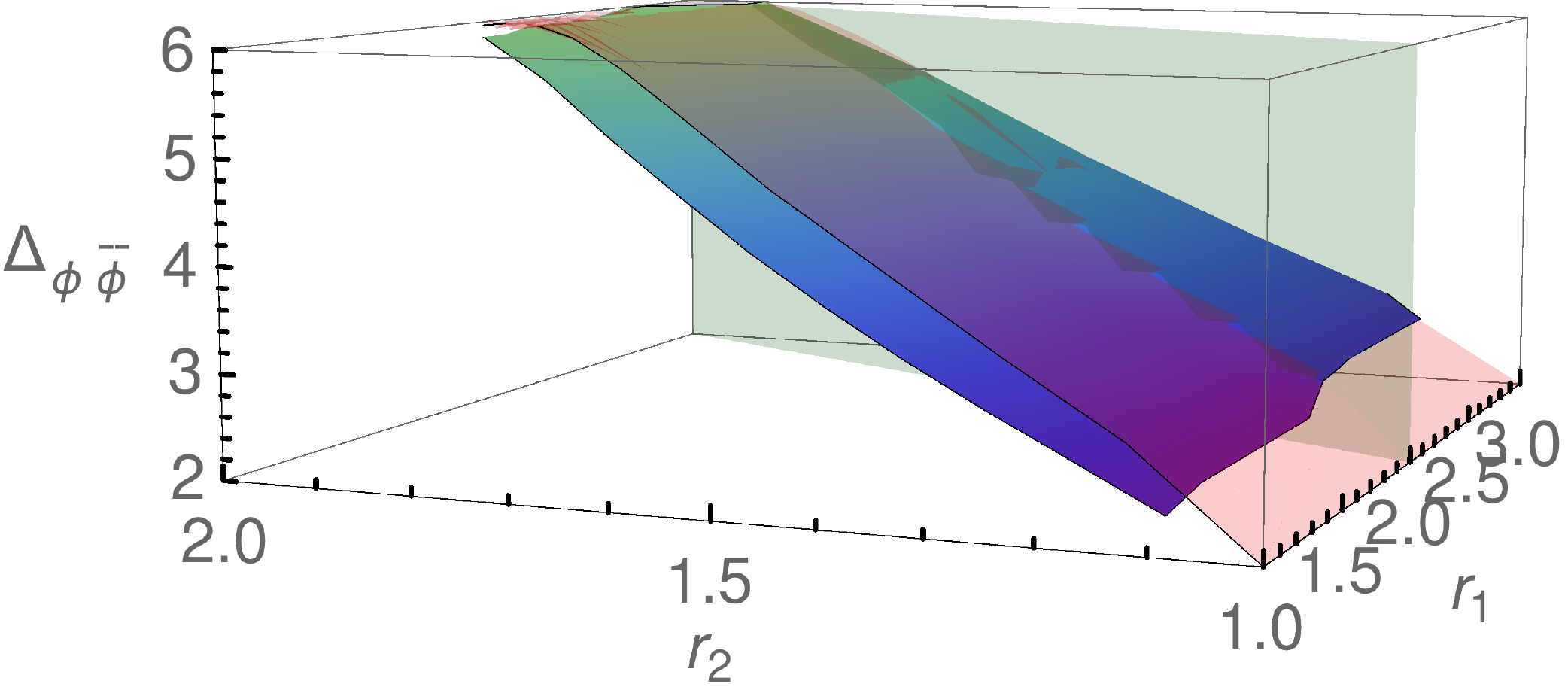}
              \caption{Upper bound on the lowest dimensional scalar long multiplet appearing in the $\phi_{r_1} \times \bar{\phi}_{-r_2}$ OPE at $\Lambda=10$. The vertical green wall marks the appearance of the short $\Em_{r_1-r_2}$ multiplet. On the right plot we have superimposed the bound coming from the single correlator bootstrap in red.}
              \label{Fig:nonchiraldim}
            \end{center}
\end{figure}

The final quantity we study is the scaling dimension of the first non-protected scalar appearing in the $\phi_{r_1} \times \bar{\phi}_{-r_2}$ OPE. The analogous single correlator bound was analyzed in \cite{Beem:2014zpa} leaving the central charge unfixed, and also fixing it to selected values. In the mixed correlator system, scaling dimension bounds are computationally intensive, unlike OPE coefficients, several searches (in the sense described in section \ref{sec:numericalimpl}) are needed in order to locate the bound.
Because of this, we present only one plot with unfixed central and for low $\Lambda$.
We obtained an upper bound for the lowest dimensional scalar long multiplet operator appearing in the $\phi_{r_1} \times \bar{\phi}_{-r_2}$ channel as a function of $r_1$ and $r_2$. This is plotted in Fig.~\ref{Fig:nonchiraldim}, together with the bounds obtained from a single correlator in \cite{Beem:2014zpa} shown as a red surface. For small external dimensions the bound is stronger than the single correlator one, as can be seen on the right side of Fig.~\ref{Fig:nonchiraldim}. While for large external dimensions it asymptotes to the single correlator bound.
As in the previous subsections, the vertical green wall is drawn at $r_1-r_2=1$, after which the multiplet $\Em_{r_1-r_2}$ is allowed, and that is responsible for the jump in the bound along this line. 
The grid is rough and almost no features are discernible in this scale. Upon closer analysis the surface does seem to have some mild ripples, which might be a hint for more interesting physics. However, a proper analysis of these features will necessitate higher $\Lambda$ and a finer grid. We therefore present this plot as a first exploratory step in the mixed correlator bootstrap of scaling dimensions, and leave a more detailed analysis for the future.

%!TEX root = ../mixedE.tex
%%%%%%%%%%%%%%%%%%%%%%%%%%%%%%%%%%%%%%%%%%%%%
\section{Conclusions}
\label{sec:conclusions}
%%%%%%%%%%%%%%%%%%%%%%%%%%%%%%%%%%%%%%%%%%%%%

In this work we have applied the numerical bootstrap program to the four-point function of two chiral operators, and their conjugates, in $\Nm=2$ SCFT. We considered a general setup in which the two chiral operators are not necessarily equal.

For the case in which the two chiral operators are equal, we have attempted to bootstrap the rank one $H_0$ theory with interesting results. We have argued that $H_0$ sits at the kink appearing in Fig.~\ref{Fig:kink}, which corresponds to the vanishing of the $\Bm$ multiplet and thus is consistent with the absence of a mixed branch. Less speculative are our results for OPE coefficients. The dimension of the chiral operator is very low for this theory: $r=\tfrac{6}{5}$, which means we are in the region where the numerics work best. By fixing the central charge to $c=\tfrac{11}{30}$ we were able to obtain the lowest spin $\Cm$ OPE coefficient to very good precision: $\lambda^2_{\Cm} = 0.469(2)$. For this result we used $\Lambda=20$ and it will only improve for higher $\Lambda$. Moreover, the whole family of $\Cm$ multiplets is separated by a gap, and similar results can be obtained for their OPE coefficients.

The rank one theories whose Coulomb branches have dimensions smaller than two are also natural candidates for the analysis of section \ref{sec:H0}, as with these dimensions we are guaranteed to be bootstrapping a generator.
While not as simple as the $H_0$ theory, the rank one $H_1$ and $H_2$ theories also have relatively low central charges, with the conjectured chiral algebras being  relatively simple \cite{Beem:2013sza}.\footnote{Similarly to the $H_0$ case, the conjectured Schur index of these theories matches the vacuum character of the conjectured chiral algebras \cite{Buican:2015ina}.}
While extrapolations of the minimum central charge for the external dimension corresponding to the $H_1$ theory are not as encouraging as the ones we have presented here (and the situation gets worse for $H_2$), fixing the central charge to the correct value, or removing higher spin currents, making use of small gaps in various channels might provide a way out. For these explorations an understanding of what characterizes the central charge bound for the dimensions corresponding to these theories would be a necessary first step. 

The study of the rank $N$ theories that have a generator of dimension $r=\tfrac{6}{5}$ might be possible even from the single correlator standpoint. We know that the left side of the kink shown in Fig.~\ref{Fig:kink_fixedc} is dominated by large central charges, and one could single out these theories by fixing the central charge, and imposing small gaps in the unprotected spectrum.

For the mixed correlator system we obtained mostly OPE coefficient bounds. The most interesting turned out to be the upper and lower bounds for the $\Em_{r_1+r_2}$ multiplet. The $r_1=r_2=2$ point in that plot constrains a component of the Riemann tensor of the conformal manifold, while the projection to the $(r_1,r_2)$ plane gives the region in which Coulomb branch relations are rigorously ruled out. The next step is to bootstrap scaling dimensions. We presented some exploratory results but much remains to be done. It would be remarkable to obtain islands in the parameter space as is the case for the $O(N)$ models \cite{Kos:2013tga}. However, a necessary assumption in order to obtain such an island is the presence of gaps in the spectrum of operators. Their presence in the $O(N)$ model was given a physical motivation, based on the number of relevant operators expected for each universality class. Similar results are harder to obtain in $\Nm=2$ SCFTs, since there is no physical motivation to justify the gaps. One could still assume them and, if an interesting feature appears, justify them a posteriori. Nevertheless, mixed correlator systems are demanding, and this would require significant computational resources.

Although in the mixed correlator section we considered generic bounds valid for all $\Nm=2$ SCFTs, zooming in on specific models is an interesting future direction. A particularly appealing class is the $(A_1,A_{2N})$ family of Argyres-Douglas theories, which only have a Coulomb branch. In particular, the rank two $(A_1,A_4)$ theory has a relatively small central charge and generator dimensions: $(c,r_1,r_2)=(\tfrac{17}{21},\tfrac{8}{7},\tfrac{10}{7})$ \cite{Xie:2012hs,Xie:2013jc}. As we have seen in this paper, it has proved fruitful to combine the numerical bounds with input from other sources, for example, input on the moduli space of vacua \cite{Xie:2014pua}. 

Another obvious continuation of this work is to study further $\Nm=2$ multiplets. An immediate example is the four-point function of stress-tensor multiplets, whose universal character makes it a natural target for the bootstrap. Another interesting system is all possible combinations of mixed correlators between $\Em$ and $\Bm$ multiplets. This will allow us to bootstrap Coulomb, Higgs and mixed branches in one consistent scheme. However, a major obstruction in this setup is the expressions for the superconformal blocks, which are not always known. The selection rules for the stress-tensor multiplet were recently calculated in \cite{Liendo:2015ofa}, but the blocks are still elusive. On the mixed correlator side, this work contains the blocks for mixed Coulomb branch multiplets $\Em$, while \cite{Nirschl:2004pa,Dolan:2004mu} contains relevant expressions for mixed Higgs branch multiplets $\Bm$. Correlators involving $\Em$ and $\Bm$ multiplets together have not yet been studied.

All in all, the $\Nm=2$ superconformal bootstrap is a long-term project, with many analytic challenges and interesting venues of exploration which we expect to revisit in future work.
%!TEX root = ../mixedE.tex

\acknowledgments

We have greatly benefited from discussions with
M. Baggio,
C. Beem,
C.-Y. Ju,
L. Rastelli,
A. Stergiou,
B. van Rees, and
D. Xie.  
The research leading to these results has received funding from the People Programme (Marie Curie Actions) of the European Union’s Seventh Framework Programme FP7/2007-2013/ under REA Grant Agreement No 317089 (GATIS).
The work of M. L. was supported in part by FCT - Portugal, within the POPH/FSE programme, through grant SFRH/BD/70614/2010, and by NSF Grant PHY-1316617. 
P. L. is supported by SFB 647 ``Raum-Zeit-Materie. Analytische und Geometrische Strukturen''.
P.L. acknowledges the hospitality of the Aspen Center for Physics, which is supported by National Science Foundation grant PHY-1066293, and the Back to the Bootstrap 2015 conference at the Weizmann Institute of Science where part of this work was performed. The authors would also like to thank the Simons Summer Workshop 2015 where part of this work was performed.

\appendix
%!TEX root = ../mixedE.tex
%%%%%%%%%%%%%%%%%%%%%%%%%%%%%%%%%%%%%%%%%%%%%
\section{Unitary representations of the \texorpdfstring{$\Nm=2$}{N=2} superconformal algebra}
\label{app:shortening}
%%%%%%%%%%%%%%%%%%%%%%%%%%%%%%%%%%%%%%%%%%%%%

%
\begin{table}[h]
\begin{centering}
\renewcommand{\arraystretch}{1.3}
\begin{tabular}{|l|l|l|}
\hline
\multicolumn{2}{|c|}{Shortening Condition} 												& Multiplet \tabularnewline 
\hline
\hline 
$\Bm^1$								&	$\Delta=2R+r\,, \quad j_1=0$						& $\Bm_{R,r(0,j_2)}$				\tabularnewline
\hline
$\bar\Bm_2$							&	$\Delta=2R-r\,, \quad j_2=0$						& $\bar{\Bm}_{R,r(j_1,0)}$		\tabularnewline
\hline 
$\Bm^1\cap\Bm^2$						&	$\Delta=r\,, \quad R=0$							& $\Em_{r(0,j_2)}$				\tabularnewline
\hline
$\bar\Bm_1\cap\bar\Bm_2$				&	$\Delta=-r\,, \quad R=0$							& $\bar \Em_{r(j_1,0)}$			\tabularnewline
\hline 
$\Bm^1\cap\bar\Bm_{2}$				&	$\Delta=2R\,, \quad j_1=j_2=r=0$					& $\hat{\Bm}_{R}$				\tabularnewline
\hline 
$\Cm^1$								&	$\Delta=2+2j_1+2R+r$								& $\Cm_{R,r(j_1,j_2)}$			\tabularnewline
\hline 
$\bar\Cm_2$							&	$\Delta=2+2 j_2+2R-r$							& $\bar\Cm_{R,r(j_1,j_2)}$		\tabularnewline
\hline
$\Cm^1\cap\Cm^2$						&	$\Delta=2+2j_1+r\,, \quad R=0$					& $\Cm_{0,r(j_1,j_2)}$			\tabularnewline
\hline 
$\bar\Cm_1\cap\bar\Cm_2$				&	$\Delta=2+2 j_2-r\,, \quad R=0$					& $\bar\Cm_{0,r(j_1,j_2)}$		\tabularnewline
\hline 
$\Cm^1\cap\bar\Cm_2$					&	$\Delta=2+2R+j_1+j_2\,, \quad r=j_2-j_1$			& $\hat{\Cm}_{R(j_1,j_2)}$		\tabularnewline
\hline
$\Bm^1\cap\bar\Cm_2$					&	$\Delta=1+2R+j_2\,, \quad r=j_2+1$				& $\Dm_{R(0,j_2)}$				\tabularnewline
\hline 
$\bar\Bm_2\cap\Cm^1$					&	$\Delta=1+2R+j_1\,, \quad -r=j_1+1$				& $\bar\Dm_{R(j_1,0)}$			\tabularnewline
\hline 
$\Bm^1\cap\Bm^2\cap\bar\Cm_2$		&	$\Delta=r=1+j_2\,, \quad r=j_2+1\,, \quad R=0$	& $\Dm_{0(0,j_2)}$ 		 \tabularnewline
\hline
$\Cm^1\cap\bar\Bm_1\cap\bar\Bm_2$	&	$\Delta=-r=1+j_1\,, \quad -r=j_1+1\,, \quad R=0$	& $\bar\Dm_{0(j_1,0)}$\tabularnewline
\hline
\end{tabular}
\par\end{centering}
\caption{Summary of the shortening conditions of the $\Nm=2$ superconformal algebra following \cite{Dolan:2002zh}.}
\label{tab:shortening}
\end{table}
In this appendix we shortly recall the classification of unitary irreducible representations of the four-dimensional $\Nm=2$ superconformal algebra \cite{Dobrev:1985qv,Dolan:2002zh}. We refer the reader to the original papers and to \cite{Beem:2014zpa} for more details.

Representations are built by acting on the superconformal primary, \ie, the operator annihilated by the $\Sm$ and $\bar{\Sm}$ supercharges, with all the $\Qm$, $\bar{\Qm}$ supercharges and the $SU(2)_R$ generators. They are labeled by the quantum numbers of the superconformal primary and by the type of shortening condition they obey, where we follow the naming conventions of \cite{Dolan:2002zh}. 
Generic \emph{long} multiplets, \ie, multiplets that obey no shortening condition, are only constrained by unitarity to obey
\be
\Am^\Delta_{R,r(j_1,j_2)}\,: \qquad \Delta\geqslant {\rm Max}(2+2j_1+2R+ r,2+2j_2+2R- r)\,.
\label{eq:long}
\ee
Then there are two basic types of shortening conditions, sometimes referred to as short and semi-short respectively:
\begin{align}
\Bm^I\, &: \qquad \Qm^I_\alpha \vert \psi \rangle =0\,, \quad & \mathrm{for}\; \alpha=1,2 \nn \\
\Cm^I\, &: \qquad \left\{ 
  \begin{array}{l l}
  \epsilon^{\alpha \beta} \Qm^I_\alpha \vert \psi \rangle_\beta =0\,, \\
  \epsilon^{\alpha \beta} \Qm^I_\alpha \Qm^I_\beta \vert \psi \rangle =0\,,
    \end{array} \right.\quad &
  \begin{array}{l l}
  \mathrm{for}\; j_1\neq 0\\
  \mathrm{for}\; j_1=0   
   \end{array} 
\end{align}
as well as the same conditions with the opposite chirality, identified by a bar.

The various possible combinations of shortening conditions are listed in Tab.~\ref{tab:shortening}, together with the quantum numbers of the superconformal primary and the name given to the multiplet.

When a long multiplet has dimension equal to the unitarity bound of \eqref{eq:long} it is no longer an irreducible representation, and it decomposes into a sum of short multiplets according to
\bea
\label{eq:AChat_dec}
\Am_{R,j_1-j_2 (j_1,j_2)}^{\Delta\to 2R+j_1+j_2+2} &\simeq & \hat\Cm_{R (j_1,j_2)} \oplus \hat\Cm_{R+\tfrac12(j_1-\tfrac12,j_2)}\oplus\hat\Cm_{R+\tfrac12(j_1,j_2-\tfrac12)}\oplus\hat\Cm_{R+1(j_1-\tfrac12,j_2-\tfrac12)}\,,\\
\label{eq:AC_dec}
\Am_{R,r(j_1,j_2)}^{\Delta\to2R+r+2+2j_1} &\simeq & \Cm_{R,r(j_1,j_2)}\oplus \Cm_{R+\tfrac12,r+\tfrac12(j_1-\tfrac12,j_2)}\,,\\
\label{eq:AbC_dec}
\Am_{R,r(j_1,j_2)}^{\Delta\to2R-r+2+2j_2}  &\simeq & \bar\Cm_{R,r(j_1,j_2)}\oplus \bar\Cm_{R+\tfrac12,r-\tfrac12(j_1,j_2-\tfrac12)}\,.
\eea
For small spin the quantum numbers on the right side can become unphysical, and the decompositions are altered. For example, the case relevant for our work is the identification
\be 
\Cm_{R,r(-\tfrac12,j_2)} = \Bm_{R+ \tfrac12, r(0,j_2)}\,.
\label{eq:CisB}
\ee
We refer the reader to \cite{Dolan:2002zh} for a full list of decompositions.

%!TEX root = ../mixedE.tex
%%%%%%%%%%%%%%%%%%%%%%%%%%%%%%%%%%%%%%%%%%%%%
\section{Superconformal blocks}
\label{app:superblocks}
%%%%%%%%%%%%%%%%%%%%%%%%%%%%%%%%%%%%%%%%%%%%%

In this appendix we give further details of the calculation of the superconformal block for the $\phi \times \bar{\phi}$ channel. We will follow the superembedding space setup of \cite{Goldberger:2011yp,Goldberger:2012xb,Fitzpatrick:2014oza}, and obtain the superblock as an eigenfunction of the Casimir operator. 
The superembedding space is defined by coordinates $X_{AB}$ and $\bar{X}^{AB}$ where $A=(\a,\ad,i)$; $\a$ and $\ad$ are Lorentz indices, and $i=1,\ldots,\Nm$ is an $SU(2)_R$ index that counts the number of supersymmetries. In this work we are interested in $\Nm=2$, but as we will see below, the $\Nm=1$ superblock can be easily obtained with this formalism. The superembedding coordinates satisfy,
\begin{align}
(X,\bar{X}) & \sim (\lambda X, \bar{\lambda} \bar{X})\,,
\\
X_{AB} & = -(-1)^{p_A p_B} X_{BA}\, , \quad p_A = \left\{
  \begin{array}{l l}
    0 \quad \text{if} \quad A=\a 
    \\
    1 \quad \text{if} \quad A=i 
  \end{array} \right. \,.
\end{align}
And also,
\be 
\bar{X}^{AB}X_{BC} = 0\, , \quad X_{[AB}X_{C\}D} = 0\, , \quad \bar{X}^{[AB}\bar{X}^{C\}D} = 0\, .
\ee
Superconformal invariants are giving by stringing products of $X$ and $\bar{X}$,
\begin{align}
\langle 1 \bar{2} \rangle & = \bar{X}_2^{AB} X_{1\, BA}\,,
\\
\langle 1 \bar{2} 3 \bar{4} \rangle & = \bar{X}_4^{AB} X_{3\, BC} \bar{X}_2^{CD} X_{1\, DA} (-1)^{p_C}\,.
\end{align}
In terms of four-dimensional coordinates,
\be 
\langle \bar{2} 1 \rangle = -2 (x_{2\, -}-x_{1\,+} + 2i\theta \sigma \bar{\theta})^2\, ,
\ee
which is the standard chiral two-point invariant.

Chiral fields are represented by a holomorphic function $\Phi(X)$ while antichiral fields correspond to anti-holomorphic functions $\Phi(\bar{X})$.
The most general four-point function consistent with superconformal invariance is,
\be 
\langle \Phi_1(X_1) \bar{\Phi}_2(\bar{X}_2) \Phi_2(X_3) \bar{\Phi}_1(\bar{X}_4) \rangle = \frac{\langle 3 \bar{2} \rangle^{\frac{\Delta_1-\Delta_2}{2}} \langle 1 \bar{4} \rangle^{\frac{\Delta_2-\Delta_1}{2}}}{\langle 1 \bar{2} \rangle^{\frac{\Delta_1+\Delta_2}{2}}\langle 3 \bar{4} \rangle^{\frac{\Delta_1+\Delta_2}{2}}} H(u,v)\,,
\ee
where $u$ and $v$ are two superconformal invariants
\be 
\frac{\langle 1 \bar{2} 3 \bar{4} \rangle}{\langle 1 \bar{4} \rangle \langle 3 \bar{2}\rangle}=\frac{-1+u+v}{4v}\, ,
\qquad 
\frac{\langle 1 \bar{2} \rangle \langle 3 \bar{4} \rangle }{\langle 1 \bar{4} \rangle \langle 3 \bar{2} \rangle } = \frac{u}{v}\, .
\ee
After acting with Casimir (see \cite{Fitzpatrick:2014oza} for details) the function $H(u,v)$ must satisfy,
\be 
C_{\Nm} H(u,v) = \left(\ell(\ell+2) + \Delta (\Delta - 4 + 2\Nm) - \frac{\Nm}{2}\alpha(\Delta_1-\Delta_2)^2\right) H(u,v)\,,
\ee
where we have kept the number of supersymmetries $\Nm$ arbitrary. The standard $r$-charge conventions are such that the coefficient $\alpha$ takes the values $\tfrac{2}{3}$ and $1$ for $\Nm=1$ and $\Nm=2$ theories respectively.
Projecting to four-dimensions and writing $H(u,v)=v^{\frac{\Delta_2-\Delta_1}{2}}\Gm_{\Delta_{12}}(u,v)$ we obtain:
\be 
\Dm_{\text{DO}} \Gm(u,v) = \left(\ell(\ell+2) + \Delta (\Delta - 4 + 2\Nm) \right) \Gm(u,v)\,,
\ee
where $\Dm_{\text{DO}}$ is the differential operator found by Dolan and Osborn in \cite{Dolan:2003hv}. Following their notation, it depends on three parameters $a$, $b$, and $c$ which in our case take the values,
\be 
a = \frac{\Delta_2-\Delta_1}{2}\, , \quad b = \frac{\Delta_2-\Delta_1+2 \Nm}{2}\, , \quad c = \Nm \,,
\ee
and the solution is
\begin{align}
\label{crossedsuperblockN}
\begin{split}
\Gm_{\Delta_{12}}(u,v) & = \frac{z \bar{z}}{z-\bar{z}}\left( k_{\Delta+\ell}(z) k_{\Delta-\ell-2}(\bar{z})-z \leftrightarrow \bar{z} \right)
\\
k_{\beta}(z) & = z^{\frac{\beta}{2}}{_2}F_1\left(\frac{\beta-\Delta_{12}}{2},\frac{\beta-\Delta_{12}+2\Nm}{2},\beta+\Nm;z\right)\,.
\end{split}
\end{align}
This solution captures the contributions of chiral fields with unequal dimensions in $\Nm=1$ and $\Nm=2$ theories, while for $\Nm=0$ reproduces the standard bosonic block:
\begin{align}
g_{\Delta,\ell}  & = \frac{z \bar{z}}{z-\bar{z}}\left( \kappa_{\Delta+\ell}(z) \kappa_{\Delta-\ell-2}(\bar{z})-z \leftrightarrow \bar{z} \right)\, , \quad 
\kappa_{\beta}(z) = z^{\frac{\beta}{2}}{_2}F_1\left(\frac{\beta-\Delta_{12}}{2},\frac{\beta-\Delta_{12}}{2},\beta;z\right)\, .
\label{eq:bos_block}
\end{align}
As a check on our result we can work out the expansion of the superconformal block in terms of bosonic blocks. Indeed $\Gm$ can be written as 
\begin{align} 
\nn
\Gm_r(u,v) & = g_{\Delta,\ell} + a_{1}g_{\Delta + 1,\ell-1}+ a_{2}g_{\Delta + 1,\ell+1}+a_{3}g_{\Delta + 2,\ell-2}+a_{4}g_{\Delta + 2,\ell}
\\
& \,+a_{5}g_{\Delta + 2,\ell+2}+a_{6}g_{\Delta + 3,\ell-1}+a_{7}g_{\Delta + 3,\ell+1}+a_{8}g_{\Delta + 4,\ell}\,,
\end{align}
where 
\allowdisplaybreaks
\begin{footnotesize}
\begin{align}
\label{acoeffs}
a_{1}  & = \frac{(\ell+2+r-\Delta)(\ell+2-r-\Delta)}{2(\ell+2-\Delta)(\ell-\Delta)}\,,
\\ \nn
a_{2}  & = \frac{(\ell+r+\Delta)(\ell+r-\Delta)}{2(\ell+2+\Delta)(\ell+\Delta)}\,,
\\ \nn
a_{3}  & = \frac{(\ell + r - \Delta) (\ell + 2 + 
   r - \Delta) (-\ell - 2 + 
   r + \Delta) (-\ell + r + \Delta)}{16 (-\ell - 
   1 + \Delta) (-\ell + \Delta)^2 \
(-\ell + 1 + \Delta)}\,,
\\ \nn
a_{4}  & = \frac{(\ell - r + \Delta) (\ell + 2 + 
   r - \Delta) (\ell + 2 - 
   r - \Delta) (\ell + r + \Delta)}{4 (-\ell - 
   2 + \Delta) (-\ell + \Delta) (\ell + \
\Delta) (\ell + 2 + \Delta)}\,,
\\ \nn
a_{5}  & = \frac{(\ell + 2 - r + \Delta) (\ell - 
   r + \Delta) (\ell + 
   r + \Delta) (\ell + 2 + r + \Delta)}{16 (\ell + 1 + \Delta) (\ell + 
   2 + \Delta)^2 (\ell + 3 + \Delta)}\,,
\\ \nn
a_{6}  & = \frac{(\ell - r + \Delta) (\ell + 
   r - \Delta) (\ell + 2 + 
   r - \Delta) (-\ell - 2 + 
   r + \Delta) (-\ell + 
   r + \Delta) (\ell + r + \Delta)}{32 (-\ell - 
   1 + \Delta) (-\ell + \Delta)^2 \
(-\ell + 
   1 + \Delta) (\ell + \Delta) (\ell + 
   2 + \Delta)}\,,
\\ \nn
a_{7}  & = \frac{(\ell + 2 - r + \Delta) (-\ell + 
   r - \Delta) (\ell + 2 + 
   r - \Delta) (-\ell - 2 + 
   r + \Delta) (\ell + 
   r + \Delta) (\ell + 2 + r + \Delta)}{32 (-\ell - 
   2 + \Delta) (-\ell + \Delta) (\ell + 
   1 + \Delta) (\ell + 
   2 + \Delta)^2 (\ell + 3 + \Delta)}\,,
\\ \nn
a_{8}  & = \frac{(\ell + 2 - r + \Delta) (\ell - 
   r + \Delta) (\ell + 
   r - \Delta) (\ell + 2 + 
   r - \Delta) (\ell + 2 - 
   r - \Delta) (\ell - 
   r - \Delta) (\ell + 
   r + \Delta) (\ell + 2 + r + \Delta)}{256 (-\ell - 
   1 + \Delta) (-\ell + \Delta)^2 \
(-\ell + 1 + \Delta) (\ell + 
   1 + \Delta) (\ell + 
   2 + \Delta)^2 (\ell + 3 + \Delta)}\,,
\end{align}
\end{footnotesize}
and where we have used the $\Nm=2$ relation $r=r_1-r_2 = \Delta_1 - \Delta_2=$ for the $r$-charge of the field being exchanged.
At the unitarity bound $\Delta =\ell + 2 + r$, the multiplet shortens
\begin{align}
\label{eq:short_scb_in_bb}
\begin{split}
\Gm_r(u,v) &= g_{\ell+2+r,\ell} + \frac{2 (1 + \ell) (1 + r + \ell)}{(2 + r + 2 \ell) (4 + r + 2 \ell)}g_{\ell+3+r,\ell+1} \\
&+\frac{(1 + \ell) (2 + \ell) (1 + r + \ell) (2 + 
   r + \ell)}{(3 + r + 2 \ell) (4 + r + 2 \ell)^2 (5 + r + 2 \ell)}g_{\ell+4+r,\ell+2}\,,
\end{split}
\end{align}
and we obtain the superconformal block associated to the $\Cm_{r(\frac{\ell}{2},\frac{\ell}{2})}$ multiplet.
Finally, for $\ell=0$ and $\Delta=r$ we obtain
\be
\nn
\Gm_r(u,v) = g_{r,0}\,,
\ee
which represents the contribution of the $\Em_{r}$ multiplet. All in all, the solution \eqref{crossedsuperblockN} for $\Nm=2$ encodes the superconformal block for all the multiplets contributing to the $\phi_{r_1} \times \bar{\phi}_{-r_2}$ channel.
%!TEX root = ../mixedE.tex
%%%%%%%%%%%%%%%%%%%%%%%%%%%%%%%%%%%%%%%%%%%%%
\section{Crossing equations}
\label{app:crossingeqs}
%%%%%%%%%%%%%%%%%%%%%%%%%%%%%%%%%%%%%%%%%%%%%

In this section we write down the crossing equations for a single correlator used in section \ref{sec:res_single}, and those used for the system of mixed correlators considered in section \ref{sec:res_mixed}.
As explained in detail in \cite{Kos:2014bka}, in order to have positivity when studying a mixed correlator of the type
\be
\langle \bar \phi_1 \phi_1 \bar \phi_2 \phi_2 \rangle\,,
\ee
we must also consider the two correlators with a single type of operator,
\be
\langle \bar \phi_i \phi_i \bar \phi_i \phi_i \rangle \quad \mathrm{with}\; i=1,2\,.
\ee
The crossing equations for a single correlator of the type $\langle \bar \phi_i \phi_i \bar \phi_i \phi_i \rangle$ were obtained in \cite{Beem:2014zpa}, and are reproduced here for convenience:
\be
\sum_{\Om \in \phi_i \phi_i} \vert \lambda_{\phi_i \phi_i \Om} \vert^2
\begin{bmatrix}
\mp (-1)^\ell F^{ii,ii}_{\pm, \Delta,\ell}(u,v)\\
0
\end{bmatrix}
+
\sum_{\Om \in \phi_i \bar \phi_i} \vert \lambda_{\phi_i \bar \phi_i \Om} \vert^2
\begin{bmatrix}
(-1)^\ell\tilde{\Fm}^{11,11}_{\pm, \Delta,\ell}\\
 \Fm^{11,11}_{-, \Delta,\ell}
\end{bmatrix} = 0\,,
\label{eq:single_corr_cross}
\ee
where the first line in the equation encodes two separate crossing equations, differing only by the signs indicated.
In the above equation we have defined
\bea
F^{ij,kl}_{\pm, \Delta,\ell} &\equiv& v^{\frac{\Delta_k + \Delta_j}{2}} g^{\Delta_{ij}, \Delta_{k,l}}_{\Delta,\ell}(u,v) \pm u^{\frac{\Delta_k + \Delta_j}{2}} g^{\Delta_{ij}, \Delta_{k,l}}_{\Delta,\ell}(v,u)\,,\\
\Fm^{ij,kl}_{\pm, \Delta,\ell} &\equiv& v^{\frac{\Delta_k + \Delta_j}{2}} \Gm^{\Delta_{ij}, \Delta_{k,l}}_{\Delta,\ell}(u,v) \pm u^{\frac{\Delta_k + \Delta_j}{2}} \Gm^{\Delta_{ij}, \Delta_{k,l}}_{\Delta,\ell}(v,u)\,,\\
\tilde{\Fm}^{ij,kl}_{\pm, \Delta,\ell} &\equiv& v^{\frac{\Delta_k + \Delta_j}{2}} \tilde{\Gm}^{\Delta_{ij}, \Delta_{k,l}}_{\Delta,\ell}(u,v) \pm u^{\frac{\Delta_k + \Delta_j}{2}}  \tilde{\Gm}^{\Delta_{ij}, \Delta_{k,l}}_{\Delta,\ell}(v,u)\,.
\eea

The full set of crossing symmetry equations for the mixed correlator system consists of \eqref{eq:single_corr_cross} with  $i=1,2$, as well as the ones given in \eqref{eq1},~\eqref{eq2},~and~\eqref{eq3}, together with the same equations with $u \leftrightarrow v$.
All in all, combining these equations we find (where for shortness we combine lines that differ only by a sign in a single line)
\allowdisplaybreaks
\begin{align}
&\sum_{\Om \in \phi_1 \phi_2} \vert \lambda_{\phi_1 \phi_2 \Om} \vert^2
\begin{bmatrix}
\mp (-1)^\ell  F^{21,21}_{\pm, \Delta,\ell}(u,v)\\
0\\
\mp  F^{12,21}_{\pm, \Delta,\ell}(u,v)\\
0\\
0\\
0\\
0\\
\end{bmatrix}
+ \sum_{\Om \in \phi_1 \phi_1} \vert \lambda_{\phi_1 \phi_1 \Om} \vert^2
\begin{bmatrix}
0\\
0\\
0\\
\mp (-1)^\ell F^{11,11}_{\pm, \Delta,\ell}(u,v)\\
0\\
0\\
0
\end{bmatrix}\nn\\
& +  \sum_{\Om \in \phi_2 \phi_2} \vert \lambda_{\phi_2 \phi_2 \Om} \vert^2
\begin{bmatrix}
0\\
0\\
0\\
0\\
0\\
\mp (-1)^\ell F^{22,22}_{\pm, \Delta,\ell}(u,v)\\
0
\end{bmatrix}
+ \sum_{\Om \in \phi_1 \bar \phi_2}  \vert \lambda_{\phi_1 \bar \phi_2 \Om} \vert^2
\begin{bmatrix}
(-1)^\ell \tilde{\Fm}^{21,21}_{\pm, \Delta,\ell}(u,v)\\
 \Fm^{12,21}_{\pm, \Delta,\ell}(u,v)\\
0\\
0\\
0\\
0\\
0\\
\end{bmatrix}\nn\\
& + \sum_{\Om \in \phi_i \bar \phi_i} \left( \lambda_{\phi_1 \bar \phi_1 \Om}^\ast \quad \lambda_{\phi_2 \bar \phi_2 \Om}^\ast \right) 
\begin{bmatrix}
\begin{pmatrix}
0 & 0 \\
0 & 0 
\end{pmatrix}\\
\begin{pmatrix}
0 & \mp \frac{1 }{2}  \Fm^{11,22}_{\pm, \Delta,\ell} (u,v) \\
\mp \frac{1 }{2}  \Fm^{11,22}_{\pm , \Delta,\ell}(u,v) & 0 \\
\end{pmatrix}\\
\begin{pmatrix}
0 & \frac{(-1)^\ell }{2}  \tilde{\Fm}^{11,22}_{\pm, \Delta,\ell} (u,v) \\
\frac{(-1)^\ell }{2}  \tilde{\Fm}^{11,22}_{\pm , \Delta,\ell}(u,v) & 0 \\
\end{pmatrix}\\
\begin{pmatrix}
(-1)^\ell\tilde{\Fm}^{11,11}_{\pm, \Delta,\ell} & 0\\
0 & 0 
\end{pmatrix}\\
\begin{pmatrix}
 \Fm^{11,11}_{-, \Delta,\ell} & 0\\
0 & 0 
\end{pmatrix}\\
\begin{pmatrix}
0 & 0\\
0 & (-1)^\ell \tilde{\Fm}^{22,22}_{\pm  , \Delta,\ell}
\end{pmatrix}\\
\begin{pmatrix}
0 & 0\\
0 &  \Fm^{22,22}_{-, \Delta, \ell} 
\end{pmatrix}
\end{bmatrix}
\begin{pmatrix}
\lambda_{\phi_1 \bar \phi_1 \Om }\\
\lambda_{\phi_2 \bar \phi_2 \Om}
\end{pmatrix}=0\,.
\label{eq:fullcrossing}
\end{align}
In the above equation we can separate the contributions of the stress tensor multiplet and the identity, which give
\be
\overrightarrow{V}_{\mathrm{fixed}} = 
\begin{bmatrix}
0\\
\mp \Fm^{11 22}_{\pm , \Delta=0,\ell=0}(u,v) \\
\tilde \Fm^{11 22}_{\pm , \Delta=0,\ell=0}(u,v) \\
\tilde \Fm^{11 11}_{\pm , \Delta=0,\ell=0}(u,v)\\
\Fm^{11 11}_{- , \Delta=0,\ell=0}(u,v) \\
\tilde \Fm^{22 22}_{\pm , \Delta=0,\ell=0}(u,v) \\
\Fm^{22 22}_{- , \Delta=0,\ell=0}(u,v)
\end{bmatrix} +
\begin{Bmatrix}
\frac{1}{72}\;\text{if}\; \Nm=1 \\
\frac{1}{6}\;\text{if}\; \Nm=2
\end{Bmatrix} 
\begin{bmatrix}
0\\
\mp \frac{\Delta_1 \Delta_2}{c} \Fm^{11 22}_{\pm , \Delta=4-\Nm,\ell=2-\Nm}(u,v)\\
(-1)^\Nm  \frac{\Delta_1 \Delta_2}{c} \tilde \Fm^{11 22}_{\pm , \Delta=4-\Nm,\ell=2-\Nm}(u,v)\\
(-1)^\Nm  \frac{\Delta_1^2}{c} \tilde \Fm^{11 11}_{\pm , \Delta=4-\Nm,\ell=2-\Nm}(u,v)\\
 \frac{\Delta_1^2}{c} \Fm^{11 11}_{- , \Delta=4-\Nm,\ell=2-\Nm}(u,v)\\
(-1)^\Nm \frac{\Delta_2^2}{c} \tilde \Fm^{22 22}_{\pm , \Delta=4-\Nm,\ell=2-\Nm}(u,v)\\
 \frac{\Delta_2^2}{c} \Fm^{22 22}_{- , \Delta=4-\Nm,\ell=2-\Nm}(u,v)
\end{bmatrix}\,.
\label{eq:id+st}
\ee
Since the stress-tensor multiplet is at the long multiplet unitarity bound, the central charge will only be completely fixed if we impose a gap in the appropriate channel. Similarly, demanding the absence of higher spin currents $\hat{\Cm}_{0,\ell}$ is only relevant if we have gaps in the corresponding channels.

\bibliography{./aux/biblio}
\bibliographystyle{./aux/JHEP}

\end{document}